\documentclass[11pt]{article}
\usepackage{amsmath,amsthm,latexsym,amssymb,amsfonts,epsfig}

\makeatletter
\@addtoreset{equation}{section}
\makeatother

\newtheorem{Theorem}{Theorem}[section]

\newtheorem{Corollary}{Corollary}[section]

\def\be{\begin{equation}}
\def\ee{\end{equation}}
\def\ba{\begin{eqnarray}}
\def\ea{\end{eqnarray}}
\def\Nl{{\mathchoice
{\setbox0=\hbox{$\displaystyle\rm N$}\hbox{\hbox to0pt
{\kern0.4\wd0\vrule height0.9\ht0\hss}\box0}}
{\setbox0=\hbox{$\textstyle\rm N$}\hbox{\hbox to0pt
{\kern0.4\wd0\vrule height0.9\ht0\hss}\box0}}
{\setbox0=\hbox{$\scriptstyle\rm N$}\hbox{\hbox to0pt
{\kern0.4\wd0\vrule height0.9\ht0\hss}\box0}}
{\setbox0=\hbox{$\scriptscriptstyle\rm N$}\hbox{\hbox to0pt
{\kern0.4\wd0\vrule height0.9\ht0\hss}\box0}}}}
\def\Zl{{\mathchoice
{\setbox0=\hbox{$\displaystyle\rm Z$}\hbox{\hbox to0pt
{\kern0.4\wd0\vrule height0.9\ht0\hss}\box0}}
{\setbox0=\hbox{$\textstyle\rm Z$}\hbox{\hbox to0pt
{\kern0.4\wd0\vrule height0.9\ht0\hss}\box0}}
{\setbox0=\hbox{$\scriptstyle\rm Z$}\hbox{\hbox to0pt
{\kern0.4\wd0\vrule height0.9\ht0\hss}\box0}}
{\setbox0=\hbox{$\scriptscriptstyle\rm Z$}\hbox{\hbox to0pt
{\kern0.4\wd0\vrule height0.9\ht0\hss}\box0}}}}
\def\Ql{{\mathchoice
{\setbox0=\hbox{$\displaystyle\rm Q$}\hbox{\hbox to0pt
{\kern0.4\wd0\vrule height0.9\ht0\hss}\box0}}
{\setbox0=\hbox{$\textstyle\rm Q$}\hbox{\hbox to0pt
{\kern0.4\wd0\vrule height0.9\ht0\hss}\box0}}
{\setbox0=\hbox{$\scriptstyle\rm Q$}\hbox{\hbox to0pt
{\kern0.4\wd0\vrule height0.9\ht0\hss}\box0}}
{\setbox0=\hbox{$\scriptscriptstyle\rm Q$}\hbox{\hbox to0pt
{\kern0.4\wd0\vrule height0.9\ht0\hss}\box0}}}}
\def\Rl{{\mathchoice
{\setbox0=\hbox{$\displaystyle\rm R$}\hbox{\hbox to0pt
{\kern0.4\wd0\vrule height0.9\ht0\hss}\box0}}
{\setbox0=\hbox{$\textstyle\rm R$}\hbox{\hbox to0pt
{\kern0.4\wd0\vrule height0.9\ht0\hss}\box0}}
{\setbox0=\hbox{$\scriptstyle\rm R$}\hbox{\hbox to0pt
{\kern0.4\wd0\vrule height0.9\ht0\hss}\box0}}
{\setbox0=\hbox{$\scriptscriptstyle\rm R$}\hbox{\hbox to0pt
{\kern0.4\wd0\vrule height0.9\ht0\hss}\box0}}}}
\def\Cl{{\mathchoice
{\setbox0=\hbox{$\displaystyle\rm C$}\hbox{\hbox to0pt
{\kern0.4\wd0\vrule height0.9\ht0\hss}\box0}}
{\setbox0=\hbox{$\textstyle\rm C$}\hbox{\hbox to0pt
{\kern0.4\wd0\vrule height0.9\ht0\hss}\box0}}
{\setbox0=\hbox{$\scriptstyle\rm C$}\hbox{\hbox to0pt
{\kern0.4\wd0\vrule height0.9\ht0\hss}\box0}}
{\setbox0=\hbox{$\scriptscriptstyle\rm C$}\hbox{\hbox to0pt
{\kern0.4\wd0\vrule height0.9\ht0\hss}\box0}}}}
\def\Hl{{\mathchoice
{\setbox0=\hbox{$\displaystyle\rm H$}\hbox{\hbox to0pt
{\kern0.4\wd0\vrule height0.9\ht0\hss}\box0}}
{\setbox0=\hbox{$\textstyle\rm H$}\hbox{\hbox to0pt
{\kern0.4\wd0\vrule height0.9\ht0\hss}\box0}}
{\setbox0=\hbox{$\scriptstyle\rm H$}\hbox{\hbox to0pt
{\kern0.4\wd0\vrule height0.9\ht0\hss}\box0}}
{\setbox0=\hbox{$\scriptscriptstyle\rm H$}\hbox{\hbox to0pt
{\kern0.4\wd0\vrule height0.9\ht0\hss}\box0}}}}
\def\Ol{{\mathchoice
{\setbox0=\hbox{$\displaystyle\rm O$}\hbox{\hbox to0pt
{\kern0.4\wd0\vrule height0.9\ht0\hss}\box0}}
{\setbox0=\hbox{$\textstyle\rm O$}\hbox{\hbox to0pt
{\kern0.4\wd0\vrule height0.9\ht0\hss}\box0}}
{\setbox0=\hbox{$\scriptstyle\rm O$}\hbox{\hbox to0pt
{\kern0.4\wd0\vrule height0.9\ht0\hss}\box0}}
{\setbox0=\hbox{$\scriptscriptstyle\rm O$}\hbox{\hbox to0pt
{\kern0.4\wd0\vrule height0.9\ht0\hss}\box0}}}}

\title{
{\sf On the Relation between}\\
{\sf Operator Constraint --, Master Constraint --,}\\
{\sf Reduced Phase Space -- and Path Integral Quantisation}
}
\author{
{\sf Muxin Han$^{1,3}$}\thanks{{\sf Muxin.Han@aei.mpg.de}}, {\sf T. 
Thiemann$^{1,2,3}$}
\thanks{{\sf 
thiemann@aei.mpg.de,tthiemann@perimeterinstitute.ca,
thiemann@theorie3.physik.uni-erlangen.de}}\\
\\
{\sf $^1$ MPI f. Gravitationsphysik, Albert-Einstein-Institut,} \\
{\sf Am M\"uhlenberg 1, 14476 Potsdam, Germany}\\
\\
{\sf $^2$ Institut f\"ur Theoretische Physik III, 
Friedrich Alexander Universit\"at Erlangen -- N\"urnberg,}\\
{\sf Staudtstrasse 7/B2, 91058 Erlangen, Germany}\\
\\
{\sf $^3$ Perimeter Institute for Theoretical Physics,}\\ 
{\sf 31 Caroline Street N, Waterloo, ON N2L 2Y5, Canada}
}
\begin{document}

\maketitle

\begin{abstract}
{\sf
Path integral formulations for gauge theories must start from the 
canonical formulation in order to obtain the correct measure. 
A possible avenue to derive it is to start from the reduced phase space
formulation. In this article we review this rather 
involved procedure in full generality. Moreover, we 
demonstrate that the reduced phase space path integral formulation 
formally agrees with the Dirac's operator 
constraint quantisation and, more specifically, with the Master 
constraint
quantisation for first class constraints. For first class
constraints with non trivial structure functions the equivalence can 
only be established by passing to Abelian(ised) 
constraints which is always possible locally in phase space. 
Generically, the correct configuration space path integral measure 
deviates from the exponential of the Lagrangian action. The corrections
are especially severe if the theory suffers from second class secondary 
constraints. In a companion paper we compute these corrections for the 
Holst and Plebanski formulations of GR on which current spin foam models 
are based.   
}
\end{abstract}

\newpage

\tableofcontents

\newpage

\section{Introduction}
\label{s1}

Path integrals for scalar Quantum Field Theories 
(QFT) on Minkowski space are supposed to compute the S -- Matrix for the 
Hamiltonian in question. A standard Folklore says that heuristically 
one should simply consider all ``paths'' between some initial and final 
scalar field configuration on a spatial hypersurfaces labelled by 
moments of time $t_i,t_f$ respectively and integrate over
the exponential of ($i$ times) the action with ``Lebesgue measure'' in 
order to obtain the evolution kernel. More specifically, let $\Omega$ be a 
(normalised) ground state (provided it exists) for the Hamiltonian $H$
on a Hilbert space $\cal H$,
let $Q$ be the configuration space of spatial scalar field 
configurations with ``configuration Lebesgue measure'' $dq$, let
${\cal Q}_{t_i,t_f}=\times_{t\in[t_i,t_f]} Q$ be the set of paths, let
$[Dq]=\prod_{t\in[t_i,t_f]} dq_t$ the ``path Lebesgue measure'',  
then 
\be \label{1.1}
<\psi_f,\;e^{i(t_f-t_i)H/\hbar}\;\psi_i>_{{\cal H}}
=\frac{\int_{{\cal Q}_{t_i,t_f}}\;[Dq]\;
\overline{\psi_f[q(t_f)]}\;\psi_i[q(t_i)]\;e^{iS_{t_i,t_f}[q,\dot{q}]/\hbar}
} 
{\int_{{\cal Q}_{t_i,t_f}}\;[Dq]\;
\overline{\Omega[q(t_f)]}\;\Omega[q(t_i)]\;e^{iS_{t_i,t_f}[q,\dot{q}]/\hbar}
}
\ee
and $S_{t_i,t_f}$ is the classical (Lorentzian) action integrated over the 
time interval $[t_i,t_f]$. Here by Lorentzian action we mean the Legendre 
transform
\be \label{1.2}  
S[q,\dot{q}]:={\rm extr}_p\;\{\int_{t_i}^{t_f}\; dt[p\dot{q}-H(p,q)]\}
\ee
where $p$ denotes the momentum conjugate to $q$ and $H$ the Hamiltonian.
This ``theorem'' is wrong for several reasons.

First of all, from the 
mathematical point of view, there is no Lebesgue measure on infinite 
dimensional spaces. Therefore one would like to consider $[DQ] 
\exp(iS/\hbar)$ as a (complex) measure on $\cal Q$ but this does not
work because the modulus of a complex measure \cite{2} is supposed to be 
normalisable which is obviously not the case here. If the Hamiltonian is 
bounded from below, it is therefore 
much more promising to consider, instead of the unitary group
$\mathbb{R}\to {\cal B}({\cal H});\; t\mapsto \exp(it H/\hbar)$ the 
contraction semigroup
$\mathbb{R}_+\to {\cal B}({\cal H});\; t\mapsto \exp(-t H/\hbar)$.
Here ${\cal B}({\cal H}$ denotes the algebra of bounded operators on 
$\cal H$.   
Under these circumstances another Folklore theorem states 
that  
\be \label{1.3}
<\psi_f,\;e^{-(t_f-t_i)H/\hbar}\;\psi_i>_{{\cal H}}
=\frac{\int_{{\cal Q}_{t_i,t_f}}\;[Dq]\;
\overline{\psi_f[q(t_f)]}\;\psi_i[q(t_i)]\;e^{-S^E_{t_i,t_f}[q,\dot{q}]/\hbar}
}  
{\int_{{\cal Q}_{t_i,t_f}}\;[Dq]\;
\overline{\Omega[q(t_f)]}\;\Omega[q(t_i)]\;e^{-S^E_{t_i,t_f}[q,\dot{q}]/\hbar}}
\ee
where now $S^E_{t_i,t_f}$ denotes the ``Euclidean'' action, that is,
the Legendre transform\footnote{Usually one obtains the Euclidian action 
by Wick rotation $t\to it$. However, we insist on this definition  
because it does not rely on an analytic structure of the fields in the 
time coordinate which is not justified anyway. Our definition is 
formally correct also in circumstances where the Hamiltonian is not
only quadratic in the momenta with constant coefficients, see below.}    
\be \label{1.4}  
S^E[q,\dot{q}]:={\rm extr}_p\;\{\int_{t_i}^{t_f}\; dt[ip\dot{q}-H(p,q)]\}
\ee
While even under these circumstances the partition function
\be \label{1.5}
Z:=\int_{{\cal Q}} \; [Dq]\; e^{-S^E/\hbar}\;  
\overline{\Omega[q(t_f)]}\;\Omega[q(t_i)]
\ee
diverges, under fortunate circumstances it is possible to assign to 
$e^{-S^E}\;[Dq]\;/Z$ a well defined measure theoretic meaning on a proper
$\sigma-$algebra $\cal Q$ (with repect to which $S^E$ is usually not even 
measurable). Whenever (\ref{1.3}) can be made rigorous, it is called the 
Feynman -- Kac formula \cite{3,4}.

However, as it is well known \cite{5}, also from the physical point of 
view, (\ref{1.1}) or (\ref{1.3}) are wrong in general. This is because
the strict derivation (see e.g. \cite{6}) of, say (\ref{1.3}) requires a 
skeletonisation of
the time interval $[t_i,t_f]$ and corresponding resolutions of the 
identity in terms of (generalised) position and momentum eigenvectors.   
That is to say, a priori one has to consider the complex hybrid action 
\be \label{1.6}
S^{\mathbb{C}}[q,p]:=\int_{t_i}^{t_f}\; dt[ip\dot{q}-H(p,q)]
\ee
which is integrated over both momentum and configuration coordinates.
If $H$ depends on $p$ only quadratically with constant coefficients, then 
one can perform the Gaussian integral and up to an (infinite) constant
which drops out in the fraction (\ref{1.3}) one arrives at the Folklore
result. However, in more general situations the result is different.
For instance, the Hamiltonian could still be quadratic in $p$ but with $q$ 
dependent coefficients which leads to a nontrivial modification of the 
``measure'' $[Dq]$. More generally, however, the Hamiltonian may not be 
quadratic or even analytic in $p$ in which case an exact configuration
space path integral representation is not available, only a saddle point 
approximation is available (plus the corresponding perturbative 
treatment of the non Gaussian corrections). Notice that the saddle point 
approximation 
and subsequent integrating out of the momentum variables reproduces
(\ref{1.4}) up to a non trivial measure factor.

So far we have only considered scalar QFT on Minkowski space and even 
here we saw that the only correct derivation of the path integral 
proceeds via the Hamiltonian formulation, as stressed for instance in 
\cite{5}. Additional technical and conceptual complications arise when we 
consider gauge theories 
and/or other background spacetime metrics. The simplest problem occurs for  
Yang -- Mills type of gauge theories: Here the 
action is gauge invariant and if the measure is anomaly free (is gauge 
invariant as well) then one should divide by the (in general infinite) 
volume of the gauge group in order to give sense to both numerator and 
denominator in (\ref{1.3}). If one considers QFT on non stationary 
background spacetimes then no natural Hamiltonian and vacuum exists 
\cite{8} and the conceptual status of the path integral as a means to 
calculate scattering amplitudes becomes veiled. Even more veiled the 
situation becomes for totally constrained systems such as General 
Relativity on spatially compact four manifolds admitting globally 
hypebolic metrics when there is no true Hamiltonian at all.
In this case certainly also the notion of a Wick rotation breaks down
which on Minkowski space allows to reconstruct the Lorentzian Wightman 
functions from the Euclidian Schwinger functions via the 
Osterwalder -- Schrader reconstruction theorem \cite{1}. Parts 
of the reconstruction theorem,
namely the construction of a Hamiltonian and a Hilbert space from a 
measure 
satisfying a natural background independent generalisation of the
OS axioms, can be generalised to background independent theories 
\cite{9}. 

It transpires that especially in the context of realistic physical 
theories, that is, General Relativity coupled to (standard) matter,
it is neither clear what the heuristic Ansatz (\ref{1.1}) or 
(\ref{1.3}) computes nor whether it is the correct formula for what it
is supposed to do.  
One possibility to deal with these problems is to try to solve the 
constraints classically and then to quantise the reduced phase space 
equipped with the (pull back of the) Dirac bracket \cite{6}. This 
can be done in two ways. The first option is to impose suitable 
gauge fixing conditions in order to 
render the sytem totally second class and then to quantise the 
corresponding pull back of the Dirac bracket together with the induced 
reduced Hamiltonian. The second option is to determine explicitly 
a sufficient number of Dirac observables and to quantise the symplectic 
structure induced by the Dirac bracket. While for rare examples 
independent means exist to determine those gauge invariants, for most 
systems the only practical way to determine a sufficient number of 
Dirac observables is via a choice of gauge fixing. Namely, as we will 
review in the next section (see also e.g. the appendix in the second 
reference of \cite{27}), there is a one to one correspondence between 
a 
choice of gauge fixing and a preferred set of gauge invariant functions
which generate the full algebra of gauge invariant functions. In that 
sense the two methods, gauge fixing and this so called relational 
approach, are completely equivalent. The method is physically very 
interesting because it not only provides a suitable algebra of gauge 
invariant objects but also a gauge invariant Hamiltonian which drives 
the time evolution of those invariants. Here the question of equivalence 
between different choices of gauge fixing arises. As we will review in 
the next section, the preferred algebras of invariants that one obtains 
via different choices of gauge fixing are isomorphic. Of course they 
differ in their physical interpretation but as Poisson algebras they are 
isomorphic, the physical quantum kinematics is not affected by the 
choice of gauge fixing. The difference arises in the physical 
Hamiltonian, that is, in the quantum dynamics. The explicit form of the 
physical Hamiultonian as a function of the invariant generators of the 
algebra of gauge invariant functions depends absolutely sensitively 
on the choice of gauge fixing and therefore even classically 
the evolution of the invariants will differ drastically from each 
other for different choices. For some choices the Hamiltonian may 
be explicitly time independent and leads to a conservative reduced 
system, for others it may not be. Even more crucial the choice of 
gauge fixing becomes in the quantum theory. Already for finite 
dimensional systems, depending on the choice of gauge fixing the 
physical Hamiltonian and other composite invariants built from the 
generators 
of the gauge invariant algebra may have discrete or continuous spectrum
\cite{Bianca}! 
Notice that we here talk about composite invariants that  
have the interpretation of a given non invariant $f$ measured in terms 
of another non invariant $T$ (the so called clock). If we change $T$
to $T'$ then its spectrum may switch from continuous to discrete or vice 
versa even though we talk about the same $f$ and about the same 
Hilbert space representation!.
In infinite dimensional situations the choice of gauge fixing has an 
even stronger influence for not only do we have to find a representation
of the generators of the algebra of observables but in addition that 
algebra should support the physical Hamiltonian. One way to read Haag's
theorem \cite{31} is that Hamiltonians with different interaction terms 
cannot be implemented on the same Hilbert space. Thus generically 
different choices of gauge fixing will force us to choose different 
representations. For instance one may want to construct a cyclic 
representation built from application of the generators to a vacuum 
(ground state of the Hamiltonian). That vacuum of course depends on the 
Hamiltonian and even for free field theories those cyclic 
representations are typically unitarily inequivalent. In case that the 
physical Hamiltonian is explicitly time dependent, one is in addition
confronted with the usual problem of QFT on curved spacetimes, namely 
that one 
has to decide at which point of time one wants to select a vacuum 
vector. 

All of this certainly strongly affects the resulting reduced phase space 
path integral because it is based on the selected Hilbert space 
representation and the transition amplitudes between physical states do 
depend on the physical Hamiltonian. For any such choice, the path 
integral
then does have the interpretation of (\ref{1.1}) or (\ref{1.3}) in terms 
of the reduced Hamiltonian. 

An additional complication that we have not 
mentioned yet is the case of a system with second class constraints.
Such a system is to be canonically quantised with respect to the Dirac 
bracket rather than the Poisson bracket. Typically the Dirac bracket
destroys the canonically conjugacy of the global coordinates of the 
phase space that one started from. Since to find representations of such 
complicated Poisson algebras is usually prohibitively difficult one 
is forced to switch to local Darboux coordinates (by means of a 
canonical transformation with respect to the original Poisson bracket)
which is always possible locally \cite{6}. Such coordinates may be very 
difficult to find in practice. Assuming this to have done nevertheless, 
one can then 
construct the reduced phase space using a choice of gauge fixing as 
already described above and after having chosen a Hilbert space 
representation subordinate to that gauge fixing, the 
transition amplitudes in terms of the induced physical Hamiltonian.

From here on then mostly one proceeds rather formally. One assumes 
that one can choose a Schr\"odinger representation based on the 
reeduced Darboux configuration space. By using well 
known skeletonisation techniques one then basically writes the 
transition 
amplitude between initial and final states $\Psi_i,\Psi_f$, as a path 
integral over the reduced Darboux phase space, replacing the reduced 
Hamiltonian operator by its classical function which results in the 
exponential of the reduced Hamiltonian Darboux action. In order to make 
contact with (\ref{1.1}) one wants to rewrite this path integral as a 
path integral over the unreduced, original configuration space and in 
terms of the original Lagrangian. As is well known, this can be 
formally done and we will review this rather involved procedure in 
section \ref{s3}. Basically one first extends the reduced Darboux phase 
space 
to the unreduced Darboux phase space thereby introducing $\delta$ 
distributions of the constraints and the gauge fixing condition as well 
as measure factors which cancel the Jacobian that arises when solving 
the $\delta$ distributions. One then observes that, in presence of the 
$\delta$ distributions 
the reduced Hamiltonian action can be written as the unreduced 
symplectic potential, in terms of the unreduced Darboux coordinates. 
Interestingly, the measure factors
and the $\delta$ distributions combine in just the right way
as to make the resulting expression independent of the gauge fixing 
condition when considered as a measure on gauge invariant functions.
This is similar to the Fadeev -- Popov theorem \cite{6} and we will
review this result in section \ref{s3}. This seems to be in 
contradiction 
to what 
we 
have said above about the dependence of the transition amplitudes 
on the gauge fixing condition. The resolution is that at this point the 
integral is {\it not} over gauge invariant functions, it is an integral 
over $\Psi_i\;\overline{\Psi_f}$ which are functions at initial and 
final points of time of the reduced Darboux coordinates which are 
not gauge invariant. More generally, in applications to scattering 
theory, we may be interested also in n-point functions so that the 
path integral is over functions of the reduced Darboux coordinates also
at intermediate times (in fact we will use the method of a generating 
functional so that there is a dependence on the reduced Daroux 
coordinates at {\it all} times). One may, in the presence of the 
$\delta$ distributions, extend the non gauge invariant, reduced Darboux 
coordinates to gauge invariant functions which use the chosen gauge 
fixing condition. However, these extended functions now display a 
complicated dependence on all unreduced Darboux coordiantes which makes 
this extension practically useless. Even if one did perform the 
extension, while one can now change the dependence on the gauge fixing 
condition in the measure, one cannot get rid of it in the gauge 
invariantly extended functions\footnote{A special situation arises 
if one considers gauge transformations that tend to the identity 
in the infinite past and future and that the only non gauge invariant 
functions in the path integral are located at the infinite past and 
future. This is not the case for the n-point functions or the generating 
functional but for the rigging kernel between two kinemtical states. 
Now the dependence on the gauge fixing formally disappears from 
the path integral, of course modulo the representation theoretic caveats 
that we have mentioned.}.  
In 
any case, one next 
performs the 
canonical 
transformation that leads from the Darboux coordinates back to the
original canonical coordinates which does not affect the symplectic 
potential and the Liouvile measure but it affects the initial and final 
states.  
Then one exponentiates the 
$\delta$ functions and, by the technique introduced in \cite{10},
gets rid of the secondary second class constraints which leads to 
further changes in the measure. Finally, one integrates out the momenta.
This is only possible if the reduced Darboux configuration coordinates, 
as 
functions 
of the original canonical coordinates, do not depend 
on the original momenta and if they do not leads in general to further 
changes in the measure while now the exponential of the covariant 
Lagrangian action appears.

The point of mentioning these in principle well known facts is twofold.
The first is that we wish to stress that even if all the assumptions 
that we have listed can be verified, the correct Langragian 
configuration space measure may differ drastically from the naive one in 
(\ref{1.1}). These deviations depend crucially on the dynamical content 
of the theory and cannot be discarded. The second point that we want to 
make is the dependence of the transition amplitudes on the chosen gauge 
fixing. This dependence is at first astonishing because one is used from 
Yang -- Mills theory that the path integral does not depend on the 
gauge fixing and it even sounds dangerous because it seems as if this 
dependence implies that gauge invariance is broken. However, this is not
the case: The dependence on the gauge fixing {\it is physically 
correct}. The reason is that in generally covariant systems the dynamics 
mixes with gauge invariance. In Yang -- Mills theory this is not the 
case, there one has a gauge invariant Hamiltonian at one's disposal 
which is not generated by a gauge fixing condition, it is simply there 
without further input. Gauge invariant functions in Yang -- Mills 
theory can also be easily constructed without ever mentioning any gauge 
fixing, for instance Wilson loops or flux tubes between quarks. The 
gauge fixing condition comes in only when cancelling an otherwise 
infinite constant. This introduces a gauge fixing $\delta$ 
distribution and a Fadeev -- Popov determinant into the measure whose 
combination is independent of the gauge fixing by construction, similar 
as in our discussion above. In contrast, in generally covariant systems 
a gauge fixing condition can be seen as {\it defining a preferred 
algebra of observables and a preferred dynamics thereof}. Gauge 
invariance is not at all broken, the dynamical system consisting 
of reduced Darboux phase space and 
reduced Hamiltonian as defined by a gauge fixing is in one to one 
correspondence with a dynamical system consting of a preferred algebra 
of Dirac observables and a gauge invariant physical Hamiltonian defined 
via the same gauge fixing (now interpreted as a choice of clocks).
The two descriptions are equivalent. The gauge fixing dependence 
comes 
in because one needs a gauge fixing in order to arrive at the very 
notion of a dynamics, or in other words, at the very notion of an 
observer. This observer dependence of the classical and quantum theory 
has already been stressed in \cite{11} and will be discussed in more 
detail in \cite{29}. Let us stress again, as we have already said, that 
similar as in Yang -- 
Mills theories the gauge fixing dependence of the {\it measure} 
disappears 
when we restrict it (as a linear functional) to gauge invariant 
functions. However, the choice of those gauge invariant functions 
themselves and the corresponding physical Hamiltonian, in other words 
the physical interpretation of the theory, induced by a choice of gauge 
fixing (clock) is what makes 
the description gauge choice dependent. In contrast, in Yang -- Mills 
theories such a 
choice of clocks is not necessary in order to arrive at useful gauge 
invariant functions. In principle, the generators of the algebra of 
gauge invariant functions for one choice of gauge fixing can be written 
as complicated functions of the generators for any other choice. 
However, this involves an infinite series of commutator functions about 
whose convergence nothing is known and which therefore is practically 
useless if not mathematically ill --defined.\\
\\
In this paper we 
want to illustrate the complications sketched above for a general theory 
which will be the first result of this 
paper. While certainly bits and pieces of our description appear in 
various places in the literature, we hope that assembling them in the 
form presented here may add a certain amount of clarity to the question 
how reduced phase space and path integral quantisation fit together.\\
\\
The second result of this paper will be to sketch how the path integral 
is related to Dirac's operator constraint quantisation \cite{7} and a 
particular incarnation of it, the so called Master Constraint Programme
\cite{12} for first class systems. As already mentioned, the reduced 
phase space rarely admits 
a global Darboux coordinate system and hence a quantisation of the 
unreduced phase space is much simpler. The price to pay is that one has 
to impose them as non -- anomalous operators on that Hilbert space in 
order 
to compute the physical Hilbert space. There are certain heuristic 
group averaging methods \cite{13} available in the literature which, as 
the name 
suggests, apply when the constraints form a Lie algebra. 
If they do not (structure functions), then not only are the constraints 
difficult to define 
without anomalies because of factor ordering difficulties but also group 
averaging is not applicable. It is for 
that reason that the Master Constraint Programme (MCP) was introduced. 
In the MCP, all constraints 
are enconded into one single Master constraint. The Master Constraint 
is a classically equivalent platform and is automatically free of 
anomalies so that group averaging (or direct integral decomposition)
methods apply.  

The central ingredient of the group averaging method is a (generalised) 
``projector'' (or rigging map) from the kinematical Hilbert space into 
the physical one, 
equipped with an associated physical inner product. It can be expressed 
in terms of a path integral which in case of a true Lie algebra is 
readily recognised as (\ref{1.1}) or (\ref{1.3}) respectively. In case 
of the Master constraint that can also be established, however, the 
proof is somewhat more involved. Not surprisingly, the key to the 
understanding of how all 
of these methods fit together is how the reduced phase space description 
arises from the constraints and a suitable gauge fixing condition which 
in turn allows for a local Abelianisation of the constraints. 
It may seem astonishing that the gauge fixing condition enters the 
interpretation of the physical Hilbert space in such a prominent way. 
The reason for why that happens is that the physical Hilbert space 
can be considered as the closure of the set of vectors that one obtains 
by applying the algebra of gauge invariant observables to a cyclic 
physical state. However, the construction of that algebra and the 
interpretation of its elements is faciliated by considering the gauge 
invariant extension of the kinematical algebra as induced by a gauge 
fixing condition. In other words, while in the operator constraint 
method one only deals with manifestly gauge invariant objects, their
interpretation again relies on a gauge fixing condition or equivalently
on a choice of rods and clocks. Different such choices result in the 
same algebra but its generators (elementary observables) differ for each
choice.\\
\\
To summarise:\\
The correct path integral formula and its interpretation can only be 
obtained by following the Hamiltonian path, otherwise one misses 
important corrections to the measure. In the context of spin foam models 
\cite{14} for Loop Quantum Gravity \cite{15} this has been pointed 
already in \cite{16} (see also \cite{16a,21}). The corrections to the 
measure are not manifestly  
covariant as first indicated in \cite{17} but seem to be required in 
order to maintain at least some form of spacetime covariance as claimed 
in \cite{18}. 
They should therefore be taken seriously in any realistic spin foam 
model for General Relativity. Work is now in progress which tries to
implement these corrections. See also \cite{21} where the covariance 
of the path integral with respect to the Bergmann -- Komar group is 
discussed.\\
\\
This article is organised as follows.\\
\\
In section two we review Dirac's analysis of gauge sytems and the 
relation between gauge fixing and a gauge invariant description.

In section three we derive the path integral from the quantisation 
of the reduced phase space based on the natural generators of the 
algebra of invariants defined by a choice of gauge fixing conditions.

In section four we derive the same path integral starting from the 
unreduced (with respect to the first class constraints) phase space
and implement the rigging map. In case that the first class algebra of 
constraints involves structure functions, using the rigging map 
technique requires to pass to new constraints that form an algebra.
This is always (locally) possible because one can always (locally) 
Abelianise constraints.

In section five we use as an alternative route the MCP and show that 
again one arrives at the same path integral. This is to be expected 
because both constraint rigging and Master constraint rigging should 
provide the generalised projector on physical states, however, the 
technical mechanism by which this works is somewhat involved.

Finally in section six we summarise and conclude.

\section{Classical Preliminaries: Gauge Fixing Versus Gauge Invariant 
Formulation}
\label{s2}

In an attempt to make this article self -- contained we start with the 
classical theory. We will neeed the corresponding notation anyway
for the path integral formulation. First we summarise the main 
ingredients of Dirac's algorithm. Then we display the relation between
the reduced phase space of gauge invariant observables and the pull 
back phase space as induced by a gauge fixing. As we will see, the 
two formulations are equivalent for suitable choices of gauge fixing. 

\subsection{Brief Review of Dirac's Algorithm}
\label{s2.1}

We consider a theory with Lagrangian $L(q^a(t),v^a(t))$
and corresponding action 
\be \label{2.1}
S=\int_{\mathbb{R}}\;dt\; L(q^a(t),v^a(t))  
\ee
Here the index $a$ takes values in a general set which may 
comprise discrete and/or continuous labels. We are interested in a theory
with gauge symmetries so that the Lagrangian will be singular, that is, 
we cannot solve all the velocities $v^a=\dot{q}^a$ for the 
canonical momenta 
\be \label{2.2}
p_a:=\frac{\partial L}{\partial v^a}
\ee
By solving a maximal number of velocities $v^\alpha$ (whose number is 
equal 
to the rank of the matrix $\partial^2/\partial v^a\partial 
v^b$) in terms of the momenta $p_a$ and the remaining velocities 
$v^i$, that is, $v^\alpha=u^\alpha(q^a,p_a;v^i)$ such that 
$(v^a)=(v^\alpha,v^i)$ (i.e. the indices $\alpha$ and $i$ take values 
in index sets that partition the index set associated with $a$) we 
obtain 
the primary constraints
\be \label{2.3}
C_i=p_i-[\frac{\partial L}{\partial v_i}]_{v^\alpha=u^\alpha}
\ee
which does not depend on the $v^i$ by assumption of the maximality of 
the the $v^\alpha$. The canonical Hamiltonian
\be \label{2.4}
H_c=[v^a p_a-L(q,v)]_{v^\alpha=u^\alpha}
\ee
always has the structure \cite{7,15}
\be \label{2.5}
H_c=H'_0(q,p)+v^i C_i(q,p)
\ee
that is, it is an affine function of the $v^i$. 

The further analysis of the system is now governed by Dirac's algorithm
\cite{7}:\\
One requires that the constraints are preserved by the Hamiltonian flow 
of $H_c$. Whenever $\{H_c,C_i\}=0$ is not satisfied on the constraint 
surface, there are two possibilities: Either 1. $\{H_c,C_i\}$ does not 
involve the velocities $v^i$ or 2. it does. In the first case we must add 
$\{H_c,C_i\}$ to the list of constraints in the second we solve all the 
equations of type 2 for some of the velocities $v^i$ (assuming that 
the system of equations is not overdetermined). Iterating like this,
one ends up, in general, with further constraints $C_I$, 
which are called secondary constraints, and the velocities are restricted 
to be of the form $v^i=v^i_0(q,p)+\lambda^m v^i_m(q,p)$. Here
$v^i=v^i_0$ solves $\{H_c,C_j\}=\{H_c,C_J\}=0$ for all $j,J$ on the 
constraint 
surface $\{C_j=C_J=0\forall j,J\}$ and $(v^i_m)_m$ is a maximal linearly 
independent set of solutions of the system  
$v^i \{C_i,C_j\}=v^i \{C_i,C_J\}=0$ for all $j,J$ (on the constraint 
surface). The coefficients $\lambda_m$ are free and phase space 
independent.  

It follows that the 
\be \label{2.6}
F_m:=v^i_m C_i
\ee
are first class constraints, i.e. they weakly (i.e. on the 
constriant surface) Poisson commute with all 
constraints. By taking linear combinations of the constraints $C_i,C_I$
(with phase space dependent coefficients) we isolate a maximal number of 
first class constraints. The constraints $F_m$ are called primary first 
class constraints, the additional ones $F_M$ are called secondary 
first class constraints. The remaining constraints among the set 
$(C_i,C_I)$ which are linearly independent of the set $(F_\mu):=(F_m,F_M)$ 
are called second class constraints and are denoted by $(S_\Sigma)$.

The canonical Hamiltonian can now be written
\be \label{2.7}
H_c=H^{\prime\prime}_0+\lambda^m F_m
\ee
where 
\be \label{2.8}
H^{\prime\prime}_0=H_0'+v_0^i C_i
\ee
is of first class by construction. It may therefore be an affine function 
of the $F_\mu$ (with phase space dependent coefficients)
\be \label{2.9}
H^{\prime\prime}_0=H_0+f^\mu F_\mu
\ee
The piece $H_0$ is referred to as the true Hamiltonian because it is not 
constrained to vanish. In totally constrained systems such as General 
Relativity it vanishes identically, that is, the canonical Hamiltonian is 
a linear combination of first class constraints. Notice that only the 
primary first class constraints appear in the canonical Hamiltonian with 
arbitrary coefficients $\lambda^m$ and so one would associate gauge 
invariance only with respect to them. However, this is in general 
inconsistent because the Poisson algebra of primary first class 
constraints generically does not close, only the full set of first class 
constraints always does. In other words, since the Poisson bracket 
between two first class functions is first class and if $O$ is weakly 
invariant under 
the $F_m$ then also $2\{F_{[m},\{F_{n]},O\}\}=\{\{F_m,F_n\},O\}$
should be weakly zero. Furthermore, the time evolution of $O$ 
with respect to $H_c$ should be gauge invariant which gives 
\be \label{2.10}
\{F_m,\{H_c,O\}\}=-\{H_c,\{O,F_m\}\}-\{O,\{F_m,H_c\}\}\approx
\{\{F_m,H^{\prime\prime}_0\},O\}\approx 0
\ee
Hence gauge invariant quantities 
should be those that weakly Poisson commute with the minimal 
subset of first class constraints generated by the Poisson brackets 
between the $\{H_c,F_m\}$ (and higher order brackets with $H_c$) and 
between the 
first class primary constraints among each other.  
For most systems of physical interest this exhausts all first class 
constraints and we will assume this to be the case here. In that 
situation the piece $H^{\prime\prime}$ of the Hamiltonian in 
(\ref{2.9}) will therefore generically contain all secondary first class 
constraints 
as well, that is, the corresponding phase space functions $f^\mu$ will 
be non -- vanishing. See \cite{6,22,22a} for a discussion when this so 
called 
Dirac conjecture can be proved. 

As far as the second class constraints are concerned, they are not 
associated with any gauge freedom. It is in fact inconsistent in general 
to require an observable to satisfy $\{S_\Sigma,O\}\approx 0$ as an 
application of the Jacobi identity reveals. This means that observables 
are not first class functions, they only have to weakly Poisson commute 
with the first class constraints, not with the second class constraints.
Hence, to solve the second class constraints we simply have to restrict
ourselves to the corresponding constraint surface. 
In other words, once we have 
computed the functions on phase space which have weakly vanishing Poisson 
brackets with all first class constraints, we should restrict them to the 
constraint surface defined by the second class constraints only. The 
induced symplectic structure between such 
observables $f,f'$ on the total constraint surface is simply the pull
back of the symplectic structure on the unconstrained phase space 
by the embedding of the constraint surface defined by the second 
class constraints into the full phase space. More precisely, let 
$\cal P$ denote the unconstrained phase space and $\overline{{\cal 
P}}:=\{m\in {\cal P};\;S_\Sigma(m)=0\;\forall \Sigma\}$ the 
constraint surface defined by the second class constraints.
Consider the corresponding embedding $J:\;\overline{{\cal P}}\to {\cal 
P}$. If $\Omega$ denotes the symplectic structure on $\cal P$ then 
$\Omega^-:=J^\ast\Omega$ denotes the pull -- back symplectic 
structure on 
$\overline{{\cal P}}$. This is again a symplectic structure because 
it is closed and non degenerate which follows from the fact that the 
matrix 
\be \label{2.9a}
\Delta_{\Sigma\Sigma'}:=\{S_\Sigma,S_{\Sigma'}\}
\ee
is non degenerate. The corresponding Poisson bracket is given by 
\be \label{2.10a}
\{J^\ast f,J^\ast f'\}^-=J^\ast \{f,f'\}^\ast
\ee
where 
\be \label{2.11}
\{f,f'\}^\ast:=\{f,f'\}-\{f,S_\Sigma\}\; (\Delta^{-1})^{\Sigma\Sigma'}\;
\{S_{\Sigma'},f'\}
\ee
denotes the Dirac bracket on the full phase space. We will prove this 
for the convenience of the reader in the next section.

The reduced phase space is defined by the Poisson algebra of gauge 
invariant observables, which are not weakly vanishing (i.e. which are 
not linear combinations of first class constraints) equipped with the 
pull-back of the Dirac bracket 
to the constraint surface defined by the second class constraints. 
Notice that the Dirac bracket generically changes the symplectic 
structure for the observables as compared to the Poisson bracket.
However, it does not change the equations of motion defined by the 
canonical Hamiltonian as the Dirac bracket 
and the Poisson bracket between two functions coincide whenever at least 
one of them is a first class function.

\subsection{Reduced Phase Space and Gauge Fixing}
\label{s2.2}

In principle the description of the previous subsection entails the 
complete information about the dynamics and the physical (gauge 
invariant) content of the theory. However, it does not provide an 
explicit description of the observables. Moreover, in totally 
constrained systems the equations of motion for the observables with 
respect to the canonical Hamiltonian are 
trivial which means that some important ingredient is missing in that 
case: A non vanishing physical Hamiltonian which drives the time 
evolution of the observables. In this section we give an explicit
construction of the reduced phase space, provide a physical Hamiltonian 
and display the relation of our framework to gauge fixing.\\
\\
We saw that we eventually obtain a constrained Hamiltonian system with
first class constraints
$F_\mu$ and second class constraints $S_\Sigma$ on a phase space with 
canonical 
pairs
$(q^a,p_a),\;a=1,..,n;\;m\le n$ with respect to the original Poisson 
bracket. As shown in \cite{22}, there always exists a local canonical 
transformation (with respect to the Poisson bracket) from the canonical 
pairs $(q^a,p_a)$ to canonical pairs
$(Q^A,P_A),\;(\phi^\mu,\pi_\mu),\;(x^\sigma,y_\sigma)$ such that
\be \label{2.12}
S_\Sigma=0 \;\;\Leftrightarrow \;\;z_\Sigma=0
\ee
where the index $\sigma$ takes half the range of that of $\Sigma$ and 
where we denoted either $x^\sigma$ or $y_\sigma$ by $z_\Sigma$ for some 
value of $\Sigma$.
It is then clear that at least weakly the Dirac bracket and the 
Poisson bracket coincide on the $(Q,P,\pi,\phi)$ and that $z$ has zero
Dirac bracket with anything. 

Next, if there is a true, gauge invariant
Hamiltonian $H_0$ in (\ref{2.9}) 
(not constrained to vanish)
enlarge the phase space by an additional canonical pair $(q^0,p_0)$
and additional first class constraint $F_0=p_0+H$. The reduced phase
space and dynamics of the enlarged system is equivalent to the original
one, hence we consider without loss of generality a system with no true
Hamiltonian (totally constrained system).
The canonical Hamiltonian of the system is then a linear combination of
the first class constraints
\be \label{2.13}
H_c=\rho^\mu F_\mu
\ee
Here we have set $\rho^M=f^M$ for secondary first class 
constraints and $\rho^m=f^m+\lambda^m$ for primary first class 
constraints where $f^\mu$ is defined in (\ref{2.9}). 

A gauge fixing is defined by a set of
gauge fixing functions $G_\mu$ with the property that the matrix
with entries $M_{\mu\nu}:=\{C_\mu,G_\nu\}$ has everywhere (on the 
unconstrained phase space) non
vanishing determinant\footnote{Ideally, the gauge $G_\mu=0$ should 
define
a unique point in each gauge orbit.}.
Notice that
we allow for gauge fixing conditions
that display an explicit time dependence. The conservation in time
of the gauge fixing conditions
\be \label{2.14}
0=\frac{d}{dt}G_\mu=\frac{\partial}{\partial t}G_\mu+\{H_c,G_\mu\}
\approx\frac{\partial}{\partial t}G_\mu+\rho^\nu M_{\nu\mu}
\ee
uniquely fixes the ``Lagrange multipliers'' 
to be the following phase space
dependent functions
\be \label{2.15}
\rho^\mu=-\frac{\partial G_\nu}{\partial t}\; 
(M^{-1})^{\nu\mu}=:\rho^\mu_0
\ee
At this point one may be puzzeled by the following issue: 
The functions $\rho^\mu$ already depend on the phase space 
through $f^\mu$. For the $\rho^m$ we can always solve (\ref{2.15}) for 
the free function $\lambda^m$. But for the $\rho^M$ the solution 
(\ref{2.15}) leads to a consistency condition on the 
already imposed gauge fixing conditions, in other words we should 
impose independent gauge fixing conditions only for the primary first 
class constraints\footnote{As an 
example, in General 
Relativity the primary 
constraints demand 
that the momenta conjugate to lapse and shift vanish, the secondary 
constraints are the spatial diffeomorphism and Hamiltonian constraints 
respectively. All constraints are first class and the canonical 
Hamiltonian is a linear combination of all of them, in particular lapse 
and shift play the role of the $f^M$ for the Hamiltonian and spatial 
diffeomorphism constraint respectively. A consistent gauge 
fixing would now be to first prescribe four functions purely built from 
the intrinsic 
metric and their conjugate momenta (independent of lapse and shift). 
Such conditions have vanishing Poisson brackets with respect to the 
primary constraints. Therefore equation (\ref{2.15}) can be computed 
and presribes lapse and shift as a function of intrinsic metric and 
conjugate momentum alone. The remaining four gauge fixing conditions for 
the velocities (Lagrange multipliers) of lapse and shift which are the 
coefficients 
of the primary constraints are now that they are the time 
derivatives (Poisson brackets with the canonical Hamiltonian) of the 
already prescribed functions for lapse and 
shift. These conditions are then consistent 
with the equations of motion, i.e. that the Lagrange multipliers are the time 
derivatives of lapse and shift. The 
corresponding matrix $\{F_\mu,G_\nu\}$ in this case is block diagonal.         
We could also have prescribed lapse and
shift in the first place as functions of intrinsic metric and conjugate 
momentum and then would have to find four additional gauge fixing 
conditions on those variables whose equations of motion lead to the 
prescribed values of lapse and shift.}. This is indeed true as far as 
fixing the free coefficients in the canonical Hamiltonian is concerned. 
However, in view of the fact that all first class constraints generate 
gauge 
transformations, one has to eventually reduce with respect to all 
their gauge motions. Therefore it is mathematically and physically  
equivalent and mathematically much more convenient to regard all 
$\rho^\mu$ as free parameters, that is, to drop  
the phase space dependence of the $f^M$. Hence  
to fix the gauge we need gauge fixing conditions 
for all first class constraints. We will see explicitly in the 
path integral formulation that one is forced to this point of view and 
that nevertheless one can restore the phase space dependence of the 
$f^M$ when eventually reducing the path integral as one over 
configuration space (rtaher than the phase space).    

By construction of the Dirac bracket, we can simply ignore the variables 
$z$ for what follows and set them equal to zero where ever they occur.
In terms of the remaining canonical pairs we can solve
$F_\mu=G_\mu=0$ for
\be \label{2.16}
\tilde{F}_\mu=\pi_\mu+\tilde{h}_\mu(Q,P)=0,\;\;
\tilde{G}_\mu=\phi^\mu-\tau^\mu(Q,P)=0
\ee
for certain functions $\tilde{h},\tau$ which generically will be 
explicitly
time dependent.
The variables $\phi,\pi$ are called the gauge degrees of freedom and
$Q,P$ are called the true degrees of freedom (although typically neither
of them is gauge invariant).

The reduced Hamiltonian $H_{{\rm red}}(Q,P)$, if it exists, is supposed
to generate the same equations of motion for $Q,P$ as the canonical
Hamiltonian does, when the constraints and the gauge fixing conditions
are satisfied and the Lagrange multipliers assume their fixed values
(\ref{2.15}), that is,
\be \label{2.17}
\{H_{{\rm red}},f\}
=\{H_{{\rm can}},f\}_{F=G=\rho-\rho_0=0}
=[\rho_0^\mu \{F_\mu,f\}]_{F=G=\rho-\rho_0=0}
\ee
for any function $f=f(Q,P)$. For general gauge fixing functions the
reduced Hamiltonian
will not exist, the system of PDE's to which (\ref{2.17}) is
equivalent to, will not be integrable.

However, a so called coordinate gauge fixing condition 
$G_\mu=\phi^\mu-\tau^\mu$
with
$\tau^\mu$ independent of the phase space always leads to a reduced
Hamiltonian as follows: We can always (locally) write the constraints in
the form (at least weakly)
\be \label{2.18}
F_\mu=M_{\mu\nu}(\pi_\nu+h'_\nu(\phi,Q,P))=:M_{\mu\nu}\;F'_\nu
\ee
where $\tilde{h}_\mu(Q,P)=h'_\mu(\phi=\tau,Q,P)$. Notice that 
the locally equivalent constraints $F'_\mu$ are actually
Abelian by a general argument \cite{6}.
Then, noticing that $M_{\mu\nu}\approx\{F_\mu,G_\nu\}$, (\ref{2.17}) 
becomes
\be \label{2.19}
\{H_{{\rm red}},f\}
=[\rho_0^\mu M_{\mu\nu} \{h_\nu,f\}]_{F=G=\rho-\rho_0=0}
=[\dot{\rho}_\mu \{h_\mu,f\}]_{G=0}
=\{\dot{\rho}_I h'_\mu,f\}
\ee
with $\tilde{h}_\mu=h_\mu(\phi=\tau,Q,P)$ and we used
that $f$ only depends on $Q,P$. This displays the reduced Hamiltonian as
\be \label{2.20}
H_{{\rm red}}(Q,P;t)=\dot{\tau}^\mu(t) h_\mu(\phi=\tau(t),Q,P))
\ee
It will be explicitly time dependent unless $\dot{\tau}_I$ is time
independent and $h_\mu$ is independent of $\phi$, that is, unless those
constraints can be deparametrised for which $\dot{\tau}_\mu\not=0$.
Hence, deparametrisation is crucial for having a conserved, reduced
Hamiltonian system.\\
\\
On the other hand, let us consider the gauge invariant point of view.
Following the general framework \cite{23,24,25,26,27,28} it is possible 
to construct a gauge invariant extension of any gauge variant function 
$f(Q,P)$ off the gauge section $\phi=\tau$ by the following formula
\be \label{2.21}
O_f(\tau)=[\exp(\beta^\mu X_\mu)\cdot f]_{\beta=\tau-\phi}
\ee
where we have denoted the Abelian Hamiltonian vector fields
$X_\mu$ by $X_\mu:=\{\pi_\mu+h_\mu,.\}$. It is easy to check that 
$\{O_f(\tau),F\mu\}\approx 0$.
Consider a one parameter family of flows $t\mapsto \tau^\mu(t)$ then
with $O_f(t):=O_f(\tau(t))$ we find
\be \label{2.22}
\frac{d}{dt} O_f(t) =
\dot{\tau}^\mu(t)\sum_{n=0}^\infty\;\;\frac{\beta^{\mu_1}..
\beta^{\mu_n}}{n!}\;\;
X_\mu X_{\mu_1} .. X_{\mu_n} \cdot f
\ee
On the other hand, consider $H_\mu(t):=O_{h_\mu}(\tau(t))$, then
\cite{26}
\ba \label{2.23}
\{H_\mu(t),O_f(t)\} &=&
=O_{\{h_\mu,f\}^\ast}(\tau(t))
=O_{\{h_\mu,f\}}(\tau(t))
=O_{X_\mu\cdot f}(\tau(t))
\nonumber\\
&=&\dot{\tau}^\mu(t)\sum_{n=0}^\infty\;\;
\frac{\beta^{\mu_1}..\beta^{\mu_n}}{n!}\;\;
X_\mu X_{\mu_1} .. X_{\mu_n} \cdot f
\ea
Here the bracket $\{.,.\}^\ast$ denotes the Dirac bracket associated 
with the second class system $(F_\mu,G_\mu)$. 
In the second step we used that neither $h_\mu$ nor $f$ depend on
$\pi_\nu$, in the third we used that $f$ does not depend on $\phi^\nu$ 
and 
in
the last we used the commutativity of the $X_\nu$. Thus the physical
Hamiltonian that drives the time evolution of the observables is simply
given by
\be \label{2.24}
H(t):=\dot{\tau}^\mu(t) h'_\mu(\tau(t),O_Q(t),O_P(t))
\ee
where we used that (\ref{2.21}) is a Poisson
automorphism \cite{26}, that is,
\be \label{2.25}
\{O_f(\tau),O_g(\tau)\}=O_{\{f,g\}}(\tau)
\ee
Here $\{f,g\}^\ast=\{f,g\}$ for functions of $Q,P$ only was exploited.       
This is exactly the same as (\ref{2.19}) under the identification
$f\leftrightarrow O_f(0)$. Hence we have shown that for suitable gauge
fixings the reduced and the gauge invariant frameworks are equivalent.
Notice that it was crucial in the derivation that $(\phi^\mu,\pi_\mu)$ 
and
$(Q^A,P_A)$ are two sets of canonical pairs. If that would not be the
case, then it would be unclear whether the time evolution of the
observables has a canonical generator.\\
\\
The power of a manifestly gauge invariant framework lies therefore not
in the gauge invariance itself. Rather, it relies on whether the gauge
fixing can be achieved globally, whether it can be phrased in terms of
separate canonical pairs, whether the observer clocks $\phi^\mu$
are such that reduced Hamiltonian system is conserved and whether they
do display
the time evolution of observables as viewed by a realistic observer.
See \cite{27,29} for a discussion of this point.

Our description sketched above shows that a useful, manifestly gauge 
invariant 
formulation implicitly also relies on a system of gauge fixing 
conditions. Namely, the gauge fixing conditions equip us first with a 
physical
interpretation of the Dirac observables and second with a physical time 
evolution: $O_f(\tau)$ has the meaning of that relational 
observable, i.e. gauge invariant quantity, 
which in the gauge $\phi=\tau$ takes the 
value\footnote{One maybe tempted to run the following contradictory
arguement: $O_f(\tau)$ obviously coincides with $f$ in the gauge 
$\phi=\tau$. Since it is also gauge invariant and since any other 
gauge can be reached from $\phi=\tau$ one may think that it takes the   
value $f$ in any other gauge, say $\phi=\tau'$ which is obviously not 
the case by inspection. The catch is that in order to reach the gauge 
$\phi=\tau'$ from $\phi=\tau$ one must apply a gauge transformation 
to $O_f(\tau)$ which maps $\phi$ to $\phi+\tau'-\tau=\phi+\delta \tau$ 
and $f$ to its
corresponding image $f+\delta f$ under this gauge transformation.
By gauge invariance we obtain $O_f(\tau)=O_{f+\delta 
f}(\tau+\delta\tau)$. Hence in the gauge $\phi=\tau'$ the observable 
takes the value $f+\delta f$ and not $f$. This is not in contradiction 
with gauge invariance because $f+\delta f$ and $f$ are evaluated at 
different 
points on the the same gauge orbit just in the right way as to give the 
same numerical value.} 
$f$.     
Its time evolution 
in terms of $\tau$ is driven by a gauge invariant Hamiltonian which 
arises by solving the constraints $F=0$ in terms of the momenta $\pi$ 
conjugate to $\phi$. The gauge fixing condition thus prominently finds 
its way into the very interpretation of the physical (reduced) phase 
space. If we would choose different clocks $\phi'$ then different
observables $O'_f(\tau)$ would result. Due to (\ref{2.25}), the algebra 
of the $O_f(\tau)$ among each other and  
of the $O'_{f'}(\tau)$ among each other respectively are isomorphic
provided that $f$ and $f'$ only depend on the respective true degrees 
of freedom.
In particular, both the $(O_{Q^A}(0),O_{P_A}(0))$ and 
$(O'_{Q^{\prime A}}(0),O'_{P'_A}(0))$ respectively provide a (local) 
system of 
coordinates on the reduced phase space and therefore one can translate 
between the two\footnote{Notice that when choosing different clock 
variables $\phi',\pi'$ we 
also have to choose different true degrees of freedom $Q',P'$. 
The algebra of the $O'_Q(0),\;O'_P(0)$ is {\it not} isomorphic 
to the one of $O_Q(0),\;O_P(0)$, rather we have 
$\{O'_P(0),O'_Q(0)\}=O'_{\{P,Q\}^{\prime\ast}}(0)$ where 
$\{.,.\}^{\prime\ast}$ denotes the Dirac bracket associated with 
$F,\phi'$. Thus, while $O_P(0),O_Q(0)$ and $O'_{P'}(0),O'_{Q'}(0)$ are 
conjugate pairs, $O'_P(0),O'_Q(0)$ are not.}.
However, their physical interpretation and physical time evolution is 
entirely different. This crucial fact will also be reflected in the 
interpretation of the path integral.

\section{Reduced Phase Space Path Integral}
\label{s3}

This section is subdivided into three parts. In the first we make some 
general remarks about scattering theory in ordinary QFT which is closely 
related to the path integral and how this applies to our case. In the 
second we formally derive the reduced phase space integral as the 
generating functional of $n-$point functions. This path integral is 
an integral over the reduced phase space. In the third section we unfold 
this path integral and integrate over the unconstrained phase space 
whereby proper gauge fixing conditions and constraints have to be 
imposed. 

\subsection{Remarks about Scattering Theory}
\label{s3.1}

The central object of interest in QFT is the scattering matrix. 
Rigorous scattering theory is in fact a difficult
subject even in ordinary QFT on Minkowski space. First of all, there is 
a notion 
of a free and interacting field $\varphi$ and $\phi$ respectively
which evolve according to the free and interacting Hamiltonian
$H_0$ and $H$ respectively. Here free means that $H_0$ does not 
contain any self -- interaction. The physical assumption is that 
in the far future $t_f\to \infty$ and far past $t_i\to -\infty$ 
any outgoing and ingoing particles respectively do not interact.
This is, of course, not really true. However, using the methods 
of local quantum physics, assuming that the theory has a mass 
gap\footnote{The four momentum squared operator should have 
a pure point point spectrum which is separated from the continuum.} one 
can 
prove that the vacuum correlators 
of 
the asymptotic fields reduce to those of the free field, where vacuum
really means the interacting vacuum. 

This means that the asymptotic 
fields generate from the interacting vacuum a Fock space ${\cal 
H}_\pm$ which in general could be a proper subspace of $\cal H$.
These states can be thought of as the rigorous substitutes for the 
states generated by the non existing asymptotic free field from the 
free vacuum. This is the famous framework of 
Haag and Ruelle, see \cite{31} and references therein. The rigorous 
S -- matrix is then defined by the scalar product between these 
asymptotic Fock states which one interprets as vector states in the 
Heisenberg picture under the free dynamics. The rigorous relation 
between 
the 
S -- matrix elements and the time ordered $n-$point functions is then 
provided by the famous LSZ formula \cite{31} which rests on the 
assumption of asymptotic completeness\footnote{More in detail, in order 
to derive the 
LSZ formulae one needs the LSZ asymptotic 
conditions which state that the matrix elements of the interacting field 
between vector states in ${\cal H}_+$ and ${\cal H}_-$ 
respectively converge to those of the free field. If asymptotic 
completeness holds, this is just weak convergence on ${\cal H}$ which is 
implied by the strong convergence of the Haag -- Ruelle theory.}, that is, 
${\cal H}={\cal H}_\pm$. 

In most textboks on QFT, the Haag -- Ruelle theory is barely mentioned. 
Rather, one somehow postulates that the free and interacting fields can 
be 
implemented on the same Hilbert space. Then one can consider Fock 
states $\psi_\pm\equiv \psi^H_\pm$ created from the free vacuum by free 
field creation 
operators 
which one considers as vectors in the Heisenberg picture in the infinite 
past and future respectively because they are time independent. To 
obtain the corresponding 
Schr\"odinger 
picture states in that limit one just has to evolve via the free 
Hamiltonian to obtain $\psi_\pm^S(t)=e^{iH_0 t/\hbar} \psi^H_\pm$ as 
$t\to \pm\infty$. To map a Schr\"odinger state from the far past to the 
far future one should however use the interacting Hamiltonian $H$
and get the evolved Schr\"odinger picture vector state $e^{iH(t_f - 
t_i)/\hbar}\psi^S_-(t_i)$. Thus the scattered Heisenberg picture state 
would be given by 
\be \label{3.1a}
\psi^H_{\rm scattered}=\lim_{t_\pm\to \pm \infty} S(t_+,t_-) \psi^H_-      
\ee
where 
\be \label{3.1b}
S(t_f,t_i):=V(t_f)^\dagger\; V(t_i)\;\;V(t)=\exp(-it H/\hbar)\exp(it 
H_0/\hbar)
\ee
The formally unitary operators $V(t)$ in principle map the evolving 
free 
Heisenberg field operators to the evolving Heisenberg field operators.
Using the differential equation for $U(t)$ and solving the resulting 
Dyson series one can formally derive the 
Gell 
-- Mann and Low magic formula \cite{31} for the scattering operator
\be \label{3.1c}
S=T\{\exp(i\int_{\mathbb{R}}\; dt\; 
[H(t)-H_0])\},\;\;H(t)=e^{-iH_0t/\hbar} \;H\; e^{i H_0 t/\hbar}
\ee
where the time ordering symbol $T$ asks to order the latest operator to 
the left. Unfortunately, all of this is mathematically ill -- defined:
A tiny subset\footnote{It is sufficient to retain the 1. uniqueness of 
the 
vacuum and 2. spatial translation invariance of the Hamiltonian (part 
of the Poincar\'e algebra) and 3. spatial translation invariance of the 
vacuum.}  
of the Haag -- 
Kastler (or Wightman) 
axioms is sufficient
to establish that the operator $V(t)$ is the identity operator (up to a
phase). 
This 
is Haag's famous theorem \cite{31}.
In other words, either there is no interaction or the magic formula is 
wrong. Indeed, (\ref{3.1c}) is ill -- defined in perturbation theory 
and needs renormalisation. In order to avoid the implication of Haag's 
theorem one can, as a regularisation, break translation invariance of 
the 
Hamiltonian in an 
intermediate step by multiplying its density by a function of compact 
support and then extend the support to infinity. This is also the 
technique underlying causal renormalisation theory \cite{32}. \\
\\
It transpires that in ordinary QFT the scattering matrix is directly
related to the time ordered n -- point functions. 
As we are interested in applications to quantum gravity, we are in a 
somewhat different situation because we do not have the axiomatic 
framework of ordinary QFT at our disposal which relies on the metric 
considered as a background field. However, one can consider a Born -- 
Oppenheimer type of approach with a representation of $\mathfrak{A}$ in 
which the three -- metric operator $q$ acts by multiplication (see 
\cite{33}
for first steps towards a technical implementation). Then, at fixed 
metric argument of the 
vector state under consideration one can consider the resulting matter 
part of the Hamiltonian and apply the techniques of QFT on curved 
(in this case ultra -- static) backgrounds \cite{8} and the 
corresponding perturbation theory \cite{35} in order to define 
scattering theory for matter. In particular, LSZ type of formulae then 
again apply. To define scattering theory for gravity
in a background independent way one should consider background 
independent semiclassical states which are concentrated on a given three
geometry and extrinsic curvature and identify their excitations with 
scattering states, see \cite{33}.

\subsection{Path Integral for n -- point functions}
\label{s3.2}

We are thus interested in the time ordered $n-$ point functions.
More in detail, suppose we have a representation 
of the $^\ast$ algebra $\mathfrak{A}$ generated by the elementary 
fields $Q^A,P_A$ (or the corresponding C$^\ast$-algebra of Weyl 
elements) on a Hilbert space ${\cal H}$ which supports the Hamiltonian 
$H$ of the (conservative) system. We will assume that $H$ is bounded 
from below and has at least one normalisable vacuum $\Omega$, i.e.
a unit vector state of minimal energy $E=\inf(\sigma(H))$) which is a 
cyclic vector for $\mathfrak{A}$. Without loss of generality we redefine 
$H$ such that $E=0$. Consider the Heisenberg 
picture operators $Q^A(t)=e^{-iHt/\hbar}\; Q^A\; e^{iHt/\hbar}$.
As motivated in the previous subsection, we are interested in the 
time ordered $n-$point functions
\be \label{3.0}
\tau^{A_1 .. 
A_n}(t_1,..,t_n):=<\Omega,T\{Q^{A_1}(t_1)..Q^{A_n}(t_n)\}\Omega>
\ee
For $n>1$ and pairwise distinct times we have 
\be \label{3.0a}
\tau^{A_1 .. 
A_n}(t_1,..,t_n)=\sum_{\pi\in S_n}\;\prod_{k=1}^{n-1}
\; [\theta(t_{\pi(k)}-t_{\pi(k+1)}) ] \;
W^{A_{\pi(1)} .. A_{\pi(n)}}(t_{\pi(1)},..,t_{\pi(n)})
\ee
where we have defined the unordered Wightman functions  
\be \label{3.0b}
W^{A_1 .. A_n}(t_1,..,t_n):=<\Omega,Q^{A_1}(t_1)..Q^{A_n}(t_n)\Omega>
\ee
We should allow for more general operator insertions but $Q^A(t)$ 
contains information about $\dot{Q}^A(0)=[H,Q^A]/(i\hbar)$ which knows 
about $P_A$, hence any scalar product between vector states in the 
dense subspace $\mathfrak{A}\Omega$ can be approximated by linear 
combinations of the functions (\ref{3.0b}). Conversely, given suitable 
positivity requirements on the Wightman functions and their 
transformation properties under time translations we can reconstruct 
${\cal H},\;\Omega,\;H$ via the GNS construction. The latter arises via 
Stone's theorem from the 
fact that we can define a strongly continuous unitary group of time 
translations.      

Using $H\Omega=0$ we may write 
\be \label{3.0c}
W^{A_1 .. 
A_n}(t_1,..,t_n):=<\Omega,e^{i(t_+ -t_1)H/\hbar} 
Q^{A_1} e^{iH(t_1-t_2)/\hbar} Q^{A_2}..Q^{A_n} 
e^{iH(t_n-t_-)/\hbar}\Omega>
\ee
for any $t_\pm$. By inserting resolutions of unity it follows that for 
suitable choices for 
$\psi_i,\psi_f$ and times $t_i,t_f$ we 
are 
interested in 
the matrix 
elements 
\be \label{3.1}
<\psi_f,U(t_f-t_i)\psi_i>_{{\cal H}},\;\;U(t)=\exp(itH/\hbar)
\ee
of the evolution operator between initial and final vectors prepared at
initial and final times $t_i,t_f$ respectively.

The path integral substitute for (\ref{3.1}) is heuristically obtained 
by skeletonisation of the time interval $[t_i,t_f]$ followed by 
insertions of unity in terms of generalised position and momentum 
eigenvectors respectively\footnote{This assumes that the operators 
$Q,P$ obey the canonical commutation relations. For more general 
algebras generalised eigenvectors may not exist because e.g.
momenta do not commute with each other. In this case one must use 
different resolutions of the identity. We will here assume that 
$\mathfrak{A}$ obeys the CCR, CAR and more general algebras can be 
treated analogously.}. Specifically, assuming that ${\cal H}$ is a 
representation 
in which the operators $Q^A$ act by multiplication, for time
steps $\epsilon=(t_f-t_i)/N$ and integration variables 
$Q_n:=Q(t_i+n\epsilon),\; P_n:=Q(t_i+n\epsilon)$ we obtain formally
\be \label{3.2}
<\psi_f,U(t_f-t_i),\psi_i>
=
\int\; \{\prod_{n=0}^N [dQ_n]\}\;\{\prod_{n=1}^N [dP_n]\}\;
\overline{\psi_f(Q_n)}\; \psi_i(Q_0) \;
[\prod_{n=1}^N \; <Q_n, e^{i\epsilon H/\hbar} P_n>\; <P_n,Q_{n-1}>]
\ee
where formally\footnote{There is no Lebesgue measure in infinite 
dimensions. However, if the Hilbert space $\cal H$ is rigorously defined 
as an $L_2$ space with a probability measure on a distributional 
extension of the classical configuration space, then (\ref{3.3}) can be 
given a meaning. We will not consider these issues for our heuristic 
purposes and confine ourselves to drawing attention to the missing steps
involved.}
\be \label{3.3}
[dQ]:=\prod_A dQ^A,\;\;[dP]:=\prod_A dP_A
\ee
The assumption is now that as $N\to\infty$ we may approximate
\be \label{3.4}
<Q_n, e^{i\epsilon H/\hbar} P_n>\approx <Q_n,P_n>\; e^{i\epsilon 
H(Q_n,P_n)/\hbar}
\ee
which can be heuristically justified by expanding the exponential in 
powers of 
$\epsilon$, ordering momentum and configuration operators to right 
and left respectively and neglecting all higher $\hbar$ corrections.
For certain Hamiltonian operators of Schr\"odinger type one can 
actually prove (\ref{3.4}) (Trotter Product formula \cite{4}) but in 
general this is a difficult subject. Making this assumption and using
the position representation of the momentum eigenfunction
\be \label{3.5}
<Q,P>=\prod_A\; \frac{\exp(-i Q^A P_A/\hbar)}{\sqrt{2\pi}}    
\ee
we obtain formally
\ba \label{3.6}
&& <\psi_f,U(t_f-t_i),\psi_i>
=
\int\; \{\prod_{n=0}^N [dQ_n]\}\;\{\prod_{n=1}^N 
[d(P_n/\sqrt{2\pi})]\}\;
\overline{\psi_f(Q_n)}\; \psi_i(Q_0) \;
\times \nonumber\\ &&
\exp(-i\frac{\epsilon}{\hbar}\sum_{n=1}^N \{ [\sum_A 
\frac{Q^A_n-Q^A_{n-1}}{\epsilon} 
P_{An}] -H(Q_n,P_n)\})
\ea
One now takes $N\to \infty$ and formally obtains 
\be \label{3.7}
<\psi_f,U(t_f-t_i),\psi_i>
=
\int\; [DQ]\; [DP/sqrt{2\pi}]
\overline{\psi_f(Q(t_f))}\; \psi_i(Q_(t_i)) \;
\exp(-\frac{i}{\hbar}\int_{t_i}^{t_f}\; dt\; \{[\sum_A \dot{Q}^A P_A]  
-H(Q,P)\})
\ee
where 
\be \label{3.8}
[DQ]=\prod_{t\in [t_i,t_f]}\; \prod_A dQ^A(t)
\ee
and similar for $[DP]$. If the Hamiltonian is at most quadratic in $P$ 
then one 
can formally perform the momentum integral. As an example, consider a 
Hamiltonian of the form 
\be \label{3.9}
H(Q,P)=\frac{1}{2} G^{AB}(Q) P_A P_B+V(Q) 
\ee
Examples of such Hamiltonians are for example the Hamiltonian constraint 
in General Relativity (neglecting the issue of gauge invariance for the 
moment) where the non trivial ``supermetric'' $G^{AB}(Q)$ is the Wheeler 
-- DeWitt metric and the potential $V(Q)$ is related to the Ricci scalar 
of the three metric $Q$. (In)famously, neiher $G$ nor $V$ are positive 
definite so that the Hamiltonian is not bounded from below in General
Relativity. 

In any case, for Hamiltonians of type (\ref{3.9}) we can formally 
perform 
the Gaussian integral and obtain 
\be \label{3.10}
<\psi_f,U(t_f-t_i),\psi_i>
={\cal N}
\int\; [DQ]\; [\sqrt{|\det(G)|}] \;
\overline{\psi_f(Q(t_f))}\; \psi_i(Q_(t_i)) \;
\exp(-i\int_{t_i}^{t_f}\; dt\; \{[\frac{1}{2} (G^{-1})_{AB} \dot{Q}^A
\dot{Q}^B-V(Q)\})
\ee
where $\cal N$ is an (infinite) numerical constant (a power of $2\pi$ 
and $\hbar$)
and  
\be \label{3.11}
[\sqrt{\det(G)}]=\prod_{t\in [t_i,t_f]} \sqrt{|\det(G)|}
\ee
is the functional determinant of the supermetric\footnote{In fact there 
is a sign factor involved which accounts for the signature of $G$. 
Equation (\ref{3.10}) is only correct if the signature of $G$ does not 
depend on $Q$.}. 

Notably, if $G$ is a non trivial function of $Q$ then it is {\it not 
true} that  
\be \label{3.12}
<\psi_f,U(t_f-t_i),\psi_i>
={\cal N}
\int\; [DQ]\; 
\overline{\psi_f(Q(t_f))}\; \psi_i(Q_(t_i)) \; 
\exp(-\frac{i}{\hbar} S[Q,\dot{Q};[t_i,t_f]])
\ee
with the classical action
\be \label{3.13}
S[Q,\dot{Q};[t_i,t_f]]:=
\int_{t_i}^{t_f}\; dt\; L(Q,\dot{Q}),\;\;
L(Q,\dot{Q})=\frac{1}{2} (G^{-1})_{AB} \dot{Q}^A
\dot{Q}^B-V(Q)
\ee
Even worse is the case that the momentum dependence of the Hamiltonian 
is higher than quadratic so that the momentum integral can no longer be 
performed exactly. In that case one can at best perform a saddle point 
approximation or one has to rely on perturbation theory.
We see that the correct path integral in general is over the phase 
space and involves the Hamiltonian action and not only over the 
configuration space involving only the Lagrangian action, so we 
will stick with 
(\ref{3.7}) in what follows. 

We still must provide a path integral formulation for the $n-$point 
functions. However, this is is easy by noting that 
\ba \label{3.13a} 
&& W^{A_1 .. A_n}(t_1,..,t_n)=\prod_{k=1}^n\; \int\; [dQ_k] \;
<\Omega,U(t_+-t_1)|Q_1> Q^{A_1}_1\;
\times \nonumber\\ &&
[\prod_{k=1}^{n-1}
<Q_k,U(t_k-t_{k+1}) |Q_{k+1}> Q^{A_{k+1}}_{k+1}]\;
<Q_n|U(t_n-t_-) | \Omega>
\ea
where $\hat{Q}^A|Q>=Q^A |Q>$ was used. Combining (\ref{3.13a}) with 
(\ref{3.7}) results in (for $t_+ >t_1>..t_n>t_-$)
\ba \label{3.13b}
W^{A_1 .. A_n}(t_1,..,t_n)
&=&
\int\; [DQ]\; [DP/\sqrt{2\pi}]
\overline{\Omega(Q(t_+))}\; \Omega(Q_(t_-)) \;
\times \nonumber\\ &&
\exp(-\frac{i}{\hbar}\int_{t_-}^{t_+}\; dt\; \{[\sum_A \dot{Q}^A P_A]  
-H(Q,P)\})\;\prod_{k=1}^n Q^{A_k}(t_k)
\ea
where 
\be \label{3.13c}
[DQ]=\prod_{t\in [t_-,t_+]}\; \prod_A dQ^A(t)
\ee
and similar for $[DP]$.

It is worth mentioning that in a rigorous 
setting \cite{1,9} one does not really consider matrix elements of 
the unitary operator $U(t)=\exp(it H/\hbar)$. Namely, consider the 
analytic continuation $t_k\mapsto it_k$ for $t_k>0$, that is, the 
Schwinger functions 
\be \label{3.13d}
S^{A_1 .. A_n}(t_1,..,t_n):=W^{A_1 .. A_n}(it_1,..,it_n)
\ee
These are correlators of the $e^{tH/\hbar} Q^A e^{-t H/\hbar}$ and now 
the same formal manipulations as before lead us to consider the   
contraction semi -- group $t\mapsto V(t)=\exp(-t H/\hbar),\;t\ge 0$.  
One now obtains instead of (\ref{3.13b}) the formula
\ba \label{3.14}
S^{A_1 .. A_n}(t_1,..,t_n)
&=&
\int\; [DQ]\; [DP/\sqrt{2\pi}]
\overline{\Omega(Q(t_+))}\; \Omega(Q_(t_-)) \;
\times \nonumber\\ &&
\exp(-\frac{1}{\hbar}\int_{t_-}^{t_+}\; dt\; \{[i\sum_A \dot{Q}^A P_A]  
+H(Q,P)\})\;\;\prod_{k=1}^n Q^{A_k}(t_k)
\ea
For Hamiltonians of the form (\ref{3.9}) with positive definite $G,V$ 
(subtract the energy gap if necessary) the formal Gaussian integration 
now gives 
\be \label{3.15}
S^{A_1 .. A_n}(t_1,..,t_n)
={\cal N}
\int\; [DQ]\; [\sqrt{\det(G)}]\; 
\overline{\Omega(Q(t_+))}\; \Omega(Q_(t_-)) \; 
\exp(-\frac{1}{\hbar} S_E[Q,\dot{Q};[t_-,t_+]])
\;\;\prod_{k=1}^n Q^{A_k}(t_k)
\ee
with the ``Euclidian'' action
\be \label{3.16}
S[Q,\dot{Q};[t_i,t_f]]:=
\int_{t_i}^{t_f}\; dt\; L_E(Q,\dot{Q}),\;\;
L_E(Q,\dot{Q})=\frac{1}{2} (G^{-1})_{AB} \dot{Q}^A
\dot{Q}^B+V(Q)
\ee
The path integral (\ref{3.15}) has better chances to be rigorously 
defined because the ``measure'' has a damping factor rather than an
oscillating one and so in the rigorous setting one {\it defines} 
(\ref{3.13b}) by backwards analytic continuation of (\ref{3.15}) (when 
possible)\footnote{It is worth mentioning that in teh axiomatic 
framework of local quantum physics \cite{31} on Minkowski space the 
Schwinger functions are automatically symmetric although the Wightman
functions are not which is a consequence of the locality axiom 
(bosonic operator valued valued distributions supported at spacelike 
separated points commute) and analyticity. In GR one does not expect
to construct a Wightman QFT due to background independence which is why
we insist on $t_k>t_{k+1}$.}. 
Equation (\ref{3.15}) (when it can be proved) is called the Feynman -- 
Kac formula \cite{1,3,4}. In what follows we therefore consider the 
Euclidian point of view.

In order to avoid 
any infinite constants we divide the contraction matrix by 
$1=<\Omega,\Omega>=<\Omega,V(t_+-t_-)\Omega>$ and obtain formally
\be \label{3.17}
S^{A_1 .. A_n}(t_1,..,t_n)
=
\frac{
\int\; [DQ]\; [DP]
\overline{\Omega(Q(t_+))}\; \Omega(Q_(t_-)) \;
\exp(-\frac{1}{\hbar}\int_{t_-}^{t_+}\; dt\; \{[i\sum_A \dot{Q}^A P_A]  
+H(Q,P)\})\prod_{k=1}^n Q^{A_k}(t_k)
}
{
\int\; [DQ]\; [DP]
\overline{\Omega(Q(t_+))}\; \Omega(Q_(t_-)) \;
\exp(-\frac{1}{\hbar}\int_{t_-}^{t_+}\; dt\; \{[i\sum_A \dot{Q}^A P_A]  
+H(Q,P)\})
}
\ee
Even if one cannot inegrate out the momenta in general, formula 
(\ref{3.17}) reveals that what we are interested in is the measure
formally given by 
\be \label{3.18}
d\mu(Q):=\frac{1}{Z}\; [DQ]\;\exp(-S_E[Q]/\hbar)\; 
\overline{\Omega(Q(t_+))}\; \Omega(Q_(t_-)) 
\ee
where 
\be \label{3.19}
\exp(-S_E[Q]/\hbar):=
\int\; [DP]\;
\exp(-\frac{1}{\hbar}\int_{t_-}^{t_+}\; dt\; \{[i\sum_A \dot{Q}^A P_A]  
+H(Q,P)\})
\ee
is the exponential of the effective Euclidian action and 
\be \label{3.20}
Z:=\int\; [DQ]\; \exp(-S_E[Q]/\hbar)\;\overline{\Omega(Q(t_+))} 
\Omega(Q(t_-))
\ee
is the partition function. None of the three quantities $[DQ],\;S_E,\;Z$ 
exists but in fortunate cases their combination can be rigorously 
defined as a measure on a suitable distributional extension of the 
space of configuration  variables $Q$. The measure $\mu$ is known if we 
know all its moments or equivalently its generating functional
\be \label{3.21}
\chi[j]:=\int\; d\mu(Q)\; e^{i\sum_A \int_{t_-}^{t_+}\;dt j_A(t) Q^A(t)}
\ee
from which the moments follow by (functional) derivation at zero
current $j$. 

The apparent drawback of the these formulae is that they involve the 
exact ground state $\Omega$ of the interacting Hamiltonian $H$ which is
difficult if not impossible to compute analytically. However, and here 
is where the Euclidian formulation again is helpful, notice that so far 
the choices for $t_\pm$ were arbitrary except that $t_k\in [t_-,t_+]$,
in particular, in the original correlator the 
dependence on $t_\pm$ is through 
$e^{-t_+ H}\Omega=\Omega$ and  
$e^{t_- H}\Omega=\Omega$. 
Now suppose in addition that $\exp(-t H)$ for $t>0$ has a positive 
integral 
kernel, i.e. maps a.e. positive functions 
to 
strictly positive functions which is usually the case. Then it follows 
from \cite{36} that $E=0$ is a simple eigenvalue and the unique (up to a 
phase) ground 
state $\Omega$ is a strictly positive 
function. It can be obtained from any a.e. positive $\Omega_0\in 
{\cal H}$ via the strong limit
\be \label{3.22}
\Omega:=\lim_{t\to \infty} \frac{e^{-t H}\Omega_0}{||e^{-t H}\Omega_0||}  
\ee
It follows that by taking the limit $t_\pm\to \pm\infty$ we can replace 
$\Omega$ by $\Omega_0$ in (\ref{3.18}) -- (\ref{3.21}) because the 
factors of $||e^{-t H}\Omega_0||$
cancel in numerator and denominator. We will assume this to 
have done for what follows. Remarkably, the choice 
of the reference vector $\Omega_0$ is rather arbitrary. \\
\\
Having justified the replacement of $\Omega$ by $\Omega_0$ in the 
Euclian regime, we analytically continue the time parameter backwards to 
define the time ordered n-point functions and thus the exponential 
becomes a pure phase. 

\subsection{Unfolding the Reduced Phase Space Path Integral}
\label{s3.3}

We would like to rewrite the path integral over the reduced phase space
coordinatised by the chosen true degrees of freedom in terms of the 
unconstrained phase space. This is of course standard, see e.g. 
\cite{6}, but we review this procedure here for the sake of 
completeness. It is, however, a rather involved procedure.

\subsubsection{Preliminary Results}
\label{s3.3.1}

The virtue of the gauge fixing conditions $G$ is that the 
system $C:=\{S,F,G\}$ is now a total second class system so that one can 
treat 
all constraints on equal footing. We will do this first in the adapted 
system of Darboux coordinates 
$(Q^A,P_A),\;(\phi^\mu,\pi_\mu)\;(x^\sigma,y_\sigma)$ which is related 
to the original system $(q^a,p_a)$ by a (local) canonical transformation 
and then show that the resulting expression is actually invariant under
canonical transformations. 
\begin{Theorem} \label{th3.1} ~\\
Let $C=\{C_{{\cal A}}\}$ be a second class system of constraints on a 
phase space with canonical coordinates $z^I$ and symplectic 
structure $\omega$ on the unconstrained phase space $\Gamma$. Denote 
the constraint surface by
$\overline{\Gamma}:=\{m\in \Gamma;\;\;C_{{\cal A}}(m)=0\;\forall\;{\cal 
A}\}$ which is a submanifold of $\Gamma$. Consider an embedding
$J:\;\hat{\Gamma}\to \Gamma$ with $J(\hat{\Gamma})=\overline{\Gamma}$
where $\hat{\Gamma}$ is a model manifold of with coordinates $x^i$ for 
$\overline{\Gamma}$. \\
i.\\
$\hat{\omega}:=J^\ast \omega$ is a symplectic structure on 
$\hat{\Gamma}$.\\
ii.\\
Let $\Omega^\ast$ be the degenerate symplectic structure on $\Gamma$
defined by the Dirac bracket corresponding to $C$. Let $f,g\in 
C^1(\Gamma)$. Then $J^\ast(\{f,g\}^\ast)=\{J^\ast f,J^\ast g\}^\wedge$
where $\{.,.\}^\wedge$ is the Poisson bracket associated with 
$\hat{\omega}$.\\
iii.\\
The relation between the Liouville measures $\mu_L$ and $\hat{\mu}_L$ on
$\Gamma$ and $\hat{\Gamma}$ respectively is 
\be \label{3.23}
\hat{\mu}_L[J^\ast f]=\mu_L[\sqrt{\det(\{C,C\})}\; \delta(C)\; f]
\ee
for any measurable function $f$. 
\end{Theorem}
We note that the right hand side of (\ref{3.23}) does not make any 
reference to the chosen embedding $J$.
\begin{proof}
~\\
i.\\
Obviously $d\hat{\omega}=J^\ast d\Omega=0$ establishes closure. Non -- 
degeneracy follows 
from the fact that $J$ has maximal rank.\\
ii.\\
Let 
\be \label{3.23a}
M_{{\cal A} {\cal B}}:=\{C_{{\cal A}}, C_{{\cal B}}\}
\ee
then\footnote{Our conventions are as follows: $i_{\chi_f}\omega+df:=0$ 
defines the Hamiltonian vector field $\chi_f$ associated to $f$ while 
$\{f,g\}:=-\chi_f[g]=-i_{\chi_f}dg=i_{\chi_f}i_{\chi_g}\omega$ defines
the Poisson bracket. The corresponding matrix is denoted by 
$\omega^{IJ}:=\{z^I,z^J\}$.} 
\be \label{3.24}
[\omega^\ast]^{IJ}=\omega^{IJ} +(M^{-1})^{{\cal A} {\cal B}} 
\omega^{IK} \omega^{JL} C_{{\cal A},K} dC_{{\cal B},L}
\ee
where $\omega^{IJ}\omega_{JK}=\delta^I_K$. Using that 
$\{f,g\}=\omega^{IJ} f_{,J} g_{,I}$ and
\be \label{3.25} 
\{J^\ast f,J^\ast g\}^\wedge
=\hat{\omega}^{ij} \;(J^\ast f)_{,i}\; (J^\ast g)_{,j} 
=\hat{\omega}^{ij}\; J^I_{,i} \;f_{,I} \;J^J_{,j} \; g_{,J} 
\ee
with $\hat{\omega}^{ij}\hat{\omega}_{jk}=\delta^j_k$,
we see that the claim is equivalent to
\be \label{3.26}
\hat{\omega}^{ij} J^I_{,i} J^J_{,j} 
=[\omega^\ast]^{IJ}
\ee
on $\overline{M}$.
To verify (\ref{3.26}) we notice that $\sigma_{{\cal A}}:=(C_{{\cal 
A},I}),\;\sigma_i:=(\omega_{IJ} J^J_{,i})$ is a linearly independent set 
of one forms on $M$ and it suffices to check (\ref{3.26}) in this basis.  
From $J^\ast C_{{\cal A}}\equiv 0$ for all $\cal A$ we immediately have
\be \label{3.27}
J^I_{,i} C_{{\cal A},I}=0
\ee
on $\overline{M}$ and by construction of the Dirac bracket it is not 
difficult to see that contraction of (\ref{3.26}) with $\sigma_{{\cal 
A}}$ results in zero on both sides. Contraction with $\sigma_i \sigma_j$
results in the identity
\ba \label{3.28}
&& \hat{\omega}^{kl}\; J^I_{,k} \; J^J_{,l}\; \sigma_{i I}\; \sigma_{j J}
=\hat{\omega}^{kl}\; \hat{\omega}_{ki}\; \hat{\omega}_{lj}
=\hat{\omega}_{ji} 
\nonumber\\
&=& [\omega^\ast]^{IJ}\; \sigma_{i I}\; \sigma_{j J}
=\omega^{IJ}\; \sigma_{i I}\; \sigma_{j J}
=\omega^{IJ}\; \omega_{IK}\; \omega_{JL} \;J^K_{,i} \;J^L_{,j}\;
\omega_{LK} \;J^K_{,i} \;J^L_{,j}
\ea
where we used (\ref{3.27}) and 
\be\label{3.28a}
\hat{\omega}_{ij}=(J^\ast \omega)_{ij}=\omega_{IJ}\; J^I_{,i} \; 
J^J_{,j}
\ee
iii.\\
Recall that for finite ($2n$-) dimensional systems the Liouville measure
is simply $\mu_L:=\wedge^n \omega={\rm Pf}(\omega) [dz]$
where ${\rm Pf}(\omega)=\sqrt{\det(\omega))}$ denotes the Pfaffian
of the matrix $\omega_{IJ}$. We adopt here the same formula for infinite 
dimensions, ignoring as ususal that the Lebesgue measure $[dz]$ does not 
exist.
Using (\ref{3.23a}) we solve the $\delta$ distribution in (\ref{3.23}) 
in terms of the embedding $J$ which we write in the form 
$z=(x,y)=J(x)=(x,Y(x))$. Here $x,y$ are separate sets of canonical
pairs so that $\omega_{IJ}$ becomes block diagonal and the block
matrices $\omega_{{\cal A} {\cal B}},\;\omega_{ij}$ are constant. We 
obtain 
\be \label{3.29}
\mu_L[\sqrt{\det(\{C,C\})}\; \delta(C)\; f]
=\int\; [dz] \; \sqrt{\det(\omega)\det(M)}(z) \delta(C(z))\;f(z)
=\int\; [dx] \; 
(\sqrt{\frac{\det(\omega)\det(M)}{[\det(c)]^2}}\; f)(J(x))
\ee
where $c_{{\cal A} {\cal B}}:=C_{{\cal A},{\cal B}}$. Here we used 
$C_{{\cal A}}(x,y)=C_{{\cal A}}(x,Y(x))+
c_{{\cal A}{\cal B}}[y-Y(x)]^{{\cal B}}+..=
c_{{\cal A}{\cal B}}[y-Y(x)]^{{\cal B}}+..$. 
We have 
\be \label{3.30}
M_{{\cal A} {\cal B}}=
\omega^{IJ} C_{{\cal A},I} C_{{\cal B},J}
=\omega^{{\cal C} {\cal D}} C_{{\cal A},{\cal C}} C_{{\cal B},{\cal 
D}} 
+\omega^{ij} C_{{\cal A},i} C_{{\cal B},j}
\ee
Equation (\ref{3.27}) takes the form
\be \label{3.30b}
C_{{\cal A},i}+C_{{\cal A},{\cal B}} Y^{{\cal B}}_{,i}=0
\ee
so that (\ref{3.30}) can be written
\be \label{3.30a}
M_{{\cal A} {\cal B}}=
c_{{\cal A} {\cal C}} c_{{\cal B} {\cal D}}
[\omega^{{\cal C} {\cal D}} 
+\omega^{ij} Y^{{\cal C}}_{,i} Y^{{\cal D}}_{,j}] 
\ee
Let us introduce the abbreviations 
\be \label{3.31}
Y^{{\cal A}}_i:=Y^{{\cal A}}_{,i},\;\;
Y_{{\cal A}}^i:=\omega_{{\cal A} {\cal B}} \omega^{ij} Y^{{\cal B}}_{,j}    
\ee
then 
\be \label{3.32}
M_{{\cal A} {\cal B}}=
c_{{\cal A} {\cal C}} c_{{\cal B} {\cal D}}
\omega^{{\cal E} {\cal D}} [\delta^{{\cal C}}_{{\cal E}}
-Y^{{\cal C}}_i Y^i_{{\cal E}}]
\ee
Consider now the matrices 
\be \label{3.33}
K^{{\cal A}}_{{\cal B}}:= Y^{{\cal A}}_i Y^i_{{\cal B}},\;
k_i^j:= Y^{{\cal A}}_i Y^j_{{\cal A}}
\ee
The key identity is now
\be \label{3.34}
\det(1-K)=\det(1-k)
\ee
To prove this we use the identity (supposing that $k$ has rank $m$) 
\be \label{3.35}
\det(1-k)=1+\sum_{l=1}^m\; (-1)^l\; 
\delta^{[i_1}_{j_1} .. \delta^{i_l]}_{j_l} \; k_{i_1}^{j_1} .. k_{i_l}^{j_l}
\ee
The same formula holds for $\det(1-K)$ just that $K$ may have a 
different rank $n$ and that summation indices are ${\cal A}$ rather than 
$i$. Now each term in the sum of (\ref{3.35}) is a polynomial in the 
the traces ${\rm tr}(k^r),\;r>0$ with a coefficient that does not depend 
on 
$m$. However, ${\rm tr}(k^r)={\rm tr}(K^r)$ for any $r$. So the only 
possible difference in the two quantities is the range of $l$. However,
notice that 
\be \label{3.36}
\delta^{[i_1}_{j_1} .. \delta^{i_l]}_{j_l} \; k_{i_1}^{j_1} .. k_{i_l}^{j_l}
=Y^{[i_1}_{{\cal A}_1} .. Y^{i_l]}_{{\cal A}_l} \; 
Y_{i_1}^{{\cal A}_1} .. Y_{i_1}^{{\cal A}_1}
=Y^{i_1}_{[{\cal A}_1} .. Y^{i_l}_{{\cal A}_l]} \; 
Y_{i_1}^{{\cal A}_1} .. Y_{i_l}^{{\cal A}_l}
\ee
is completely skew in either set of indices, hence the sum anyway 
extends to $\min(m,n)$ only.

We conclude with $\det(\omega)=\det((\omega_{ij}))\det((\omega_{{\cal 
A}{\cal B}}))$ that 
\be \label{3.37}
\frac{\det(\omega)\det(M)}{\det(c^2)}=\det((\omega_{ik}))\;
\det((\delta_j^k-k_j^k))
=
\det((\omega_{ij}-\omega_{{\cal A}{\cal B}} Y^{{\cal A}}_{,i} Y^{{\cal 
B}}_{,j})=\det(\hat{\omega}_{ij})
\ee
\end{proof}
\begin{Corollary} \label{col3.1} ~\\
The measure $\mu_G$ on $\Gamma$ defined by (\ref{3.23}) 
in terms of a gauge fixing condition $G$, as a linear functional  
is in fact independent of the gauge fixing condition
when restricted to gauge invariant functions $f$.
\end{Corollary}
\begin{proof}
By definition of a gauge fixing condition $G$ for a first class 
constraint set $\{F\}$, it defines a section of the first class 
constraint surface (i.e. it defines a hypersurface that intersects 
each gauge orbit in precisely one point) and it can be reached from any 
point on the same gauge orbit. Hence any two gauge fixings $G,G'$ are 
related by a gauge transformation $\varphi$ which can be written as a 
composition of canonical transformations of the form 
$\exp(\beta^\mu\{F_\mu,.\})$ for real valued (phase space 
independent) parameters. By the first class property, there exist 
matrices $L,M,N$ such that 
$\varphi\cdot F_\mu=L_\mu^\nu F_\nu$ and 
$\varphi\cdot S_\Sigma=M_\Sigma^{\Sigma'} S_{\Sigma'}+N_\Sigma^\mu 
F_\mu$ where $L,M$ are non -- singular\footnote{At least for 
$\beta^\mu$ close to zero.}. In matrix 
notation
$\varphi \cdot F=L\cdot F,\; \varphi \cdot S=M\cdot S+N\cdot F$.
This can be inverted 
\be \label{3.38}
F=(L^{-1})\cdot (\varphi\cdot F),\;
S=(M^{-1})\cdot[(\varphi\cdot S)-N\cdot (L^{-1})\cdot (\varphi\cdot F)]
\ee
By assumption, $f$ is (weakly) gauge invariant, 
$f(m)\approx f(\varphi\cdot m)$ 
and the 
Liouville measure is invariant under canonical transformations (since 
the symplectic structure is), $d\mu_L(\varphi\cdot m)=d\mu_L(m)$.

We exhibit the dependence of the measure (\ref{3.23}) on $G$ by
$\mu_G$. Notice that in terms of $\{C\}=\{F,G,S\}$ we have 
\ba \label{3.39}
\det(\{C,C\})_{C=0}
&=&\det \left( \begin{array}{ccc}
\{F_\mu,F_\nu\} & \{F_\mu,G_\nu\} & \{F_\mu,S_{\Sigma'}\} \\
\{G_\mu,F_\nu\} & \{G_\mu,G_\nu\} & \{G_\mu,S_{\Sigma'}\} \\
\{S_\Sigma,F_\nu\} & \{S_\Sigma,G_\nu\} & \{S_\Sigma,S_{\Sigma'}\} 
\end{array}
\right)_{C=0}
=\det \left( \begin{array}{ccc}
 0  & \{F_\mu,G_\nu\} & 0 \\
\{G_\mu,F_\nu\} & \{G_\mu,G_\nu\} & \{G_\mu,S_{\Sigma'}\} \\
0 & \{S_\Sigma,G_\nu\} & \{S_\Sigma,S_{\Sigma'}\} 
\end{array}
\right)_{C=0}
\nonumber\\
&=&\{[\det(\{F,G\})]^2\; \det(\{S,S\})\}_{C=0}
\ea
Using the automorphism property of canonical transformations 
$[\varphi \cdot f](m)=f(\varphi\cdot m)$ etc. and (\ref{3.38}) we have
{\small
\ba \label{3.40}   
&& \mu_{\varphi\cdot G}(f) 
=
\int_M \; d\mu_L(m)\;\delta(S(m))\;\delta([\varphi\cdot G](m))\;
\delta(F(m))\; |\det(\{F,\varphi\cdot G\}(m))| \; 
\sqrt{\det(\{S,S\}(m))}\; f(m)    
\nonumber\\
&=&
\int_M \; d\mu_L(m)\;
\delta((M^{-1}[\varphi\cdot S-N L^{-1} \varphi\cdot F])(m))\;
\delta(G(\varphi\cdot m))\;
\delta((L^{-1} [\varphi\cdot F])(m))\; 
|\det(\{L^{-1} \varphi\cdot F,\varphi\cdot G\}(m))|
\times\nonumber\\ &&
\; \sqrt{\det(\{M^{-1}[\varphi\cdot S-N L^{-1} \varphi\cdot F],
M^{-1}[\varphi\cdot S-N L^{-1} \varphi\cdot F]\}(m))}\; 
f(\varphi\cdot m)    
\nonumber\\
&=&
\int_M \; d\mu_L(m)\;|\det(M)(m)|\; |(\det(L)(m))|
\delta([\varphi\cdot S](m))\;
\delta(G(\varphi\cdot m))\;
\delta([\varphi\cdot F](m))\; 
|\det((L^{-1} \{\varphi\cdot F,\varphi\cdot G\})(m))| 
\;
\times\nonumber\\ &&
\sqrt{\det(M^{-1}\{\varphi\cdot S,\varphi\cdot S\}(M^{-1})^T
-N L^{-1} \{\varphi\cdot F,\varphi\cdot S\}(M^{-1})^T}
\nonumber\\&&
\overline{-M^{-1} \{\varphi\cdot S,\varphi\cdot F]\} (N L^{-1})^T
+N L^{-1} \{\varphi\cdot F,\varphi\cdot F]\} (N L^{-1})^T)(m))}
\; f(\varphi\cdot m)    
\nonumber\\ 
&=&
\int_M \; d\mu_L(m)\;|\det(M)(m)|\; |(\det(L)(m))|
\delta([\varphi\cdot S](m))\;
\delta(G(\varphi\cdot m))\;
\delta([\varphi\cdot F](m))\; 
|\det((L^{-1} \{\varphi\cdot F,\varphi\cdot G\})(m))| 
\;
\times\nonumber\\ && 
\sqrt{\det(M^{-1}\{\varphi\cdot S,\varphi\cdot S\}(M^{-1})^T)(m))}
f(\varphi\cdot m)    
\nonumber\\
&=&
\int_M \; d\mu_L(m)\;
\delta([\varphi\cdot S](m))\;
\delta(G(\varphi\cdot m))\;
\delta([\varphi\cdot F](m))\; 
|\det((\{\varphi\cdot F,\varphi\cdot G\})(m))| 
\;
\times\nonumber\\ &&
\sqrt{\det(\{\varphi\cdot S,\varphi\cdot S\}(m))}
f(\varphi\cdot m)    
\nonumber\\
&=&
\int_M \; d\mu_L(\varphi\cdot m)\;
\delta(S(\varphi\cdot m))\;
\delta(G(\varphi\cdot m))\;
\delta(F(\varphi\cdot m))\; 
|\det((\{ F, G\})(\varphi\cdot m))| 
\;
\times\nonumber\\ &&
\sqrt{(\det(\{S,S\})(\varphi\cdot m))}
f(\varphi\cdot m)    
\nonumber\\
&=& \mu_G(f)
\ea
}
where in the third step we used that Poisson brackets with $L,M,N$ 
do not contribute since the $\delta-$distributions have support at 
$\varphi\cdot F=\varphi\cdot S=0$, in the fourth we used the first class 
property and again the support of the $\delta$ distributions, in the 
fifth we cancelled the determinants of the matrices $L,M$, in the sixth 
we exploited the Poisson automorphism property of $\varphi$ as well as 
the invariance of the Liouville measure and in the last we performed a 
trivial relabelling. 
\end{proof}
The statements of theorem \ref{th3.1} and corollary \ref{col3.1} 
show that the measure $\mu_G$ (\ref{3.23}) is the correct extension to 
the 
full phase space of the pull -- back measure defined by a gauge fixing 
condition and that correlators among gauge invariant functions 
are actually independent of the gauge fixing condition. For instance, in 
terms of 
the gauge invariant observables $O^{(G)}_f$, where we have exhibited 
the dependence on $G$, we have 
$\mu_{G'}[O^{(G)}_f]=\mu_{G}[O^{(G)}_f]$ for any $G'=\varphi\cdot G$.

This can also be understood geometrically: Given two gauge fixing
conditions $G,G'$ we obtain 
$\hat{\omega}_G=J_G^\ast \omega,\;\hat{\omega}_{G'}=J_{G'}^\ast 
\omega$ from the corresponding embeddings 
$J_G:\;\hat{M}\to \overline{M}_G,\;J_{G'}:\;\hat{M}\to 
\overline{M}_{G'}$. Now clearly\footnote{In abuse of notation
we write $\varphi\cdot m=\varphi(m)$ i.e. we identify the action of 
the exponential map with the corresponding diffeomorphism.} 
\ba \label{3.41}
\overline{M}_{\varphi\cdot G}
&=& \{m\in M;\;S(m)=F(m)=\varphi^\ast G(m)=0\}
\nonumber\\
&=& \{m\in 
M;\;M^{-1}(m)[S(\varphi(m))-N(m)L^{-1}(m)F(\varphi(m))]
L^{-1}F(\varphi(m))=G(\varphi(m))=0\}
\nonumber\\
&=& \{m\in 
M;\;S(\varphi(m))=F(\varphi(m))=G(\varphi(m))=0\}
\nonumber\\
&=& \{\varphi^{-1}(\varphi(m))\in 
M;\;S(\varphi(m))=F(\varphi(m))=G(\varphi(m))=0\}
\nonumber\\
&=& \varphi^{-1}(\overline{M}_G)
\ea
so that
\be \label{3.42}
J_{\varphi\cdot G}=\varphi^{-1}\circ J_G
\ee
and therefore from the fact that $\varphi$ is canonical 
$\varphi^\ast\omega=\omega$
\be \label{3.43}
\hat{\omega}_{\varphi^\ast G}=
J_{\varphi^\ast G}^\ast \omega=
J_{G}^\ast \omega=\hat{\omega}_G
\ee
~\\
Remark:\\
The fact that $\varphi$ is canonically generated by first class 
constraints featured crucially into this argument. This has the 
following relevance:\\
Suppose we are given a system which as gauge symmetry has spatial 
diffeomorphism invariance in $D$ spatial directions. Suppose that the 
field content consists, possibly among other things, of GR minimally 
coupled to $D$ scalar fields $\phi_1,..,\phi_n$. From the curvature 
of the metric and higher derivatives we can also form $D$ algebraically  
independent scalars $R_1,..,R_n$. Suppose that at least locally they 
define a coordinate system so that $x\mapsto 
\phi(x):=(\phi_1(x),..,\phi_n(x))$ and 
$x\mapsto R(x)$ defines a (local) diffeomorphism. Pick 
any fixed diffeomorphism $\varphi_0$. Then both $G=\phi-\varphi_0$ and 
$G'=R-\varphi_0$ are bona fide gauge fixing conditions. However, there 
does not exist any canonically generated diffeomorphism 
$\varphi_\xi=\exp(\{\int\;d^Dx\; \xi^a(x) C_a(x),.\})$ with phase space 
indendent $\xi$ such that $\varphi_\xi\cdot G=G'$. The reason is that 
the spatial diffeomorphism constraint does not mix field species.
It is true that we can find a phase space dependent function 
$\hat{\xi}[\phi,R]$ defined by $\varphi_\xi \circ \phi=R$ such that 
$[\varphi_\xi\cdot G]_{\xi=\hat{\xi}}=G'$, however, due to the phase 
space dependence of $\hat{\xi}$ it is not true that 
$[\varphi_\xi\circ]_{\xi=\hat{\xi}}=\varphi_{\hat{\xi}}\circ$.
The latter is also a canonical transformation with generator
$\int\; d^Dx \;\hat{\xi}^a C_a$ but it does not generate the searched 
for field dependent diffeomorphism, provided it exists at all. 
Notice that corollary \ref{col3.1} remains true for field dependent 
$\hat{\xi}$, just the matrices $L,M,N$ look different, this is not the 
point, the point is that it is not clear that a canonical transformation 
exists which maps $\phi$ to $R$.
It may therefore be true that gauge fixings seperate into equivalence 
classes depending on whether such phase space dependent gauge 
transformations exist or not. If that was the case, then it would not be 
true that the measure (\ref{3.23}) as a linear functional on gauge 
invariant functions would be independent of the gauge fixing condition,
it would depend at least on the equivalence class.

\subsubsection{From reduced Darboux coordinates to unreduced Darboux 
coordinates}
\label{s3.3.2}

In order to combine the results of sections \ref{s3.2} and \ref{s3.3.1} 
we notice that the parameter manifold $\hat{M}$ (which is the same 
for any gauge fixing) can be identified with 
the manifold equipped with Darboux coordinates $\{Q^A,P_A\}$. 
These are adapted to our choice of $G$ such that $F=G=0$ or 
equivalently $F'=G=0$ can be 
solved for 
$\{\phi^\mu,\pi_\mu\}$ in terms of $\{Q^a,\P_A\}$ which also
defines the embedding $J_G$. In 
particular, if 
$\hat{f}$ 
only depends on $\{Q^A,P_A\}$ then we can form our preferred observables 
$O^{(G)}_{\hat{f}}$ and due to the identity
$\hat{f}=J_G^\ast O^{(G)}_{\hat{f}}$ we find from (\ref{3.23})
\be \label{3.44}
\hat{\mu}_L[\hat{f}]=\mu_G[O^{(G)}_{\hat{f}}]=
\mu_{\varphi\cdot G}[O^{(G)}_{\hat{f}}]
%=\mu_{\varphi\cdot G}[O^{(\varphi\cdot G)}_{\hat{f}}]
\ee
where corollary \ref{col3.1} was used. 
Of course, for practical calculations the precise expression for 
$O^{(G)}_{\hat{f}}$ in terms of $Q^A,P_A,\phi^\mu,\pi_\mu$ is rather 
cumbersome to use. However, due to the $\delta$ distribution 
$\delta(G)$ involved in $\mu_G$ obviously 
\be \label{3.45a}
\hat{\mu}_L[\hat{f}]=\hat{\mu}_G[O^{(G)}_f]=\mu_G[\hat{f}]
\ee
so that we can drop the gauge invariant extension under the path 
integral at the price of having to to keep the G dependence in $\mu_GG$ 
because $\hat{f}$ is not gauge inavariant so that corollary \ref{col3.1}
does not apply. Even if we keep $O^{(G)}_{\hat{f}}$ rather than 
$\hat{f}$, still the G dependence does not disappear because while we 
can drop it from $\mu_G$, it remains in $O^{(G)}_{\hat{f}}$ which is a 
specific type of Dirac observable which uses the structure $G$. This 
is in accordance with what we said in the introduction. 

We are now ready to extend the reduced Darboux coordinate phase space 
path integral of section \ref{s3.2} to all Darboux coordinates: The 
Liouville measure used there is precisely given by $\hat{\mu}_L$ 
because in Darboux coordinates $\det(\hat{\omega})=1$. Furhermore, 
for our choice of gauge fixing $G^\mu=-\phi^\mu+\tau^\mu(t)$ and 
$F'_\mu=\pi_\mu+h'_\mu(\phi,Q,P)$ we have $|\det(\{F',G\})|=1$ and since 
$S'_\sigma=z_\Sigma=(x^\sigma,y_\sigma)$ in Darboux coordiantes are 
canonical pairs we have $\det(\{S',S'\})=1$. It is therefore trivial to 
write the generating functional of $n-$point functions as a path 
integral over the entire phase space by simply using formula 
(\ref{3.23}) at each point of time
\ba \label{3.45}
\chi[j] &:= & \frac{Z[j]}{Z[0]}
\nonumber\\
Z[j] &:=&
\int\; [DQ\;DP\;D\phi\;D\pi\;Dx\;Dy] 
\;\delta[G]\;\delta[S']\;\delta[F']\; 
|\det[\{F',G\}|\;
\sqrt{\det[\{S',S'\}]}\;
\overline{\Omega_0(Q(+\infty))}\; \Omega_0(Q(-\infty)) \;
\times \nonumber\\ &&
\exp(-i\frac{1}{\hbar}\int_{\mathbb{R}}\; dt\; \{[\sum_A \dot{Q}^A P_A]  
-H_{{\rm red}}(Q,P;t)\})\;e^{i\int_{\mathbb{R}}\;dt\; j_A(t) Q^A(t)}
\ea
where for instance
\be \label{3.47}   
\delta[F']=\prod_t\; \delta(F'(t)),\;\;\det[\{S',S'\}]:=\prod_t\;
\det(\{S'(t),S'(t)\})
\ee
and 
\be \label{3.48}
S^{A_1 .. A_n}(t_1,..,t_n)=
i^{-n} [\frac{\delta^n \chi[j]}{\delta j_{A_1}(t_1).. 
j_{A_n}(t_n)}]_{j=0}
\ee
Here we have explicitly kept $\det(\{S',S'\})=1$ because we will see 
that (\ref{3.45}) is covariant under changing to equivalent constraints.
To remind the reader, we 
recall that the possibly explicitly time dependent reduced Hamiltonian
is given by 
\be \label{3.49} 
H_{{\rm red}}(Q,P;t)=\dot{\tau}^\mu(t) h'_\mu(\phi=\tau(t),Q,P))
\ee
where $F_\mu$ at $S=0$ or equivalently $S'=0$ was brought into the 
equivalent 
form 
$F'_\mu=\pi_\mu+h'_\mu(\phi,Q,P)$ which motivated the use of a gauge 
fixing of the form $G^\mu=\tau^\mu(t)-\phi^\mu$.  

Formula (\ref{3.45}) achieves the goal to extend the reduced phase space 
path integral to the full phase space, albeit in the specific, local  
Darboux
coordinates that were picked by motivations from quantum 
theory\footnote{Due to the second class constraints, the use of such 
coordinates is mandatory because otherwise the representation theory of 
the reduced symplectic structure becomes too difficult.} and the 
constraint structure of the theory and in terms of the convenient 
equivalent constraints $S',F'$. 

\subsubsection{Restoring the Original Canonical Coordinates and 
Constraints}
\label{s3.3.3}

The 
next step will be to restore the original Darboux coordinates 
$(q^a,p_a)$ as well as 
the original constraints $S,F$ rather than $S',D'$. 
To that end we notice the identity
\ba \label{3.54}
&& \int\; dt\; [P_A \dot{Q}^A -H_{{\rm red}}(\tau;Q,P)]
\nonumber\\
&=& \int\; dt\; [P_A \dot{Q}^A -\dot{\tau}^\mu h_\mu(\tau;Q,P)]
\nonumber\\
&=& \int\; dt\; [P_A \dot{Q}^A+\pi_\mu \dot{\phi}^\mu+
\pi_\mu [\dot{\tau}^\mu-\dot{\phi}^\mu]
 -\dot{\tau}^\mu [\pi_\mu+h_\mu(\tau;Q,P)]]
\nonumber\\
&=& \int\; dt\; [P_A \dot{Q}^A+\pi_\mu \dot{\phi}^\mu+
\pi_\mu G^\mu
 -\dot{\tau}^\mu \tilde{F}_\mu]
\ea
Since the path integral is supported at 
$G^\mu=\tau^\mu-\phi^\mu=0,\;F'_\mu=\pi_\mu+h'_\mu=0,\;
S'_\Sigma:=z_\Sigma=(x^\sigma,y_\sigma)=0$ we can 
rewrite
(\ref{3.54}) {\it under the integral} in the form 
\be \label{3.55}
\int\; dt\; 
[P_A\dot{Q}^A+\pi_\mu\dot{\phi}^\mu+y_\sigma \dot{y}^\sigma]
\ee
Now for certain, phase dependent, non singular matrices $M,N$ we have 
$F'=M\cdot F,\;S'=N\cdot S$. But then 
\be \label{3.54c}
\delta(F')|\det(\{F',G\})|=\delta(F)|\det(\{F,G\})|,\;\;  
\delta(S')\sqrt{\det(\{S',S'\})}=\delta(S)\sqrt{\det(\{S,S\})}
\ee
is covariant under change to equivalent constraints. This allows us to 
immediately restore the original constraints in (\ref{3.45}) altough 
everything is still written in terms of the unreduced and adapted 
darboux coordinates $(Q,P),(\phi,\pi),(x,y)$. However, that system 
of coordinates originates from the original system of canonical pairs 
$(q^a,p_a)$ by a a canonical transformation \cite{6}. Accordingly, by 
applying the inverse canonical transformation $\alpha$, we can restore 
the system
of coordinates $(q,p)$ which leaves the Liouville measure in 
(\ref{3.45}) invariant, which leaves (\ref{3.55}) invariant up to a 
total differential which we assume to vanish at $t_\pm$, which 
reexpresses 
$F,G,S$ in terms Darboux coordinates in terms of the original 
coordinates and finally is covariant with respect to the Poisson 
brackets involved because e.g. 
$\alpha(\{F,G\})=\{\alpha(F),\alpha(G)\}$. Therefore, (\ref{3.45} can be 
rewritten as 
\ba \label{3.54b}
\chi[j] &:= & \frac{Z[j]}{Z[0]}
\nonumber\\
Z[j] &:=&
\int\; [Dq\;Dp] 
\;\delta[G]\;\delta[S]\;\delta[F]\; 
|\det[\{F,G\}|\;
\sqrt{\det[\{S,S\}]}\;
\overline{\Omega_0(Q[q,p](+\infty))}\; \Omega_0(Q[q,p](-\infty)) \;
\times \nonumber\\ &&
\exp(-\frac{i}{\hbar}\int_{\mathbb{R}}\; dt\; [\sum_a \dot{q}^a p_a])  
\;e^{i\int_{\mathbb{R}}\;dt\; j_A(t) Q^A[q,p](t)}
\ea
Notice that the effect of the reduced Darboux coordinates did not 
completely 
disappear: The initial and final state depend on $Q$ as well as the 
exponential involving the current $j$. But $Q=Q[q,p]$ may be a 
complicated function of the original canonical coordinates $q,p$.\\
\\
Remark:\\ 
Notice that at this stage we can formally get rid of the gauge fixing 
condition
in (\ref{3.54b}) by the ``Fadeev -- Popov trick'' if we pay a price: As 
we have already remarked before, 
due to the presence 
of $\delta[G]$ we may replace everywhere the non gauge invariant $Q$ by 
$O^{(G)}_Q$. Then the exponent, as a symplectic potential  
and 
the measure $[Dq]\; [Dp]$, which is formally the Liouville measure 
associated with $\omega$, is gauge invariant since gauge transformations 
are canonical transformations. Also $\{S,S\},\;F$ are weakly gauge 
invariant due to the first class property and since canonical 
transformations preserve Poisson brackets. Hence, after 
the gauge invariant extension of $Q$, the only non -- 
gauge invariant ingredient of the integrand of $Z[j]$ in (\ref{3.45}) is 
$\delta[G]$. In fact, $\alpha_\beta(G)=G-\beta$ where we have introduced 
the gauge 
transformations $\alpha_\beta:=\exp(\beta^\mu\{F'_\mu,.\}$ which,
since the constraints $F'$ are Abelian, have the Abelian group 
$\mathfrak{G}$  
structure $\alpha_\beta\circ \alpha_{\beta'}=\alpha_{\beta+\beta'}$.
Since the 
remaining ingredients are all gauge invariant, we may replace $G$ by 
$\alpha_\beta(G)$ for any $\beta$. Now extend both numerator $Z[j]$ and
denominator $Z[0]$ by the infinite ``gauge volume'' constant $\int 
[D\beta]$ with ``Haar measure'' $[D\beta]$. We can then trivially 
integrate out the 
$\delta[\alpha_\beta(G)]=\delta[G-\beta]$ and find 
\ba \label{3.51}
\chi[j] &:= & \frac{\tilde{Z}[j]}{\tilde{Z}[0]}
\nonumber\\
Z[j]&:=&
\int\; [Dq]\; [Dp] \;\delta[S]\;\delta[F]\;
\sqrt{\det[\{S,S\}]}\;
\overline{\Omega_0(O^{(G)}_Q[q,p](+\infty))}\; 
\Omega_0(O^{(G)}_Q[q,p](-\infty)) 
\;
\times \nonumber\\ &&
\exp(-\frac{i}{\hbar}\int_{\mathbb{R}}\; dt\; \sum_a\; \dot{q}^a \;p_a)
\;e^{i\int_{\mathbb{R}}\;dt\; j_A(t) O^{(G)}_{Q^A}[q,p](t)}
\ea
The price that we have to pay is that we have to replace $Q^A$ by 
$O^{(G)}_{Q^A}=: O_{Q^A}$ etc. which is an even more complicated 
function 
of 
$q,p$. This makes this method of getting rid of the gauge fixing 
condition useless in practice. The only exception is when we consider
zero current $j=0$ and restrict to gauge transformations that are the 
identity in the infinite past and future. Then the gauge invariant 
extension in the argument of $\Omega_0$ and more general boundary states 
is not necessary and the argument just displayed goes through. 
The restriction to such asymptotically trivial gauge transformations 
however means that we set the constraints to zero by hand on the 
kinematical Hilbert space.
%Now recall that 
%\be \label{3.52}
%O^{(G)}_f(t)=[\alpha_\beta(f)]_{\beta=\tau(t)-\phi}
%\ee
%If the $\tilde{F}$ deparametrise, that is, $h_\mu$ is 
%independent of $\phi$ then on functions $f=f(Q,P)$ 
%the map $t\mapsto 
%O^{(G)}_{(.)}(t)$ in 
%fact reduces to  
%\be \label{3.53}  
%O^{(G)}_f(t)=
%\exp(\{[\tau^\mu(t)-\phi^\mu(t)] h_\mu,.\})\cdot f =:\gamma^{(G)}_t 
%\cdot f
%\ee
%The canonical transformations $\gamma^{(G)}_t$ can then be extended to 
%the full phase space. However, $O^{(G)}_{(.)}(t)$ is not a canonical 
%transformation on the full phase space with respect to the Poisson 
%bracket, rather it is a canonical transformation with respect to the 
%Dirac bracket and only on the part of the phase space not involving 
%$\phi$ because $O_\phi(t)=\tau(t)$ which has zero Poisson brackets with 
%everything. Thus, since $\gamma_t$ really involves $h$ rather than 
%$\tilde{F}$ we cannot get rid of $O_Q,O_P$ since the masure is not 
%invariant under this canonical transformation. Furthermore, since 
%eventually we would like to rewrite the path integral as an integral 
%involving the covariant action, we are forced to reintroduce $F$. The 
%relation between $F,\tilde{F}$ is given by $F=M\tilde{F}$ and we infer 
%on the constraint surface that $\{F,G\}=M\{\tilde{F},G\}=-M$ so that 
%reintroducing $F$ reintroduces $|\det(\{F,G\})|$. For those reasons we
%keep the gauge fixing condition in the measure as in (\ref{3.45}). 
%However, the version (\ref{3.51}) will be important in making contact 
%with the Dirac constraint quantisation.  

\subsubsection{Restoring the Lagrangian}
\label{s3.3.4}

The idea is now to exponentiate the constraints and to rewrite the 
total resulting exponent in terms of the classical action.
Thus we introduce Lagrange multipliers $\lambda^\mu$ for all first class 
constraints and $\mu^\Sigma$ for all second class constraints 
and write 
\ba \label{3.54a}
\chi[j] &:= & \frac{Z[j]}{Z[0]}
\nonumber\\
Z[j]&:=&
\int\; [Dq]\; [Dp]\;[D\lambda]\;[D\mu] 
\;\delta[G]\; 
|\det[\{F,G\}|\;
\sqrt{\det[\{S,S\}]}\;
\overline{\Omega_0(Q[q,p](+\infty))}\; \Omega_0(Q[q,p](-\infty)) \;
\times \nonumber\\ &&
\exp(-\frac{i}{\hbar}\int_{\mathbb{R}}\; dt\; 
[\sum_a \dot{q}^a p_a - \sum_\mu \lambda^\mu F_\mu-\sum_\Sigma 
\mu^\Sigma S_\Sigma)])\;e^{i\int_{\mathbb{R}}\;dt\; j_A(t) 
Q^A[q,p](t)}
\ea
The final task is to remove the secondary constraints so that the 
action appears in covariant form after Legendre transformation, i.e.
with primary constraints only. The technique for doing this is  
well known \cite{10}. We will not treat the general case with 
secondary constraints of arbitrary high but finite order N (i.e.
one has secondary, tertiary, .. N-ary constraints). For a systematic 
classification of such systems and a convenient choice of basis of those 
constraints see e.g. \cite{38} and references therein. Here we pick 
a typical situation of particular interest for General Relativity. 
The general case will be even more complicated with even more 
modifications to the measure than we encounter below:\\
We assume that the canonical Hamiltonian has the following structure
\be \label{3.55a}
H=H_0'+v_f\cdot F^{(1)}+v_s\cdot S^{(1)},\;\;H^{(0'})=-q_f\cdot 
\tilde{F}^{(2)}
\ee
As the notation suggests, $F^{(1)}$ is a vector with entries consisting 
of primary first class constraints while $S^{(1)}$ is a vector with 
entries consisting of primary second class constraints. The vector 
$\tilde{F}^{(2)}$ is related to a set of secondary second class 
constraints that we will derive shortly. Usually $F^{(1)},S^{(1)}$ 
simply express the fact that the momenta $p_f,p_s$ conjugate to 
$q_f,q_s$ respectly vanish because the Lagrangian does not depend on the 
velocities $v_f,v_s$. It is also usually the case that $\tilde{F}^{(2)}$
does not depend on $q_f,p_f,p_s$ but on $q_s$. Thus we assume that (we 
do not denote indices) 
\be \label{3.56}
\{F^{(1)},F^{(1)}\}=\{F^{(1)},S^{(1)}\}=\{S^{(1)},S^{(1)}\}
=\{F^{(1)},\tilde{F}^{(2)}\}=0
\ee
while the terms not proportional to $F^{(1)},S^{(1)}$ of 
\be \label{3.57}
S^{(2)}:=\{\tilde{F}^{(2)},S^{(1)}\}
\ee
defines a vector of secondary second class constraints. We abuse 
notation by identifying that vector with (\ref{3.57}) in order not to 
have to introduce indices. Notice that 
\be \label{3.58}
\{H,F^{(1)}\}=\{H_0',F^{(1)}\}=\tilde{F}^{(2)}
\ee
thus justifying the name secondary first class constraint.

It is also often the case that the 
$\tilde{F}^{(2})$ close on themselves, that is
\be \label{3.58a}
\{\tilde{F}^{(2)},\tilde{F}^{(2)}\}\propto \tilde{F}^{(2)}
\ee
which we also will assume. Correspondingly, 
\be \label{3.59}
\{H,\tilde{F}^{(2)}\}=\{H_0',\tilde{F}^{(2)}\}+ \{v_s \cdot 
S^{(1)},\tilde{F}^{(2)}\}\propto
\tilde{F}^{(2)},S^{(2)}
\ee
does not produce tertiary constraints. These assumptions imply by 
the Jacobi identity that 
\be \label{3.60}
\{F^{(1)},S^{(2)}\}=-\{\tilde{F}^{(2)},\{S^{(1)},F^{(1)}\}\}-
\{S^{(1)},\{F^{(1)},\tilde{F}^{(2)}\}\}=0
\ee  
Finally 
\be \label{3.61}
\{H,S^{(2)}\}=q_f\cdot 
\{\tilde{F}^{(2)},S^{(2)}\}+v_s\cdot\{S^{(1)},S^{(2)}\}
\ee
and we assume that the matrix $\{S^{(1)},S^{(2)}\}$ is invertible. Hence
the Dirac algorithmus does not produce any tertiary constraints and the 
velocity $v_s$ must be fixed in order to equate (\ref{3.61}) to zero.
Accordingly the Hamiltonian becomes
\be \label{3.62}
H=q_f\cdot[\tilde{F}^{(2)}- 
\{\tilde{F}^{(2)},S^{(2)}\} [\{S^{(1)},S^{(2)}\}]^{-1}\cdot S^{(1)}]
+v_f\cdot F^{(1)}=:q_f\cdot F^{(2)}+v_f\cdot F^{(1)}
\ee
a linear combination of first class constraints. Thus in terms of the 
previous notation, the first class constraints $F_\mu$ comprise 
$F^{(1)},F^{(2)}$, the second class constraints 
$S_\Sigma$ comprise $S^{(1)},S^{(2)}$ and finally the primary 
constraints $C_i$ comprise $F^{(1)},S^{(1)}$. 

This is a simple but non 
trivial situation often encountered in concrete models and this concrete 
form now enables us to explicitly carry out the steps outlined in 
\cite{10}: In (\ref{3.54}) by an obvious change of notation we write 
\ba \label{3.63}
\chi[j] &:= & \frac{Z[j]}{Z[0]}
\nonumber\\
Z[j] 
&:=&
\int\; [Dq]\; [Dp]\;[D\lambda_1]\;[D\lambda_2]\;[D\mu_1]\;[D\mu_2] 
\;\delta[G]\; 
|\det[\{F,G\}|\;
\sqrt{\det[\{S,S\}]}\;
\overline{\Omega_0(Q[q,p](+\infty))}\; \Omega_0(Q[q,p](-\infty)) \;
\times \nonumber\\ &&
\exp(-\frac{1}{\hbar}\int_{\mathbb{R}}\; dt\; \{
[i(\sum_a \dot{q}^a p_a - \lambda_1\cdot F^{(1)}-\lambda_2\cdot F^{(2)}
-\mu_1 \cdot S^{(1)}-\mu_2 \cdot S^{(2)})]\})\;
e^{i\int_{\mathbb{R}}\;dt\; j_A(t) Q^A[q,p](t)}
\nonumber\\
&:=&
\int\; [Dq]\; [Dp]\;[D\lambda_1]\;[D\lambda_2]\;[D\mu_1]\;[D\mu_2] 
\;\delta[G]\; 
|\det[\{F,G\}|\;
\sqrt{\det[\{S,S\}]}\;
\overline{\Omega_0(Q[q,p](+\infty))}\; \Omega_0(Q[q,p](-\infty)) \;
\times \nonumber\\ &&
\exp(-\frac{1}{\hbar}\int_{\mathbb{R}}\; dt\; \{
[i(\sum_a \dot{q}^a p_a - \lambda_1\cdot F^{(1)}-\lambda_2\cdot 
\tilde{F}^{(2)}
-\mu_1 \cdot S^{(1)}-\mu_2 \cdot S^{(2)})]\})\;
e^{i\int_{\mathbb{R}}\;dt\; j_A(t) Q^A[q,p](t)}
\ea
where in the second step we have shifted the integration variable 
$\mu_1$ in order to absorb the contribution $\mu_0\cdot 
S^{(1)}=F^{(2)}-\tilde{f}^{(2)}$. 

We now perform a canonical 
transformation with generator $\mu_2\cdot S^{(1)}/\lambda_2'$ at each 
time 
$t\in [t_-,t_+]$ which we assume to become the identity at $t_\pm$.
Here $\lambda_2'$ is the unique component of $\lambda_2$ such 
that $\lambda_2\cdot\{\tilde{F}^{(2)},S^{(1)}\}=\lambda_2' S^{(2)}$ 
modulo terms proportional to $F^{(1)},S^{(1)}$.
This transformation preserves the Liouville measure, the symplectic 
potential $\int dt p_a \dot{q}^a$ and
$\overline{\Omega_0(Q[q,p](+\infty))}\; \Omega_0(Q[q,p](-\infty)) \;
e^{i\int_{\mathbb{R}}\;dt\; j_A(t) Q^A(t)}$ since in this example 
under consideration $Q^A$ is 
among the coordinates independent of $q_s$. If we assume that 
$G=G^{(1)},G^{(2)}$ do not involve $q_s$ then also $G$ is invariant.
Indeed, one can choose $G^{(1)},G^{(2)}$ to be gauge fixing conditions 
on $q^{(1)}_f:=q_f$ and $q^{(2)}_f$ respectively which are both 
independent of $q_s=q_s^{(1)}$ so that 
$q_s^{(1)},q_s^{(2)},q^{(1)}_f,q^{(2)}_f,Q^A$ 
comprises a 
complete system of configuration coordinates. Furthermore, clearly 
$F^{(1)},S^{(1)}$ are invariant. But, denoting the canonical 
transformation by $\alpha$, we have  
\ba \label{3.64}
&& \alpha(\lambda_2\cdot \tilde{F}^{(2)})=\lambda_2\cdot \tilde{F}^{(2)}
-\mu_2\cdot S^{(2)}-\frac{1}{2\lambda_2'}\mu_2\cdot 
\{S^{1},S^{(2)}\}\cdot \mu_2+O(\mu_2^3,F^{(1)},S^{(1)})
\nonumber\\
&& \alpha(\mu_2\cdot S^{(2)})=\mu_2\cdot S^{(2)}+\frac{1}{\lambda_2'}
\mu_2\cdot \{S^{1},S^{(2)}\}\cdot \mu_2 +O(\mu_2^3)
\ea
Since integrating again over $\lambda_1,\mu_1$ enforces 
$F^{(1)}=S^{(1)}=0$ we can drop terms proportional to $F^{(1)},S^{(1)}$.

Next, $\alpha(\{F,G\})=\{\alpha(F),\alpha(G)\}=\{\alpha(F),G\}$ under 
the assumptions made. This will in general depend non trivially on 
$\mu_2$ through $\alpha(\tilde{F}^{(2)})$. Likewise $\alpha(\{S,S\})=
\{\alpha(S),\alpha(S)\}$ will in general depend non trivially on 
$\mu_2$ through $\alpha(S^{(2)})$. Consider 
$|\det(\{F,G\})|,\;\sqrt{\det(\{S,S\})}$ as expanded in powers of 
$\mu_2$. 
Also, since 
\be \label{3.65}
\alpha(\lambda_2\cdot \tilde{F}^{(2)}+\mu_2\cdot S^{(2)})=
\lambda_2\cdot \tilde{F}^{(2)}
+\frac{1}{2\lambda_2'}\mu_2\cdot\{S^{(1)},S^{(2)}\}\cdot \mu_2
+O(\mu_2^3)
\ee
let us power expand $\alpha(\delta(F))\delta(S))$ around that quadratic 
term and perform the Gaussian integral. Since 
$\det(\{S,S\})=\det(\{S^{(1)},S^{(2)}\})^2$ this yields 
\ba \label{3.66}
Z[j]&=&
\int\; [Dq]\; [Dp]\;[D\lambda_1]\;[D\lambda_2 
|\lambda_2'|^{1/2}]\;[D\mu_1] 
\;\delta[G]\; 
|\det[\{F,G\}|\;
{\root 4 \of{\det[\{S,S\}]}}\; V\;
\overline{\Omega_0(Q[q,p](+\infty))}\; \Omega_0(Q[q,p](-\infty)) \;
\times\nonumber\\ &&
\exp(-\frac{1}{\hbar}\int_{\mathbb{R}}\; dt\; \{
[i(\sum_a \dot{q}^a p_a - \lambda_1\cdot F^{(1)}-\lambda_2\cdot 
\tilde{F}^{(2)}
-\mu_1 \cdot S^{(1)})]\})\;
e^{i\int_{\mathbb{R}}\;dt\; j_A(t) Q^A[q,p](t)}
\ea
where the local factor $V$ accounts for the additional contributions 
just 
mentioned. The integral over $p_f=F^{(1)},p_s=S^{(1)}$ produces 
$\delta(\lambda_1-v_f)\delta(\mu_1-v_s)$ and cancels the integral over 
$\lambda_1,\mu_1$. Denoting 
$\{q^{\prime \alpha}\}=\{q^{(2)}_f,q^{(2)}_s,Q^A\},  
\;\;\{p'_\alpha\}=\{q^{(2)}_f,q^{(2)}_s,Q^A\}$ we are left with 
\ba \label{3.67}
Z[j] &=&
\int\; [Dq]\; [Dp']\;\;[D\lambda_2 
|\lambda_2'|^{1/2}] 
\;\delta[G]\; 
|\det[\{F,G\}|\;
{\root 4 \of{\det[\{S,S\}]}}\; V\;
\overline{\Omega_0(Q[q,p](+\infty))}\; \Omega_0(Q[q,p](-\infty)) \;
\times \nonumber\\ &&
\exp(-\frac{1}{\hbar}\int_{\mathbb{R}}\; dt\; \{
[i(\sum_a \dot{q}^{\prime a} p'_a -\lambda_2\cdot 
\tilde{F}^{(2)}\})\;
e^{i\int_{\mathbb{R}}\;dt\; j_A(t) Q^A[q,p](t)}
\ea
Now, by definition (see also sectoion \ref{s2.1}), solving 
$p'_\alpha=\partial L(q^{\prime\beta},q_f,q_s;v^{\prime \beta})/\partial 
v^{\prime \alpha}$ for $v^{\prime 
\alpha}$ 
yields $v^{\prime }=u^{\prime \alpha}(q^{\prime 
\beta},q_f,q_s;p'_\beta)$ 
and 
\be \label{3.68}
H_0'=q_f\tilde{F}^{(2)}(q^s,q^{\prime \beta};p'_\beta)=
[v^{\prime \alpha}p'_\alpha-L(q^{\prime \beta},q_f,q_s;v^{\prime 
\beta})]_{v'=u'}
\ee
As is well known, the inverse of this Legendre transformation is 
\be \label{3.69}
L=[v^{\prime \alpha}p'_\alpha-H_0'(q^{\prime 
b},q_f,q_s;p'_\beta)]_{p'=\partial 
L/\partial v'}
\ee
Therefore a saddle point expansion about the extremum $p'=\partial 
L/\partial v'$ 
of the exponent in (\ref{3.67}) yields 
\ba \label{3.70}
Z[j] &=&
\int\; [Dq]\;\;\;[D\lambda_2 
|\lambda_2'|^{1/2}] 
\;\delta[G]\; 
|\det[\{F,G\}|\;
{\root 4 \of{\det[\{S,S\}]}}\; V'\;
\overline{\Omega_0(Q[q](+\infty))}\; \Omega_0(Q[q](-\infty)) \;
\times \nonumber\\ &&
\exp(\frac{i}{\hbar}\int_{\mathbb{R}}\; dt\; 
L(q_f=\lambda_2,q_s,q^{\prime b};v^{\prime b}))\;
e^{i\int_{\mathbb{R}}\;dt\; j_A(t) Q^A[q](t)}
\ea
where $V'/V$ accounts for the additional modifications that come from 
the saddle point approximation. {\it Notice that a possible 
dependence on $p$ in $Q^A$ prohibits the saddle point 
approximation beyond its zeroth order term!} Assuming that $Q^A$ is 
independent of $p$ as it is the case in this example under
consideration and assuming 
that $G^{(1)}$ 
really is a 
coordinate condition on $q_f$ and noticing that $F,S,G^{(2)},V'$ do not 
depend on $q_f$ we have after relabelling $\lambda_2\to q_f$
\ba \label{3.71}
Z[j] &=&
\int\; [Dq]\;\;\;[|q_f'|^{1/2}] 
\;\delta[G^{(2)}]\; 
|\det[\{F,G\}|\;
{\root 4 \of{\det[\{S,S\}]}}\; V'\;
\overline{\Omega_0(Q[q](+\infty))}\; \Omega_0(Q[q](-\infty)) \;
\times \nonumber\\ &&
\exp(\frac{i}{\hbar}\int_{\mathbb{R}}\; dt\; 
L(q_f,q_s,q^{\prime b};v^{\prime b}))\;
e^{i\int_{\mathbb{R}}\;dt\; j_A(t) Q^A[q](t)}
\ea
which is our final result.\\
\\
To summarise:\\
The path integral can be brought into a form
only involving a configuration integral and the exponent of the 
covariant action, but there is a non trivial measure factor depending on 
$S,F,G,V'$ which accounts for the correct implementation of the 
dynamics. Missing that factor means quantising an entirely different 
system. The measure is not covariant with respect to the Lagrangian
symmetries, however, by construction it is covariant with respect to 
the Hamiltonian symmetries generated by the first class constraints 
\cite{21}.
As is well known from classical Noether theory, these two symmetries 
coincide 
only on shell, that is, when the equations of motion hold, i.e. in the 
semiclassical sector of the path integral (critical points of the 
action). But that is hardly 
surprising. The quantum effects, that is, the fluctuations and higher 
correlations will receive corrections coming from the measure factor
and one completely misses them if one postulates the naive 
covariant measure $[dq]\exp(iS)$. Notice also that the path integral 
remembers the gauge fixing condition explicitly through the dependence
of the wave functions, as well as the exponential of the current, on 
$Q^A$ only, rather than all of $q^a$, which are adapted to $G$.

\section{Operator Constraint Quantisation Path Integral}
\label{s4}

As already mentioned, in the presence of second class 
constraints, operator constraint quantisation is in general
impossible if one does not pass to local Darboux coordinates with 
respect to the Dirac bracket because otherwise one does not find 
representations of the canonical commutation relations. Hence 
we assume that we have passed from the global conjugate pairs 
$(q^a,p_a)$ 
to local conjugate pairs 
$(z_\Sigma)=(x^\sigma,y_\sigma),\;(\phi^\mu,\pi_\mu),\;(Q^A,P_A)$
in terms of which the second and first class constraints respectively 
can be reformulated as  
$S_\Sigma=0\;\Leftrightarrow\; z_\Sigma=0$ and $F_\mu=0\;\Leftrightarrow
\; F'_\mu=\pi_\mu+h'_\mu(\phi;Q,P)=0$. The $F'_\mu$ are Abelian
$\{F'_\mu,F'_\nu\}=0$ and first class $\{F'\mu,z_\Sigma\}=0$ while
$\{y_\sigma,x^{\sigma'}\}=\delta_\sigma^{\sigma'}$ and thus the 
Dirac bracket on functions of $\phi,\pi,Q,P$ reduces to the Poisson 
bracket. 

On the assumption that the constraints $F'_\mu$ can be quantised without 
anomalies on the kinematical Hilbert space ${\cal H}_{{\rm kin}}=L_2(dQ 
d\Phi)$, that is, $[F'_\mu,F'_\nu]=0$ we define a rigging map 
heuristically as (one has to be careful with domains and ranges and 
$\eta$ should be defined as anti -- linear map, however we do not need 
to enter the discussion of these niceties here, see 
\cite{13} for further information) 
\be \label{4.1}
\eta:\;{\cal H}_{{\rm kin}} \to {\cal H}_{{\rm phys}};\;
\psi\mapsto [\eta(\psi)](\phi,Q)=
\int\; [d\beta/(2\pi)]\; [e^{i\beta^\mu 
F'_\mu} 
\psi](\phi,Q)
\ee
In the case at hand we can easily bring (\ref{4.1}) into a form from 
which 
it is obvious that it solves the constraints $F'\mu=0$. First of all we 
notice that (\ref{4.1}) can be formally written as 
\be \label{4.2}
\eta(\psi)=\prod_\mu \delta(F'_\mu) \psi
\ee
where the order of the $\delta$ distributions is irrelevant due to the 
Abelianess of the constraints. This is not the case for the $F_\mu$ 
which is why the heuristic projector defined in 
\cite{40} does not solve the constraints. This is already a hint that
(\ref{4.1}) indeed solves the constraints. To actually prove it we 
notice that $e^{i\beta^\mu \pi_\mu}\psi(\phi)=\psi(\phi-\beta)$ and so
(we suppress the $Q$ argument in what follows)
\be \label{4.3}
[\eta(\psi)](\phi)=
\int\; [d\beta/(2\pi)]\; V(\beta)\; \psi(\phi-\beta)
=\int\; [d\beta/(2\pi)]\; V(\phi-\beta)\; \psi(\beta)
\ee
where
\be \label{4.4}
V(\beta)=e^{\beta^\mu[\pi_\mu+h'_\mu(\phi)]} \;e^{-i\beta^\mu \pi_\mu}
\ee
Using
\be \label{4.5}
e^{i\beta^\mu \pi_\mu} \; h'_\nu(\phi) e^{-i\beta^\mu \pi_\mu}= 
h'_\nu(\phi-\beta)
\ee
we derive 
\be \label{4.6}
\frac{1}{i}\frac{\partial V(\beta)}{\partial 
\beta^\mu}=V(\beta)h'_\mu(\phi-\beta)
\ee
Denoting $\beta(t):=\beta_1+t(\beta_2-\beta_1)$ it follows 
\be \label{4.7}
V(\beta_2)-V(\beta_1)=\int_0^1\;dt_1\;\frac{d}{dt_1}V(\beta(t_1))
=i\int_0^1 \; dt_1\; V(\beta(t_1))\; \dot{\beta}^\mu(0) 
h'_\mu(\phi-\beta(t_1))
\ee 
where we noticed that 
$\dot{\beta}(t)=\dot{\beta}(0)=\beta_2-\beta_1=$const.
Equation (\ref{4.7}) can be iterated into a Dyson series: We need 
\be \label{4.8}
\beta_{t_1}(t_2)=\beta_1+t_2(\beta(t_1)-\beta_1)=\beta_1+t_1 
t_2(\beta_2-\beta_1)=\beta(t_1 t_2)
\ee
so that    
\ba \label{4.9}
V(\beta(t_1))-V(\beta_1) 
&=& \int_0^1\;dt_2\;\frac{d}{dt_2}V(\beta_{t_1}(t_2))
=i\int_0^1 \; dt_2\; V(\beta(t_1 t_2))\; t_1 \dot{\beta}^\mu(0) 
h'_\mu(\phi-\beta(t_1 t_2))
\nonumber\\
&=& i\int_0^{t_1} \; dt_2\; V(\beta(t_2))\; \dot{\beta}^\mu(0) 
h'_\mu(\phi-\beta(t_2))
\ea
Accordingly we obtain for any $\beta_2,\;\beta_1$ 
\be \label{4.10}
V(\beta_1)^{-1}\; V(\beta_2)=T_l\exp(i\int_0^1\; 
dt\;[\beta_2-\beta_1]^\mu\;h'_\mu(\phi-\beta_1-t(\beta_2-\beta_1)))
\ee
where the path ordering symbol $T_l$ orders the earliest time to the 
left.
Choosing $\beta_1=\phi,\;\beta_2=\phi-\beta$ we find 
\be \label{4.11}
V(\phi-\beta)=V(\phi)\;T_l\exp(-i\int_0^1\;dt\;\beta^\mu 
\;h'_\mu(t\beta)) 
=:V(\phi) U(\beta)
\ee
where, using again (\ref{4.10}) with $\beta_2=\phi,\beta_1=0$ and 
noticing from the definition (\ref{4.4}) that $V(0)=1$
\be \label{4.12}
V(\phi)=T_l\exp(i\int_0^1\; 
dt\;\phi^\mu\;h'_\mu(\phi(1-t)))=T_r\exp(i\int_0^1\; 
dt\;\phi^\mu\;h'_\mu(\phi t))
\ee  
where we have performed the change of variables $t\mapsto 1-t$ 
which switches $T_l$ to $T_r$ which orders the earliest time to the 
right. For later use we notice the identity
\be \label{4.12a}
U(\beta)=V(\beta)^{-1}=V(\beta)^\dagger
\ee
which establishes unitarity of $V(\beta)$ (as an operator on 
$L_2(dQ)$) and can 
easist be demonstrated 
by writing $V(\beta)$ in the form 
\be \label{4.12b}
V(\beta)=\lim_{N\to \infty} 
e^{\frac{i}{N}\beta^\mu h'_\mu(\beta)} \;
e^{\frac{i}{N}\beta^\mu h'_\mu(\frac{N-1}{N}\beta)} \;
e^{\frac{i}{N}\beta^\mu h'_\mu(\frac{N-2}{N}\beta)} \; ... \;
e^{\frac{i}{N}\beta^\mu h'_\mu(\frac{1}{N}\beta)} 
\ee
The point of these manipulations is that we can now write
\be \label{4.13}
[\eta(\psi)](\phi,Q)=V(\phi)[\eta'(\psi)](Q),\;\;
[\eta'(\psi)](Q)=\int\;[d\beta/(2\pi)]\;V(\beta)^{-1}\;\psi(\beta,Q)
\ee
Obviously $\eta'(\psi)$ no longer depends on $\phi$ so that the rigging
map essentially produces functions whose $\phi$ dependence is restricted 
to be of the form $V(\phi)\Psi(Q)$ for suitable $\Psi\in L_2(dQ)$. In 
order to show that such functions really solve $F'_\mu=0$ it is very 
crucial that $[F'_\mu,F'_\nu]=0$, otherwise this doe not hold.
Essentially, the proof boils down to showing (we again suppress the 
$Q$ dependence)
\be \label{4.14}
[\frac{\partial}{\partial \phi^\mu}-ih'_\mu(\phi)]V(\phi)=0
\ee
This almost looks like a parallel transport equation on $\phi$ space 
with respect to a one form $-ih'_\mu(\phi)$ with values in a Lie 
algebra 
of (anti -- self adjoint) operators on $L_2(dQ)$ defined on a common 
dense and 
invariant domain and $V(\phi)$ looks like its holonomy along the 
path $t\mapsto t \phi$. The 
difference 
with the parallel equation is of course that the latter is an ODE
while (\ref{4.14}) is a system of PDE's so that the issue of 
integrability arises and so the following theorem is not trivial
(notice tha we {\it do not} need to assume $\partial_{[\mu} 
h'_{\nu]}=0$). Its validity rests on the fact that 
\be \label{4.14a}
[F'_\mu,F'_\nu]
=-\{2\partial_{[\mu} (-i h'_{\nu]}+[(-i h'_\mu),(-i h'_\nu)]\}=0
\ee
i.e. that the curvature of the connection $-ih'_\mu$ vanishes.
\begin{Theorem} \label{th4.1} ~\\
Equation (\ref{4.14}) holds pointwise in $\phi$ space on a dense set of 
analytic vectors\footnote{A vector $\psi$ is called analytic for an 
operator $A$ if $||A^n \psi||<\infty$ for all $n$ and 
$\sum_{n=0}\;t^n\;||A^n 
\psi||/(n!)<\infty$ for some $t>0$.}  
for the operator $h'(\phi)=\phi^\mu h'_\mu(\phi)$.
\end{Theorem}
\begin{proof}
Let $V_0(\phi):=1$ for $N\in \mathbb{N}_0$ and for $N>0$
\be \label{4.15}
V_N(\phi)=1+\sum_{n=1}^N\; i^n\; \int_0^1\; dt_1\; \phi^{\nu_1} 
h'_{\nu_1}(t_1 \phi)\;...\;\int_0^{t_{n-1}}\;dt_n\; \phi^{\nu_n} \;
h'_{\nu_n}(t_n \phi)
\ee
Clearly $\lim_{N\to \infty} V_N(\phi)=V(\phi)$ converges on  
analytic 
vectors for $h'(\phi):=\phi^\mu h'_\mu(\phi)$. We define for $N>0$ the 
remainder 
\be \label{4.16}
R_N(\phi)=-i^{N-1}\sum_{n=1}^N
\int_0^1\; dt_1\; \phi^{\nu_1} h'_{\nu_1}(t_1 \phi)\;...\;
\int_0^{t_{n-1}}\; dt_n\; t_n \phi^{\nu_1}\{[h'_\mu,h'_{\nu_1}]\}(t_n 
\phi)\;...\;
\int_0^{t_{N-1}}\;dt_N\; \phi^{\nu_N} \;
h'_{\nu_N}(t_N \phi)
\ee
and prove by induction for $N>0$ that 
\be \label{4.17}
\partial_\mu V_N(\phi)=i h'_\mu(\phi)\; V_{N-1}(\phi)+R_N(\phi)
\ee
By definition, on analytic vectors of $h'(\phi)$ the norm of the 
remainder converges 
(pointwise in 
$\phi$ space) to zero (it is of order $1/[(N-1)!]$)  
so that once that (\ref{4.17}) is established, the proof is complete.

For $N=1$ we obtain
\ba \label{4.18}
\partial_\mu V_1(\phi) 
&=& 
i \int_0^1\; dt\; [h'_\mu(t\phi)+t\phi^\nu(\partial_\mu h'_\nu)(t\phi)]
\nonumber\\
&=& 
i \int_0^1\; dt\; [\frac{d}{dt}[t \;h'_\mu(t\phi)]+2 
t\phi^\nu(\partial_{[\mu} 
h'_{\nu]})(t\phi)]
\nonumber\\
&=& ih'_\mu(\phi)- \int_0^1\; dt\; \{ 
t\phi^\nu\;([h'_\mu,h'_\nu])(t\phi)]\}
\nonumber\\
&=& ih'_\mu(\phi) V_0(\pi) +R_1(\phi)
\ea
where in the third step we used 
\be \label{4.19}
[F'_\mu,F'_\nu]=0\;\;\Leftrightarrow\;\; 2 i \partial_{[\mu} 
h'_{\nu]}+[h'_\mu,h'_\nu]=0
\ee
Assuming that (\ref{4.17}) holds up to $N$ we compute 
{\small
\ba \label{4.20}
&&\partial_\mu(V_{N+1}-V_N) 
=
i^{N+1} \partial_\mu \int_0^1\; dt_1\;\phi^{\nu_1} h'_{\nu_1}(t_1 
\phi)\;...\;\int_0^{t_N}\; dt_{N+1}\;\phi^{\nu_{N+1}} 
h'_{\nu_{N+1}}(t_{N+1} \phi)
\nonumber\\
&=&
i^{N+1} \sum_{n=1}^{N+1} \int_0^1\; dt_1\;\phi^{\nu_1} h'_{\nu_1}(t_1 
\phi)\;...\;\int_0^{t_{n-1}}\;dt_n[h'_\mu(t_n\phi)+t_n 
\phi^{\nu_n}(\partial_\mu h'_{\nu_n})(t_n\phi)]\;...\;
\int_0^{t_N}\; dt_{N+1}\;\phi^{\nu_{N+1}} 
h'_{\nu_{N+1}}(t_{N+1} \phi)
\nonumber\\
&=&
i^{N+1} \sum_{n=1}^{N+1} \int_0^1\; dt_1\;\phi^{\nu_1} h'_{\nu_1}(t_1 
\phi)\;...\;\int_0^{t_{n-1}}\;dt_n[\frac{d}{dt_n}(t_n 
h'_\mu(t_n\phi))+2 t_n 
\phi^{\nu_n}(\partial_{[\mu} h'_{\nu_n]})(t_n\phi)]\;...\;
...\nonumber\\ &&
\int_0^{t_N}\; dt_{N+1}\;\phi^{\nu_{N+1}} 
h'_{\nu_{N+1}}(t_{N+1} \phi)
\nonumber\\
&=&
i^{N+1} \sum_{n=1}^{N+1} 
\int_0^1\; dt_1\;\phi^{\nu_1} h'_{\nu_1}(t_1 \phi)
\;...\;
\int_0^{t_{n-2}}\; dt_{n-1}\;\phi^{\nu_{\nu_{n-1}}} 
h'_{\nu_{n-1}}(t_{n-1} 
\phi)\;
\int_0^{t_{n-1}}\;dt_n[\frac{d}{dt_n}(t_n 
h'_\mu(t_n\phi))]\; 
\times \nonumber\\ &&
\times \int_0^{t_n} dt_{n+1} \phi^{\nu_{n+1}} 
h'_{\nu_{n+1}}(t_{n+1}\phi)\;...\;
\int_0^{t_N}\; dt_{N+1}\;\phi^{\nu_{N+1}} 
h'_{\nu_{N+1}}(t_{N+1} \phi)
\nonumber\\
&&
-i^N \sum_{n=1}^{N+1} \int_0^1\; dt_1\;\phi^{\nu_1} h'_{\nu_1}(t_1 
\phi)\;...\;\int_0^{t_{n-1}}\;dt_n\;t_n 
\phi^{\nu_n}([h'_\mu,h'_{\nu_n}])(t_n\phi)]\;...\;
\int_0^{t_N}\; dt_{N+1}\;\phi^{\nu_{N+1}} 
h'_{\nu_{N+1}}(t_{N+1} \phi)
\nonumber\\
&=& R_{N+1}+ih'_\mu(\phi)[V_N(\phi)-V_{N-1}(\phi)]
\nonumber\\
&& + i^{N+1} \sum_{n=2}^{N+1} 
\int_0^1\; dt_1\;\phi^{\nu_1} h'_{\nu_1}(t_1 \phi)
\;...\;
\int_0^{t_{n-2}}\; dt_{n-1}\;\phi^{\nu_{\nu_{n-1}}} 
h'_{\nu_{n-1}}(t_{n-1} \phi)\;t_{n-1} \;h'_\mu(t_{n-1}\phi))\; 
\times\nonumber\\ &&
\times
\int_0^{t_{n-1}} dt_{n+1} \phi^{\nu_{n+1}} 
h'_{\nu_{n+1}}(t_{n+1}\phi)\;...\;
\int_0^{t_N}\; dt_{N+1}\;\phi^{\nu_{N+1}} 
h'_{\nu_{N+1}}(t_{N+1} \phi)
\nonumber\\
&& -i^{N+1} \sum_{n=1}^{N} 
\int_0^1\; dt_1\;\phi^{\nu_1} h'_{\nu_1}(t_1 \phi)
\;...\;
\int_0^{t_{n-2}}\; dt_{n-1}\;\phi^{\nu_{\nu_{n-1}}} 
h'_{\nu_{n-1}}(t_{n-1} \phi)\;
\int_0^{t_{n-1}}\;dt_n \;t_n 
h'_\mu(t_n\phi)\; \phi^{\nu_n} 
h'_{\nu_n}(t_n\phi) \;
\times\nonumber\\ &&
\times \int_0^{t_n}\; dt_{n+2}\;\phi^{\nu_{\nu_{n+2}}} 
h'_{\nu_{n+2}}(t_{n+2} \phi)\;
\;...\;
\int_0^{t_N}\; dt_{N+1}\;\phi^{\nu_{N+1}} 
h'_{\nu_{N+1}}(t_{N+1} \phi)
\nonumber\\
&=& R_{N+1}+ih'_\mu(\phi)[V_N(\phi)-V_{N-1}(\phi)]
\nonumber \\
&& -i^{N-1} \sum_{n=1}^N 
\int_0^1\; dt_1\;\phi^{\nu_1} h'_{\nu_1}(t_1 \phi)
\;...\;
\int_0^{t_{n-1}}\; dt_n\;\phi^{\nu_n} 
h'_{\nu_n}(t_n \phi)\;t_n \;h'_\mu(t_n\phi))\; 
\;...\;
\int_0^{t_{N-1}}\; dt_N\;\phi^{\nu_N} 
h'_{\nu_N}(t_N \phi)
\nonumber\\
&& +i^{N-1} \sum_{n=1}^N 
\int_0^1\; dt_1\;\phi^{\nu_1} h'_{\nu_1}(t_1 \phi)
\;...\;
\int_0^{t_{n-1}}\;dt_n \;t_n 
h'_\mu(t_n\phi)\; \phi^{\nu_n} 
h'_{\nu_n}(t_n\phi) \;
\;...\;
\int_0^{t_{N-1}}\; dt_N\;\phi^{\nu_N} 
h'_{\nu_N}(t_N \phi)
\nonumber\\
&=& R_{N+1}+ih'_\mu(\phi)[V_N(\phi)-V_{N-1}(\phi)]-R_N
\ea
}
In the fourth step we have separated two contributions and 
the second is easily recognized as the definition of $R_{N+1}$.
The non trivial step was the fifth one where we performed an integration
by parts in the first contribution which produces two sums.
We have set $t_0=1$ in the first sum and in the second in the 
last term the integral over $t_{N+2}$ etc. is just unity. In 
the sixth step we have 
relabelled in the first sum in the n-th term $t_{n+1}\to 
t_n,..,t_{N+1}\to t_{N}$
and then reset the summation range to $n=1,..,N$.
In the second sum in the n-th term we have relabelled $t_{n+2}\to 
t_{n+1},..,t_{N+1}\to t_N$ which combines the two sums to $-R_N$.

Thus, by assumption (\ref{4.17})
\be \label{4.21}
\partial_\mu V_{N+1}=[\partial_\mu V_N-ih'_\mu V_{N-1}-R_N]
+ih'_\mu V_N+R_{N+1}
=ih'_\mu V_N+R_{N+1}
\ee
\end{proof}
Having shown that the rigging map is well defined and produces solutions 
to the constraints $F'_\mu$ we can compute the physical inner product
between states $\eta(\psi)$ defined by (we drop the factors $1/(2\pi)$ 
as the physical inner product is defined only up to a scale)
\ba \label{4.22}
<\eta(\psi),\eta(\psi')>_{{\rm phys}}
&:=& <\psi,\eta(\psi')>{{\rm kin}}
\nonumber\\
&=&\int\; [d\phi]\;\int\;[dQ]\; 
\overline{\psi(\phi,Q)}\;[\eta(\psi')](\phi,Q)
\nonumber\\
&=&\int\; [d\phi]\;\int\;[dQ]\; 
\overline{\psi(\phi,Q)}\;V(\phi)[\eta'(\psi')](Q)
\nonumber\\
&=&\int\; [d\phi]\;<\psi(\phi,.),V(\phi) \eta'(\psi')>_{L_2(dQ)}
\nonumber\\
&=&<\int\; [d\phi]\;V(\phi)^{-1}\;\psi(\phi,.),\eta'(\psi')>_{L_2(dQ)}
\nonumber\\
&=&<\eta'(\psi),\eta'(\psi')>_{L_2(dQ)}
\ea
where $\eta'(\psi)$ was defined in (\ref{4.13}).  
This calculation demonstrates that the physical Hilbert space can be 
identified with the Hilbert space ${\cal H}_{{\rm red}}:=L_2(dQ)$ which 
we obtained also in 
the reduced phase space approach. The identification is established by
\be \label{4.23} 
W:\;{\cal H}_{{\rm red}} \to {\cal H}_{{\rm phys}}\;
\Psi(Q)\mapsto V(\phi)\Psi(Q)
\ee
${\cal H}_{{\rm phys}}$ can also be 
recognised as the (closure of the) set of equivalence classes of 
vectors in ${\cal H}_{{\rm kin}}$ where $\psi\sim \psi'$ iff 
$\eta'(\psi)=\eta'(\psi')$ are the same $L_2(dQ)$ functions. Notice that 
$\eta'$ is not a projector,
$[\eta']^2$ is ill defined. 

It is worthy pointing out the importance of 
the knowledge of the map (\ref{4.23}): Often one only knows a path 
integral expression for $<\eta(\psi),\eta(\psi')>_{{\rm phys}}$ in terms 
of the boundary states $\psi,\psi'$ which, however, lack any physical
interpretation, they are not gauge invariant. The vectors $\eta(\psi)$ 
are gauge invariant, however, the path integral expression which 
we will also derive below is not in terms of $\eta(\psi)$ but in terms 
of $\psi$, $\eta(\psi)$ is often not not known explicitly. In the case 
considered here, $\eta(\psi)$ is known explicitly: Neglecting about the 
details of the domains of the maps we have $\eta({{\cal H}}_{{\rm 
kin}})=W({\cal H}_{{\rm red}})$ and since $W$ just operates by a unitary 
operator with a specific $\phi$ dependence, all the non trivial physical
information is contained in ${\cal H}_{{\rm red}}$.  

To make the link with the path integral formulation now does not require 
much further work. For any $\Psi,\Psi'\in {\cal H}_{{\rm red}}$ pick 
$\psi,\psi'\in {\cal H}_{{\rm kin}}$ with $\eta'(\psi)=\Psi,\;
\eta'(\psi')=\Psi'$. Any such $\Psi$ is generated from the cyclic 
vacuum vector $\Omega$ (a ground state vector under the time evolution, 
i.e. a stationary vector under $H_{{\rm red}}(t)$ for some fixed value 
of $t$; in the case of a conservative system, the choice of this $t$ is 
irrelevant) by 
operating 
with 
(limits of) polynomials $f$ of the  
operators $Q^A$. On the other hand, from the point of view of 
${\cal H}_{{\rm phys}}$ the operators $Q^A$ are ill defined because they 
are not gauge invariant, or in other words $Q^A \eta(\psi)$ is not 
annihilated by the $F'_\mu$. The following operators, however, preserve
${\cal H}_{{\rm phys}}$
\be \label{4.24}
\tilde{Q}^A=[\exp(i\beta^\mu F'_\mu) Q^A \exp(-i\beta^\mu 
F'\mu)]_{\beta=\phi}
\ee
which is the quantisation of the corresponding classical formula
(\ref{2.21}) upon replacing $\{F'_\mu,Q^A\}_{(n))}$ by 
$[F'_\mu,Q^A]_{(n)}/i^n$. To show that 
$[F'_\mu,\tilde{Q}^A]=0$ we notice that since 
$[\pi_\mu,Q^A]=0$ we have with the definition of $V(\beta)$ (\ref{4.4})
\ba \label{4.24a}
\tilde{Q}^A &=& 
[\exp(i\beta^\mu F'_\mu)\; e^{-i\beta^\mu \pi_\mu}\;Q^A 
e^{i\beta^\mu \pi_\mu}\;
\exp(-i\beta^\mu F'\mu)]_{\beta=\phi}
\nonumber\\
&=& V(\phi) Q^A V(\phi)^{-1} 
\ea
Notice that $\tilde{Q}^A$ is self-adjoint on ${\cal H}_{{\rm phys}}$ if 
$Q^A$ is on ${\cal H}_{{\rm red}}$.
Since any physical state is of the form $V(\phi) \Psi(Q)$ it is obvious 
that $\tilde{Q}^A$ preserves ${\cal H}_{{\rm phys}}$ by theorem 
\ref{th4.1}. We conclude
\be \label{4.25}
<\Psi,\Psi'>_{{\rm red}}=<\Omega,f(Q)^\dagger\; 
f'(Q)\Omega>_{{\rm red}}= 
<W\Psi,W\Psi'>_{{\rm phys}}= 
<W\Omega,f(\tilde{Q})^\dagger f'(\tilde{Q}) W\Omega>_{{\rm phys}}
\ee
We see that the physical scalar product can be directly related to the  
reduced Hilbert space inner product. Now we just need to relate the 
latter to the n-point functions already derived in the previous section.
But this is easy: Evidently (\ref{4.25}) is a finite linear combination 
of monomials of the form 
\be \label{4.26}
<\Omega,Q^{A_1}\;...\;Q^{A_n}\Omega>_{{\rm red}}
\ee
which is the coincidence limit of an n-point function 
\be \label{4.27}
\lim_{t_1,..,t_n\to t} 
<\Omega,Q^{A_1}(t_1)\;...\;Q^{A_n}(t_n)\Omega>_{{\rm 
red}}
\ee
for arbitrary $t$. In interacting Wightman QFT's it is expected that 
such equal time correlators are too singular \cite{31}. On the other 
hand, 
if the theory can be canonically quantised at all then such limits 
must exist as otherwise the notion of equal time commutation relations 
is meaningless and therefore presumably violates at 
least one of the 
Wightman axioms, e.g. the uniqueness of the vacuum. In any case, we 
derived a path integral formula for the 
right hand side of (\ref{4.27}) in terms of a path integral for the 
generating functional. 

There is also a more direct derivation for a path integral formula 
for $<\eta(\psi),\eta(\psi')>_{{\rm phys}}$ for which, however, the 
relation to the reduced phase space path integral is less clear.
On the other hand that alternative derivation makes the connection to 
the Master constraint path integral clearer. We will thus display it 
here for completeness. We start from the definition of the rigging map 
(\ref{4.1}), choose some arbitrary but fixed reference vector 
$\Omega_0$ and normalise the physical inner product by asking that the 
norm of $\eta(\psi)$ be unity. Thus we have to divide (\ref{4.1}) by a 
constant up to which the inner product is anyway undetermined and obtain
\be \label{4.28}
<\eta(\psi),\eta(\psi')>_{{\rm phys}}=
\frac{
\int\;[d\beta]\;<\psi,e^{i\beta^\mu F'_\mu}\psi'>_{{\rm kin}}
} 
{
\int\;[d\beta]\;<\Omega_0,e^{i\beta^\mu F'_\mu}\Omega_0>_{{\rm kin}}
}
\ee
Notice that (\ref{4.28}) is not a path integral over $\beta$, it is just
an integral at fixed time of the Lagrange multipliers $\beta^\mu$. In 
order to introduce a path integral of Lagrange multipliers we introduce 
an arbitrary time parameter $T$ which we will eventually send to 
$\infty$ and multiply both numerator and denominator of (\ref{4.28}) by 
the infinite constant
\be \label{4.29}
C=\int\;[D\lambda]\;\prod_\mu\;\delta(\int_{-T}^T\;dt \lambda^\mu(t))
\ee
which is a path integral over paths $t\mapsto \lambda(t),\;t\in[-T,T]$.
By shifting the integration variable 
$\lambda(t)=\lambda'(t)-\frac{1}{2T}\beta$ for any constant path 
$\beta/(2T)$ we see that $C$ can also be written
\be \label{4.30}
C=\int\;[D\lambda]\;\prod_\mu\;\delta(\int_{-T}^T\; dt 
\lambda^\mu(t)-\beta^\mu)
\ee
where $\beta$ is arbitrary. Inserting this into (\ref{4.28}) and 
interchanging the $[D\lambda],[d\beta]$ integrals we obtain
\ba \label{4.31}
<\eta(\psi),\eta(\psi')>_{{\rm phys}} 
&=&
\frac{
\int\;[d\beta]\;<\psi,e^{i\beta^\mu F'_\mu}\psi'>_{{\rm kin}}\;
[\int\;[D\lambda]\;\prod_\mu\;\delta(\int_{-T}^T\; dt 
\lambda^\mu(t)-\beta^\mu)]
} 
{
\int\;[d\beta]\;<\Omega_0,e^{i\beta^\mu F'_\mu}\Omega_0>_{{\rm kin}}\;
[\int\;[D\lambda]\;\prod_\mu\;\delta(\int_{-T}^T\; dt 
\lambda^\mu(t)-\beta^\mu)]
}
\nonumber\\
&=&
\frac{
\int\;[D\lambda]\;
\;<\psi,e^{i[\int_{-T}^T\;dt \lambda^\mu(t)] F'_\mu}\psi'>_{{\rm kin}}\;
[\int\;[d\beta]\;\prod_\mu\;\delta(\int_{-T}^T\;dt 
\lambda^\mu(t)-\beta^\mu)]
} 
{
\int\;[D\lambda]\;
\;<\Omega_0,e^{i[\int_{-T}^T\;dt \lambda^\mu(t)] F'_\mu}\Omega_0>_{{\rm 
kin}}\;
[\int\;[d\beta]\;\prod_\mu\;\delta(\int_{-T}^T\;dt 
\lambda^\mu(t)-\beta^\mu)]
}
\nonumber\\
&=&
\frac{
\int\;[D\lambda]\;
\;<\psi,e^{i[\int_{-T}^T\;dt \lambda^\mu(t)] F'_\mu}\psi'>_{{\rm kin}}\;
} 
{
\int\;[D\lambda]\;
\;<\Omega_0,e^{i[\int_{-T}^T\;dt \lambda^\mu(t)] F'_\mu}\Omega_0>_{{\rm 
kin}}\;
}
\ea
By writing 
\be \label{4.32}
[\int_{-T}^T\;dt \lambda^\mu(t)] F'_\mu=
\lim{N\to \infty}\frac{1}{2N}\sum_{n=-N}^{N-1}\lambda^\mu(nT/N) F'_\mu
\ee
we finally obtain using the usual skeletonisation techniques
\be \label{4.33}
<\eta(\psi),\eta(\psi')>=
\frac{
\int\;[DQ\;DP\;D\phi\;D\pi\;D\lambda]\;
\overline{\psi(Q_T,\phi_T)}\;\psi'(Q_{-T},\phi_{-T})\;
e^{i\int_{-T}^T\;dt[P_A\;\dot{Q}^A+\pi_\mu\;\dot{\phi}^\mu
-\lambda^\mu F'_\mu(Q,P,\phi,\pi)]}
}
{
\int\;[DQ\;DP\;D\phi\;D\pi\;D\lambda]\;
\overline{\psi(Q_T,\phi_T)}\;\psi'(Q_{-T},\phi_{-T})\;
e^{i\int_{-T}^T\;dt[P_A\;\dot{Q}^A+\pi_\mu\;\dot{\phi}^\mu
-\lambda^\mu F'_\mu(Q,P,\phi,\pi)]}
}
\ee
Notice that all canonical coordinates and Lagrange multipliers are 
integrated over paths in time
in the interval $[-T,T]$ and that the operator $F'_\mu$ has been 
replaced by the classical function in (\ref{4.32}). In this expression 
the parameter $T$ is arbitrary and we can take $T\to 
\infty$.

In order to invoke the gauge fixing conditions 
$G_\mu=\tau^\mu(t)-\phi^\mu$ we will make use 
of the 
Fadeev -- Popov procedure: Let 
$\alpha_{\gamma(t)}=\exp(\gamma^\mu(t)\{F'_\mu(t),.\})$ where 
$F'_\mu(t)=F'_\mu(u(t))$ is the constraint on the copy of the phase 
space at time $t$ and $u(t)=(Q(t),P(t),\phi(t),\pi(t))$. Then 
$\alpha_{\gamma(t)}(G^\mu(t))=G^\mu(t)+\gamma^\mu(t)$ so that 
in this case trivially
\be \label{4.34}
\int\;[D\gamma]\;\prod_{t,\mu}\; \delta(\alpha_{\gamma(t)}(G^\mu(t))=1
\ee
We multiply both numerator and denominator of (\ref{4.33}) by this 
unity.
Assuming $\lim_{T\to \infty} \gamma(\pm T)=0$ the kinetic term in the 
exponential of (\ref{4.33}) is invariant (being a symplectic potential),
$F'_\mu$ is invariant due to the Abelianess and the Liouville measure 
at time $t$ is 
invariant under the canonical transformations $\alpha_\gamma(t)$. Thus 
after a change of variables from $u\to \alpha_\gamma(u)$
and since $[\alpha_\gamma(G)](u)=G(\alpha_\gamma(u))$ 
nothing depends on $\gamma$ anymore and the integral over $D\gamma$ can 
be dropped. We obtain
\be \label{4.35}
<\eta(\psi),\eta(\psi')>=
\frac{
\int\;[DQ\;DP\;D\phi\;D\pi]\;
\overline{\psi(Q_T,\phi_T)}\;\psi'(Q_{-T},\phi_{-T})\;\delta[F']\;\delta[G]\;
e^{i\int_{-T}^T\;dt[P_A\;\dot{Q}^A+\pi_\mu\;\dot{\phi}^\mu}]
}
{
\overline{\Omega_0(Q_T,\phi_T)}\;\Omega_0(Q_{-T},\phi_{-T})
\;\delta[F']\;\delta[G]\;
e^{i\int_{-T}^T\;dt[P_A\;\dot{Q}^A+\pi_\mu\;\dot{\phi}^\mu}]
}
\ee
where we have also integrated over $\lambda$. 

Finally, in order to invoke the second class constraints in the form
$z_\Sigma=(x^\sigma,y_\sigma)=0$ we simply insert a $\delta$ 
distribution $\delta[z]$ and integrate over $z$. This yields
\be \label{4.36}
<\eta(\psi),\eta(\psi')>=
\frac{
\int\;[DQ\;DP\;D\phi\;D\pi\;Dx\;Dy]\;
\overline{\psi(Q_T,\phi_T)}\;\psi'(Q_{-T},\phi_{-T})\;\delta[F']\;\delta[G]\;
\delta[z]\;
e^{i\int_{-T}^T\;dt[P_A\;\dot{Q}^A+\pi_\mu\;\dot{\phi}^\mu
+y_\sigma\dot{x}^\sigma]}
}
{
\int\;[DQ\;DP\;D\phi\;D\pi\;Dx\;Dy]\;
\overline{\Omega_0(Q_T,\phi_T)}\;\Omega_0(Q_{-T},\phi_{-T})\;
\delta[F']\;\delta[G]\;
\delta[z]\;
e^{i\int_{-T}^T\;dt[P_A\;\dot{Q}^A+\pi_\mu\;\dot{\phi}^\mu
+y_\sigma\dot{x}^\sigma]}
}
\ee
Next we observe that $\det(\{F',G\}),\det(\{z,z\})$ are constant in 
the system of coordinates chosen so we can multiply numerator and 
denominator of (\ref{4.36}) by these constants. As established in 
section \ref{s3},
the expression
\be \label{4.37}
\delta[F']\;\delta[G]\;\delta[z]\;|\det(\{F',G\})|\;\sqrt{\det(\{z,z\}}
\ee
is invariant under any mapping $(F',G,z)\mapsto (F,G',S)$ as long as 
both triples reduce to the same gauge cut of the same constraint 
surface. We may therefore restore the original first and second 
class constraints $F,S$ while keeping $G=G'$ provided we keep the 
determinant factors in (\ref{4.37}). Finally we can restore the original 
system of coordinates $q^a,p_a$ which arise from 
$(q^{\prime a},p'_a):=(Q^A,P_A),(\phi^\mu,\pi_\mu),(x^\sigma,y_\sigma)$ 
by a canonical 
transformation $\alpha$ because the symplectic potential in the exponent 
of 
(\ref{4.36}) as well as the Liouville measure remain invariant 
and the the Poisson brackets are simply expressed in the new 
coordinates, e.g. 
\be \label{4.38}
\{S,S\}(q',p')=\{S,S\}(\alpha(q,p))=\{S\circ\alpha,S\circ \alpha\}(q,p)
\ee
(by $S$ we denote the original $S$ expressed in whatever canonical 
coordinates). Accordingly
\be \label{4.39}
<\eta(\psi),\eta(\psi')>=
\frac{
\int\;[Dq\;Dp]\;
\overline{\psi(Q_T,\phi_T)}\;\psi'(Q_{-T},\phi_{-T})\;\delta[F]\;\delta[G]\;
\delta[S]\;|\det(\{F,G\})|\;\sqrt{\det(\{S,S\})}\;
e^{i\int_{-T}^T\;dt\;p_a \dot{q}^a}
}
{
\int\;[Dq\;Dp]\;
\overline{\Omega_0(Q_T,\phi_T)}\;\Omega_0(Q_{-T},\phi_{-T})\;
\delta[F]\;\delta[G]\;
\delta[S]\;|\det(\{F,G\})|\;\sqrt{\det(\{S,S\})}\;
e^{i\int_{-T}^T\;dt\;p_a \dot{q}^a}
}
\ee
Notice that due to the gauge fixing condition $G(t)=\tau(t)-\phi(t)$ the 
integral over $\phi$ is anyway concentrated at the fixed path $\tau(t)$
so that it is allowed to assume that $\psi,\psi',\Omega$ are actually 
independent of $\phi$. In this sense the final result (\ref{4.39}) 
precisely agrees with (\ref{3.54a}) with the understanding that 
$\psi,\psi'$ in (\ref{4.39}) can be generated from the generating 
functional (\ref{3.54a}) by suitable functional differentiation with 
respect to the current $j$ at $j=0$ at coincident points of time $\pm T$ 
in the limit $T\to \infty$.

\section{Master Constraint Path Integral}
\label{s5}

The Master Constraint Programme (MCP) was originally designed precisely
in order to be able to cope with gauge systems whose classical first 
class 
constraint algebra involves structure functions \cite{12} and for which 
therefore group averaging techniques do not work. It is true 
that locally the first class constraints $F$ can be replaced by 
equivalent ones whose algebra is Abelian and we have made heavy use of 
that fact in the two previous sections. However, for the case of 
interest, namely General Relativity, in vacuum the Abelian constraints 
are rather non local on the spatial manifold, algebraically difficult to 
deal with and 
not explicitly known even classically \cite{41}. Even with standard 
matter this is true. It is for this reason that in \cite{27,28} non 
standard matter (Brown -- Kucha{\v r} Dust \cite{42}) was used in order 
to achieve the Abelianisation in a local form and such that the 
resulting expressions remain practically managable. The MCP does not 
rely on Abelianisation and thus is both more global (on phase space) in 
character and 
does not require any special type of matter. In principle it does not 
even require that the constraints are quantised without anomalies and 
even second class constraints can be treated by the MCP \cite{12}.
Since the Master Constraint is a weighted sum of squares of the first 
class constraints, we expect that its kernel is empty when the 
constraints are not quantised without anomalies. In that case one could 
consider the Hilbert ``subspace'' corresponding to the lowest 
``eigenvalue'' as the suitable substitute for the anomaly free 
situation. See \cite{12} for further discussion. In that sense the MCP 
may be considered as a much more flexible approach to constrained 
systems with structure functions.

While for a wide range of models the MCP has been tested versus the 
more traditional operator constraint method \cite{12}, its equivalence 
with the latter is so far lacking. On the one hand, the equivalence 
seems to be obvious since both the Master constraint and the individual 
constraints are supposed to define the same (common) kernel. On the 
other hand, the equivalence is rather not obvious because the formulae 
for defining the physical inner product or equivalently the rigging 
map are totally different. For the individual constraints in Abelianised 
form the rigging map is defined in (\ref{4.1}) while for the MCP it is 
heuristically defined by\footnote{Again there are subtle domain issues 
which we neglect here and 
moreover one should switch to to a direct integral representation of 
${\cal H}_{{\rm kin}}$ subordinate to $M$; see \cite{12} for details.}
\be \label{5.1}
\eta_M:\;{\cal H}_{{\rm kin}}\to {\cal H}^M_{{\rm phys}};\;
\psi\mapsto\int_{\mathbb{R}}\;\frac{dt}{2\pi}\;e^{itM}\psi
\ee
where the Master constraint is defined by
\be \label{5.2}
M=\sum_{\mu,\nu}\; F_\mu^\dagger\; K_{\mu\nu} \; F_\nu
\ee
The symmetric (possibly operator valued) matrix $K$ should be so chosen 
such 
that $M$ is positive and such that it arises from a classically positive 
definite matrix valued function on phase space. There is great 
flexibility in the choice of $K$ and while all (sufficiently 
differentiable) positive 
definite classical matrices are equivalent, in quantum theory this 
flexibility must be exploited in order to arrive at well defined master 
Constraint Operators \cite{12}. Normally we require that $F_\mu$ is 
quantised as a self adjoint operator but in the case of structure 
functions this must be relaxed \cite{12} which is why we included the 
adjoint in (\ref{5.2}).

The task of the present section is to connect with the results of the 
previous two sections. Those sections made use of the Abelian 
constraints $F'_\mu$ and we will therefore use those in order to build 
our Master Constraint. We assume as in sections \ref{s3} and \ref{s4} 
that $F'_\mu$ is self -- adjoint since the $F'_\mu$ are supposed to be 
quantised without anomalies. As in the previous section we choose a 
reference vector $\Omega_0$ and define the Master Constraint physical 
inner product by
\be \label{5.3}
<\eta_M(\psi),\eta_M(\psi')>^M_{{\rm phys}}:=
\frac{
\int_{\mathbb{R}}\; dt\; <\psi,e^{it M} \psi'>_{{\rm kin}}
}
{
\int_{\mathbb{R}}\; dt\; <\Omega_0,e^{it M} \Omega_0>_{{\rm kin}}
}
\ee
To see that (\ref{4.1}) and (\ref{5.3}) formally coincide, recall 
\cite{4} that 
for any self -- adoint operator $A$ on a (separable\footnote{In LQG the 
Hilbert space is not separable but the operator M preserves the 
separable subspaces into which the Hilbert space decomposes}) Hilbert 
space $\cal 
$ there 
exists a unitary transformation (generalised Fourier transform)
\be \label{5.4}
U:\;{\cal H}\to {\cal H}^\oplus:=\int_{{\rm 
spec}(A)}\;d\mu(\lambda)\;{\cal H}^\oplus_\lambda;\;\;\psi\mapsto 
(\tilde{\psi}(\lambda))_{\lambda\in {\rm spec}(A)}
\ee
from $\cal H$ to a direct integral of Hilbert spaces ${\cal 
H}^\oplus_\lambda$ (possibly with different dimensions for each 
$\lambda$ but in a maeasurable way, hence more 
general than a Hilbert bundle) with 
respect to a probability measure $\mu$ on the spectrum spec$(A)$ of $A$.
Here $\tilde{\psi}(\lambda)\in {\cal H}^\oplus_\lambda$.
The correspondence between the inner products is 
\be \label{5.5}
<\psi,\psi'>_{{\cal H}}=<\tilde{\psi},\tilde{\psi}'>_{{\cal H}^\oplus}
:=\int_{{\rm 
spec}(A)}\;d\mu(\lambda)\;
<\tilde{\psi}(\lambda),\tilde{\psi'}(\lambda)>_{{\cal H}^\oplus_\lambda} 
\ee
The point of this spectral decomposition is that 
$[UAU^{-1}\tilde{\psi}](\lambda)=\lambda\tilde{\psi}(\lambda)$, i.e. 
$A$ acts by multipliction by $\lambda$ on ${\cal H}^\oplus_\lambda$
If (the spectral projections of) two self -- adjoint operators 
$A,B$ commute 
then $UBU^{-1}$ preserves ${\cal H}^\oplus_\lambda$ and we may apply 
the just quoted theorem which then tells us that there exists a joint 
probability measure $d\mu(\lambda_A,\lambda_B)$ on the joint spectum 
${\rm spec}(\{A,B\})={\rm spec}(A)\times{\rm spec}(B)$ of 
$A,B$ and a representation of $\cal H$ as a direct integral of Hilbert 
spaces ${\cal H}^\oplus_{\lambda_A,\lambda_B}$ on which $A,B$ 
respectively act by multiplication by $\lambda_A,\lambda_B$ 
respectively. 

Iterating like that we obtain the statement that for a 
(countable) family of mutually commuting self -- adjoint operators 
$F'_\mu$ there exists a unitary operator $U$ from ${\cal H}_{{\rm 
kin}}$ 
to ${\cal H}^\oplus$ which is the direct integral with respect to a 
measure $\mu$ on the joint spectrum of the $F'_\mu$ of Hilbert spaces 
${\cal H}^\oplus_{\{\lambda_\mu\}_\mu}$ on which $U F'_\mu U^{-1}$ acts 
by multiplication by $\lambda_\mu$. This is the key to link (\ref{5.3}) 
and (\ref{4.1}). Namely we formally obtain for (\ref{4.1})
\ba \label{5.6}
<\eta(\psi),\eta(\psi')>_{{\rm phys}}
&=& 
\frac{
\int\;[d\beta]\; <\psi,e^{i\beta^\mu F'_\mu}\;\psi'>_{{\rm kin}} 
}
{
\int\;[d\beta]\; <\Omega_0,e^{i\beta^\mu F'_\mu}\;\Omega_0>_{{\rm kin}} 
}
\nonumber\\
&=& 
\frac{
\int_{{\rm spec}(\{F'\})}\;d\mu(\{\lambda\})\;
<\tilde{\psi}(\{\lambda\}),\tilde{\psi'}(\{\lambda\})>_{{\cal 
H}^\oplus_{\{\lambda\}}}\;
[\int\;[d\beta]\; e^{i\beta^\mu \lambda_\mu}]
}
{
\int_{{\rm spec}(\{F'\})}\;d\mu(\{\lambda\})\;
<\tilde{\Omega_0}(\{\lambda\}),\tilde{\Omega_0}(\{\lambda\})>_{{\cal 
H}^\oplus_{\{\lambda\}}}\;
[\int\;[d\beta]\; e^{i\beta^\mu \lambda_\mu}]
}
\nonumber\\
&=& 
\frac{
\int_{{\rm spec}(\{F'\})}\;d\mu(\{\lambda\})\;
<\tilde{\psi}(\{\lambda\}),\tilde{\psi'}(\{\lambda\})>_{{\cal 
H}^\oplus_{\{\lambda\}}}\;\delta(\{\lambda\})
}
{
\int_{{\rm spec}(\{F'\})}\;d\mu(\{\lambda\})\;
<\tilde{\Omega_0}(\{\lambda\}),\tilde{\Omega_0}(\{\lambda\})>_{{\cal 
H}^\oplus_{\{\lambda\}}}\;\delta(\{\lambda\})
}
\nonumber\\
&=& 
\frac{
\rho(\{0\})\;
<\tilde{\psi}(\{0\}),\tilde{\psi'}(\{0\})>_{{\cal 
H}^\oplus_{\{0\}}}
}
{
\rho(\{0\})\;
<\tilde{\Omega_0}(\{0\}),\tilde{\Omega_0}(\{0\})>_{{\cal 
H}^\oplus_{\{0\}}}
}
\nonumber\\
&=& 
\frac{
<\tilde{\psi}(\{0\}),\tilde{\psi'}(\{0\})>_{{\cal 
H}^\oplus_{\{0\}}}
}
{
<\tilde{\Omega_0}(\{0\}),\tilde{\Omega_0}(\{0\})>_{{\cal 
H}^\oplus_{\{0\}}}
}
\ea
where formally $d\mu(\{\lambda\})=:\rho(\{\lambda\})[d\lambda]$. Notice 
that $\rho(\{\lambda\})$ can have distributional contributions if 
the spectrum has a pure point part, see \cite{12,19}.
Of course there are measure theoretic issues such as: if $0$ lies in the 
continuous spectrum of some $F'_\mu$ then $\{0\}$ has $\mu$ measure zero 
and ${\cal H}^\oplus_{\{0\}}$ is not well defined without further 
assumptions spelled out in \cite{12}. For the purposes of this paper 
we take a formal attitude and simply let the formal cancellation of the 
$\rho(\{0\})$ in numerator and denominator of (\ref{5.6}) take place as 
indicated. For a more careful definition see \cite{19}. 

On the other hand we have 
\ba \label{5.7}
<\eta_M(\psi),\eta_M(\psi')>^M_{{\rm phys}}
&=& 
\frac{
\int\;dt\; <\psi,e^{i tM}\;\psi'>_{{\rm kin}} 
}
{
\int\;dt\; <\Omega_0,e^{i t M}\;\Omega_0>_{{\rm kin}} 
}
\nonumber\\
&=& 
\frac{
\int_{{\rm spec}(\{F'\})}\;d\mu(\{\lambda\})\;
<\tilde{\psi}(\{\lambda\}),\tilde{\psi'}(\{\lambda\})>_{{\cal 
H}^\oplus_{\{\lambda\}}}\;
[\int\;dt\; e^{it\sum_{\mu,\nu} K^{\mu\nu}\lambda_\mu \lambda_\nu}]
}
{
\int_{{\rm spec}(\{F'\})}\;d\mu(\{\lambda\})\;
<\tilde{\Omega_0}(\{\lambda\}),\tilde{\Omega_0}(\{\lambda\})>_{{\cal 
H}^\oplus_{\{\lambda\}}}\;
[\int\;dt\; e^{it\sum_{\mu,\nu} K^{\mu\nu}\lambda_\mu \lambda_\nu}]
}
\nonumber\\
&=& 
\frac{
\int_{{\rm spec}(\{F'\})}\;d\mu(\{\lambda\})\;
<\tilde{\psi}(\{\lambda\}),\tilde{\psi'}(\{\lambda\})>_{{\cal 
H}^\oplus_{\{\lambda\}}}\;
\delta(\sum_{\mu,\nu} K^{\mu\nu}\lambda_\mu \lambda_\nu)
}
{
\int_{{\rm spec}(\{F'\})}\;d\mu(\{\lambda\})\;
<\tilde{\Omega_0}(\{\lambda\}),\tilde{\Omega_0}(\{\lambda\})>_{{\cal 
H}^\oplus_{\{\lambda\}}}\;
\delta(\sum_{\mu,\nu} K^{\mu\nu}\lambda_\mu \lambda_\nu)
}
\nonumber\\
&=& 
\frac{\rho(\{0\})\;J(\{0\})\;{\rm Vol}(S)\;
<\tilde{\psi}(\{0\}),\tilde{\psi'}(\{0\})>_{{\cal 
H}^\oplus_{\{0\}}}\;
}
{
\rho(\{0\})\;J(\{0\})\;{\rm Vol}(S)\;
<\tilde{\Omega}_0(\{0\}),\tilde{\Omega}_0(\{0\})>_{{\cal 
H}^\oplus_{\{0\}}}\;
}
\nonumber\\
&=& 
\frac{
<\tilde{\psi}(\{0\}),\tilde{\psi'}(\{0\})>_{{\cal 
H}^\oplus_{\{0\}}}\;
}
{
<\tilde{\Omega}_0(\{0\}),\tilde{\Omega}_0(\{0\})>_{{\cal 
H}^\oplus_{\{0\}}}\;
}
\ea
where $J(\{\lambda\})$ is the Jacobian that arises by switching from 
$\{\lambda\}$ to polar coordinates adapted to the radius squared
$r^2:=\sum_{\mu,\nu} K^{\mu\nu}\lambda_\mu \lambda_\nu$. Of course we 
have assumed that $K_{\mu\nu}$ is just a complex valued positive 
definite matrix. Vol$(S)$ is the volume of the corresponding sphere.
Fo countably many $F'_\mu$ the volume of the infinite dimensional sphere
vanishes as well as the Jacobian at zero. To justify (\ref{5.7}) less 
formally one has to take a limit as the number $N$ of $F'_\mu$ 
approaches 
infinity so that Vol$(S^{N-1})$ is finite and one also has to regularise 
$\delta(M)$ 
by 
$\delta(M-\epsilon^2)$ and take $\epsilon\to 0$ as to make 
$J(\epsilon)$ finite. See \cite{19} for the details and also
(\ref{5.10}) below for a sketch.\\
\\
Hence (\ref{5.6}) and (\ref{5.7}) agree with each other modulo formal 
manipulations and thus give rise to the same path integral formulation.
Our method of ``proof'' above used spectral theory. We will now 
provide a more direct (but also formal) ``proof'' using 
only path integral techniques. The idea is the same as at the end of 
section \ref{s4} and was already sketched in \cite{12}. First of all we 
use the same technique as used between (\ref{4.28}) and (\ref{4.33}) in 
order to write (\ref{5.3}) as 
\ba \label{5.8}
&& <\eta_M(\psi),\eta_M(\psi')>^M_{{\rm phys}}
=
\frac{
\int[D\lambda]\;
<\psi,e^{i[\int_{-T}^T\; dt\lambda(t)] M} \psi'>_{{\rm kin}}
}
{
\int[D\lambda]\;
<\Omega_0,e^{i[\int_{-T}^T\; dt\lambda(t)] M} \Omega_0>_{{\rm kin}}
}
\nonumber\\
&=&
\frac{
\int[DQ\;DP\;D\phi\;D\pi\;D\lambda]\;\overline{\psi(Q_T,\phi_T)}\;
\psi'(Q_{-T},\phi_{-T})\;
e^{i\int_{-T}^T\; dt[(P_A\dot{Q}^A+\pi_\mu\dot{\phi}^\mu)(t)-
\lambda(t) M(Q(t),P(t),\phi(t),\pi(t)]} 
}
{
\int[DQ\;DP\;D\phi\;D\pi\;D\lambda]\;\overline{\Omega_0(Q_T,\phi_T)}\;
\Omega_0(Q_{-T},\phi_{-T})\;
e^{i\int_{-T}^T\; dt[(P_A\dot{Q}^A+\pi_\mu\dot{\phi}^\mu)(t)-
\lambda(t) M(Q(t),P(t),\phi(t),\pi(t)]} 
}
\nonumber\\
&=&
\frac{
\int[DQ\;DP\;D\phi\;D\pi]\;\overline{\psi(Q_T,\phi_T)}\;
\psi'(Q_{-T},\phi_{-T})\;
[\prod_{t\in [-T,T]}\;\delta((M(t))]\;
e^{i\int_{-T}^T\; dt[(P_A\dot{Q}^A+\pi_\mu\dot{\phi}^\mu)(t)]}
}
{
\int[DQ\;DP\;D\phi\;D\pi]\;\overline{\Omega_0(Q_T,\phi_T)}\;
\Omega_0(Q_{-T},\phi_{-T})\;
[\prod_{t\in [-T,T]}\;\delta((M(t))]\;
e^{i\int_{-T}^T\; dt[(P_A\dot{Q}^A+\pi_\mu\dot{\phi}^\mu)(t)]}
}
\ea
where $T$ is again an arbitrary parameter which we take to $\infty$ 
eventually. If in (\ref{4.33}) we perform the integral over 
$\lambda$ then the only difference between (\ref{4.33}) and 
(\ref{5.8}) is that instead of $\delta[F']$ the distribution 
$\delta[M]$ appears in both numerator and denominator.
But clearly the two distributions have the same support 
$\pi=-h'(\phi,Q,P)$. Let us therefore explicitly do the integral
in both (\ref{4.33}) and (\ref{5.8}) and compare the results. It 
suffices to do this at fixed $t$ because both $\delta$ distributions 
factorise over $[-T,T]$. We consider $\delta(M)$
as the limit $N\to\infty,\epsilon\to 0$ of  
\be \label{5.9}
\delta_{N,\epsilon}(M):=\delta(\sum_{\mu,\nu\le N}\; K^{\mu\nu} 
\;F'_\mu\;F'_\nu -\epsilon^2)
\ee 
Let $f=f[\pi]$ be any functional of $\pi_\mu,\; \mu=1,..,N$. The 
$N\times N$ submatrix $K^{\mu\nu}_N=K^{\mu\nu};\;\mu,\nu\le N$ is also
positive definite on the corresponding vector subspace. Hence its square 
root and inverse is well defined. Thus, by shifting the integration 
variable and switching to radial $r$ and polar coordinates $\varphi$ 
respectively we obtain with the unit vector $x_\mu/r=n_\mu(\varphi)$
\ba \label{5.10}
&& 
\int_{\mathbb{R}^N}\; d^N\pi\; \delta_{N,\epsilon}(M)\;f(\pi)
\nonumber\\
&=& 
\int_{\mathbb{R}^N}\; d^Nx\; \delta(x^T\; K_N\; x-\epsilon^2)\;f(-h'+x)
\nonumber\\
&=& 
\frac{1}{\sqrt{\det(K_N)}}\int_{\mathbb{R}^N}\; d^Nx\; 
\delta(x^T\;x-\epsilon^2)\;
f(-h'+K_N^{-1/2}x)
\nonumber\\
&=& 
\frac{1}{\sqrt{\det(K_N)}}\int_{\mathbb{R}_+}\; r^{N-1}\; dr\; 
\delta(r^2-\epsilon^2)\;\int_{S^{N-1}}\;d{\rm Vol}(\varphi)\;
f(-h'+K_N^{-1/2}r n(\varphi))
\nonumber\\
&=& 
\frac{\epsilon^{N-2}}{2\sqrt{\det(K_N)}} 
\;\int_{S^{N-1}}\;d{\rm Vol}(\varphi)\;
f(-h'+K_N^{-1/2}\epsilon n(\varphi))
\ea
In the limit $\epsilon\to 0$ this approaches 
\be \label{5.11}
\frac{\epsilon^{N-2}}{2\sqrt{\det(K_N)}} \;{\rm Vol}(S^{N-1})\;
f(-h')
\ee
and in that sense we may write 
\be \label{5.12}
\delta_{N,\epsilon}(M)=
\frac{\epsilon^{N-2}}{2\sqrt{\det(K_N)}} \;{\rm Vol}(S^{N-1})\;
\delta_N(F'),\;\;\delta_N(F')=\prod_{\mu\le N}\; \delta(F'_\mu)
\ee
Since $K_N$ is a phase space independent constant, when inserting 
(\ref{5.12}) into (\ref{5.8}), the prefactor cancels in both numerator 
and denominator and we arrive at (\ref{4.33}) in the limit $\epsilon\to 
0$ and $N\to \infty$.

\section{Conclusions and Outlook}
\label{s6}

The three tasks accomplished in the present paper are:
\begin{itemize}
\item[1.] We have demonstrated that within the limits of the formal
nature of the manipulations that are usually employed when dealing with 
path integrals, three canonical quantisation methods, namely the 
reduced phase space --, the operator constraint and the Master 
Constraint quantisation all lead to the same path integral formulation
for the physical inner product. In order that rigging map techniques 
can be employed to the operator constraint approach, in the case of 
structure functions one has to pass to Abelianised constrants.
\item[2.] The resulting path integral can be written in terms of the 
classical Lagrangian from which the classical theory descends. However,
the correct measure to be used is not the naive Lebesgue measure 
on path space, rather this measure must be corrected by factors that
depend on the first and second class constraints as well as the 
gauge fixing condition.
\item[3.] The gauge fixing condition is in one to one correspondence 
with the choice and interpretation of a convenient choice of an algebra 
of physical observables and a physical Hamiltonian. It is possible 
to do without gauge fixing conditions provided one finds alternative 
methods to construct an algebra of Dirac observables. However, the 
resulting algebra is almost surely algebraically more complicated, more 
difficult to quantise, lacks an a priori physical interpretation and 
is not equipped with a preferred physical time evolution. In particular,
if one wants to talk about the scattering matrix between physical 
states, the dependence on the gauge fixing is unavoidable because it 
determines the physical time evolution of the chosen ``basis'' of gauge 
invariant operators.
\end{itemize}
As we have already stated in the introduction, certainly not all the 
results and 
techniques derived and used in the 
present paper are new, bits and pieces of it 
are already 
in the literature. However, we believe we have assembeled the material
in a new and fruitful way in order to better understand the relations 
between the four quantisation methods discussed in this paper. Also 
we think that the mathematical and physical influence of the gauge 
fixing condition hasbeen described in this paper from a new angle.

As we have seen explicitly, both methods of proof in section 
\ref{s5} actually 
relied on the 
fact that the matrix $K$ is a constant function on phase space. However,
this is not the case for the concrete Master Constraint for General 
Relativity
studied in \cite{12}. Namely, there one considered an expression of the 
form \be \label{5.13}
M=\int_\sigma\;d^3x\; \frac{C^2}{\sqrt{\det(q)}}
\ee
where $C$ is the Hamiltonian constraint and $q$ is the intrinsic three
metric of the hypersurface $\sigma$. The ``matrix'' 
$K(x,y)=\delta(x,y)/\sqrt{\det(q)(x)}$ is chosen here in order to make 
(\ref{5.13}) invariant under spatial diffeomorphisms and is clearly 
a non -- trivial function on phase space. In view of the analysis of the 
previous section, rather than the Hamiltonian constraint in its original 
form $C$, in the presence of the dust matter one would choose it in the 
locally equivalent form $C'(x)=\pi(x)+h'(q(x),P(x))$ where $q,P$ are the 
gravitational degrees of freedom and $\pi$ is one of the dust momenta. 
Notice that for this type of matter $h'$ does not depend on the dust
configuration fields $\phi$ and therefore dust deparametrises the 
system and leads to a conserved physical Hamiltonian.
However, also $C'$ is a scalar density and thus to make the 
corresponding Master constraint spatially diffeomorphism invariant, 
one would again need a phase space dependent matrix of the type 
considered above. Thus it appears as if the analysis of the present 
section does not apply to GR.

However, this is not the case: Namely, the dust offers the possibility
to completely abelianise the full constraint algebra including spatial 
diffeomorphisms. Thus in contrast to the usual situation in which the 
spatial diffeomorphisms form a subalgebra of the constraint algebra but 
not an ideal, it is possible to completely solve the spatial 
diffeomorphism constraint before solving the Hamiltonian constraint. 
In particular it is possible to perform a canonical transformation 
to coordinates such that $C'$ only depends on spatially diffeomorphism 
invariant fields \cite{27}. It is therefore no longer necessary to choose a 
density weight minus one matrix $K$. We can simply take an orthonormal 
basis $b_\mu$ of $L_2(\sigma, d^3x)$ and consider the 
$F'_\mu:=<b_\mu,C'>$. Then one chooses any phase space independent 
matrix $K^{\mu\nu}$ subject to certain fall off conditions in index 
space (typically $K$ should be trace class \cite{12}). The fact that
$C'$ has density weight one ensures that $C'$ can be quantised on 
the unique \cite{43} LQG Hilbert space \cite{44} as was shown 
explicitly in \cite{28}. That quantisation, however, is most probably
too naive in order guarantee anomaly freeness and must be improved.
Yet, since the anomaly is an $\hbar$ correction to the classical result,
the relation between the MCP (which also works in the anomalous case) 
and the path integral formulation derived in the previous section, 
remains correct in the semiclassical limit. An alternative to working
with $C'$ already reduced with respect to the spatial diffeomorphism 
constraint is to keep the unreduced 
$C'$ and the unreduced Abelianised spatial diffeomorphism 
constraints $C'_j$ \cite{27}. The caveat in LQG to quantising the 
classical 
generator of spatial diffeomorphisms which arises due to strong 
discontinuity of the one paranmeter unitary subgroups of spatial 
diffeomorphisms on the LQG Hilbert space 
is circumvented because $C'_j$ is not 
a density one covector but a density one scalar and thus can be 
quantised on the LQG Hilbert space \cite{46}, albeit it is difficult, 
similar to $C'$, to achieve anomaly freeness.\\
\\
This paper has been the starting point for further analysis. In 
\cite{16a}
we have computed the correct measure for the Holst action and have 
checked explicitly that it is consistent with the analysis of \cite{16}
for the Plebanski action. In \cite{19} the relation between 
the Master Constraint Programme and the operator constraint programme 
for Abelian and anomaly free constraints and with phase space 
independent matrix $K$ was analysed with higher mathematical precision 
at the level of the 
canonical theory and it is shown that under certain technical 
assumptions the two methods lead to the same result, thus partly 
removing the 
formal character of the analysis of section \ref{s5}. Finally, 
in \cite{20}
it was formally checked by using available semiclassical techniques 
\cite{45} that the Master Constraint Programme for General 
Relativity leads also to the 
expected path integral formula up to a local measure factor when one 
considers phase space dependent matrices $K$ and non -- Abelian 
constraints. However, the results here
are less strong (more formal) than in the Abelianised case.  

Many further questions arise from the present paper:\\
Since the Master Constraint can in principle also accomodate 
(sums of squares of) second class constraints if one subtracts a 
suitable normal
ordering constant \cite{12}, one could ask whether the separate 
treatment of first and second class constraints could be unified and if 
yes how the corresponding path integral would look like. Secondly,
in applications to path integral formulations of LQG one should really 
take the unavoidable measure factor derived in \cite{16a} and following 
the general theory summarised here seriously and define a corresponding 
spin foam model. Work is in progress in order to achieve that. Next, due 
to the measure factor the theory lacks 
manifest spacetime diffeomorphism invariance. On the other hand it 
should be manifestly invariant under the gauge transformations generated 
by the first class constraints which in General Relativity corresponds 
to the Bergmann -- Komar ``group'' \cite{47} (more precisely it is 
the enveloping 
algebra generated by the secondary first class constraints of GR).
The two groups are known to coincide when the classical equations of 
motion hold and this is the reason why the Lagrangian and Hamiltonian 
descriptions are equivalent classically. 
However, off shell there is no particular relation between these two
``groups'' and it is consistent with the classical theory that the 
spacetime 
diffeomophism group is not a symmetry of the quantum theory. In 
\cite{21} it is further analysed in which sense the Bergmann -- Komar 
group is a symmetry of the Hamiltonian path integral. It 
seems that the attempt to construct a spacetime covariant path integral 
of GR has no chance to be derived from a canonical platform which is 
the only systematic starting point that we have and it would 
be interesting to understand better the implications of this conclusion. 
In some sense it is clear that spacetime diffeomorphism invariance is 
far from sufficient in order to guarantee that one has a correct 
quantisation of a given classical theory. Many Lagrangians are spacetime 
diffeomorphism covariant (e.g. higher derivative theories) but all of 
them have different Hamiltonian constraints (even different numbers of 
degrees of frredom). The effect of this will show, in particular, in the 
local measure factor that we have exhibited.  \\
\\
\\
{\large Acknowledgements}\\
\\
T.T. thanks Kristina Giesel and Sergeij Alexandrov for illuminating 
discussions and comments. We also would like to thank Jonathan Engle for 
many in depth discussions.
The part of the research performed at the Perimeter Institute for 
Theoretical Physics was supported in part by funds from the Government of  
Canada through NSERC and from the Province of Ontario through MEDT.
%to the Perimeter Institute for Theoretical Physics.
%This research project was supported in part by DOE-Grant
%DE-FG02-94ER25228 to Harvard University.

%\begin{appendix}

%\end{appendix}

\end{document}